# U.S. Long-Term Earnings Outcomes by Sex, Race, Ethnicity, and Place of Birth

Kevin L. McKinney, *U.S. Census Bureau*

John M. Abowd, *U.S. Census Bureau and Cornell University*

Hubert P. Janicki, *U.S. Census Bureau*

December 6, 2021

## Abstract

This paper is part of the Global Income Dynamics Project cross-country comparison of earnings inequality, volatility, and mobility. Using data from the U.S. Census Bureau's Longitudinal Employer-Household Dynamics (LEHD) infrastructure files we produce a uniform set of earnings statistics for the U.S. From 1998 to 2019, we find U.S. earnings inequality has increased and volatility has decreased. The combination of increased inequality and reduced volatility suggest earnings growth differs substantially across different demographic groups. We explore this further by estimating 12-year average earnings for a single cohort of age 25-54 eligible workers. Differences in labor supply (hours paid and quarters worked) are found to explain almost 90% of the variation in worker earnings, although even after controlling for labor supply substantial earnings differences across demographic groups remain unexplained. Using a quantile regression approach, we estimate counterfactual earnings distributions for each demographic group. We find that at the bottom of the earnings distribution differences in characteristics such as hours paid, geographic division, industry, and education explain almost all the earnings gap, however above the median the contribution of the differences in the returns to characteristics becomes the dominant component.

Any opinions and conclusions expressed herein are those of the authors and do not represent the views of the U.S. Census Bureau or other sponsors. All results have been reviewed to ensure that no confidential information is disclosed (DRB clearance numbers CBDRB-FY21-CED002-B002, CBDRB-FY21-168, and CBDRB-FY22-049). This research uses data from the U.S. Census Bureau's Longitudinal Employer-Household Dynamics Program, which was partially supported by NSF grants SES-9978093, SES-0339191, and ITR-0427889; National Institute on Aging grant AG018854; and grants from the Alfred P. Sloan Foundation.

# I. Introduction

This paper is part of the Global Income Dynamics Project cross-country comparison of earnings inequality, volatility, and mobility. Using data from the U.S. Census Bureau's Longitudinal Employer-Household Dynamics (LEHD) infrastructure files from 1998 to 2019 we find U.S. earnings inequality has increased and volatility has decreased. Taken together, these two results suggest inequality differences are both larger and more persistent at the individual worker level post Great Recession than prior to the Great Recession, which leads into the second part of the paper where we document significant long-term real earnings differences both across and within sex, race, ethnicity, and place of birth demographic groups. For each demographic group, we follow a single cohort of eligible workers ages 25 to 54 in 2004 for 12 years. Substantial differences exist across groups. Native-born Black and Hispanic/Latino male workers earn 18% to 77% less than a similar White male would have earned over the same 12-year period (including zero-earnings years). Similarity is defined as being at the same point in the group earnings distribution.

A large body of work focused on individual lifecycle earnings dynamics prompted the development of the Global Income Dynamics project. Guvenen et al. (2021) and Altonji, Hynsjo, and Vidangos (2021) are two of the more recent papers summarizing the existing literature.[1] Quantifying earnings disparities by race and sex characterizes this research. The documentation and study of earnings disparities by race and sex has likewise been a source of considerable research. Altonji and Blank (1999) provide a thorough review of the early literature. Daly, Hobijn, and Pedtke (2017) provide a more recent summary of the basic trends for average wages. In the 20 years since the publication of Altonji and Blank, many of the headline findings remain true. The Black/White male wage gap has barely changed over the past 4 decades and the Black/White female wage gap has widened for the past 35 years.

While earlier studies focused on differences in average earnings or wages, Bayer and Charles (2018) compare earnings levels by percentile and rank in the earnings distribution among men from 1940-2014. They find that most of the historical reduction of the Black/White earnings gap at the median during the "great compression" that occurred between 1950-1970 has now been undone. That is, the Black and White earnings gap at the median is now as large as it was in the 1950s. Bayer and Charles found that Median Black male earnings were at the 27th percentile of the White distribution after the Great Recession, virtually the same as our result for 2004-2015, and at the 24th percentile in 1940.

---

[1] We focus our discussion on the U.S., but the cross-country nature of the Global Income Dynamics project allows for a comparison to other countries such as Argentina that are characterized by macroeconomic instability. See Blanco et al. (2021).

Most statistics on earnings inequality in the U.S. are based on household surveys. Indeed, the U.S. Census Bureau produces an annual report that documents changing trends in income and earnings inequality by demographic characteristics based on the Current Population Survey-Annual Social and Economic Supplement (CPS ASEC) (see, for example, Semega et al. (2020)). Complementary recent work uses administrative data to expand the literature on race and sex earnings differentials. Kopczuk, Saez, and Song (2010) use data from the Social Security Administration to look at mobility and earnings inequality in the U.S. since 1937. They find that the sex wage gap, rather than the impact of immigration or racial earnings disparities, has the most important empirical relation to overall mobility measures. Gideon, Heggeness, Murray-Close, and Myers (2017) find that when data from the Social Security Administration (SSA) Detailed Earnings Record are linked with record-level CPS data, estimates of the Black/White earnings gap at the mean increases. Chetty, Hendren, Jones, and Porter (2020) use federal income tax data linked to Decennial Census and American Community Survey (ACS) data to study intergenerational earnings differences. Blacks have lower rates of upward mobility and higher rates of downward mobility than whites. In comparison, White and Hispanic children have similar rates of intergenerational mobility.

What accounts for these earnings discrepancies across race and ethnicity? Cajner, Radler, Ratner, and Vidangos (2017) find that observables such as education, age and experience have little effect in explaining differential Black/White labor market outcomes such as unemployment. They also find that the inability to increase hours worked is an important impediment to earnings growth that varies by race. Denning, Jacob, Lefgren, and vom Lehn (2019) find that at least half of the gender earnings gap can be explained by hours differences (conditioning on race) when occupation-specific tasks are considered. Chetty, Hendren, Jones, and Porter (2020) find that conditioning on parents' income, Black/White income differentials for men are entirely explained by employment and wages with only a small contribution from marital status, education, and wealth. This suggests that the Black/White earnings gap is driven in part by differences in job opportunities.

Our main findings from the Global Income Dynamics Project indicators show uneven earnings growth across earnings percentiles over time and an increase in earnings inequality. These patterns motivate a deeper analysis of earnings across demographic groups. To focus on these demographics, we summarize time series changes in earnings by using a broad-based measure of long-term average earnings that captures active years, partially active years and zero earnings years. We find stark disparities among workers by comparing percentiles conditional on demographic group. When compared to our reference group of native-born White Non-Hispanic male workers, we find low-earning Black and Hispanic workers face larger earnings differentials than those with higher earnings. For example, Black men at the 10th percentile earn 18 percent of corresponding White male earnings. At the 90th percentile, Black men earn 54 percent of corresponding White male earnings. Similar differences by percentile of the reference distribution persist across most native- and foreign-born groups as well as sex. An interesting exception is foreign-born Black Non-Hispanic females who see smaller earnings differentials (compared to our reference group of White males) at lower earnings percentiles compared to higher earnings percentiles.

The differences in long-term average earnings across percentiles by demographic group reflect differences in labor market participation, age, education, human capital, geography and industry of employment. A basic regression analysis explains much of these earnings gaps, however, to better understand how these factors account for differences across the earnings distribution we perform a quantile regression decomposition as proposed in Machado and Mata (2005). We find that most of the earnings differentials for low earners in each demographic group are due to differences in observable characteristics. For example, more than 90 percent of the earnings differentials between Black and White Non-Hispanic males below the median can be accounted for by differences in observables. A similar pattern holds for most other demographic groups. Earnings differences among higher earning workers are largely not accounted for by differences in observable characteristics. Rather, these differences are due to model coefficients—the differences in estimated labor market returns to observable factors specific to each demographic group.

Our decomposition represents a substantial contribution to the literature. While quantile decompositions, such as those proposed by Machado and Mata (2005), have been applied to the study of earnings and wage inequality in the U.S, few focus on differences across race or ethnicity. Autor, Katz, and Kearney (2005) present a broad decomposition of wage inequality. While they do not focus explicitly on race and ethnicity, their results are consistent with our findings. They find that earnings differentials at the upper end of the distribution are accounted for by model coefficients rather than observable characteristics. We expand on their work by decomposing earnings differences by demographic group, an exercise generally not feasible given the sample size limitations of survey data. The quantile decomposition in Bayer and Charles (2018) is one recent study that does focus on racial earnings gaps. They analyze the historical gains and losses of Black males relative to White males at select quantiles. Their decomposition attempts to quantify earnings differences due to skills and those due to prices. These categories are broadly comparable to the differences that we attribute to characteristics and returns to covariates despite methodological differences in our approach. Like our study, they find a role for both components, where the price component is responsible for a large fraction of the historical earnings gains among high earners and a small fraction among low earners. While our study does not provide an analysis of historical trends, we do provide an analysis by more detailed demographic groups. Our long-term average earnings measure is a novel contribution that is difficult to construct with other available longitudinal data sources and allows us to incorporate repeated spells of nonemployment that are more prominent at the lower quantiles of the distribution.

The remainder of the paper proceeds as follows. In section II we describe the sources of earnings data used for our analysis. Section III summarizes the inequality and mobility statistics for the cross-county component of the Global Income Dynamics Project. Section IV examines inequality by demographic groups, where we focus specifically on disparities in long-term earnings by sex, race, ethnicity, and place of birth. Section V concludes.

## II. Data

The empirical work in this paper is based on job-level earnings information from the Longitudinal Employer-Household Dynamics (LEHD) infrastructure files, developed and maintained by the U.S. Census Bureau.[2] In the LEHD infrastructure, a "job" is the statutory employment of a worker by a statutory employer as defined by the Unemployment Insurance (UI) system in each state. Mandated reporting of UI-covered wage and salary payments between one statutory employer and one statutory employee is governed by the state's UI system. Reporting covers private employers and state and local government. There are no self-employment earnings unless the proprietor draws a salary, which is indistinguishable from other employees in this case. [3]

---

[2] See Abowd et al. (2009) for a detailed summary of the construction of the LEHD infrastructure.

The LEHD program is based on a voluntary federal-state partnership. When a state becomes a member of the partnership, current as well as all available historical data for that state are ingested into the LEHD internal database. By 2004, LEHD data represent the complete universe of statutory jobs covered by the UI system in the United States. However, studying job-level inequality, the task for which having a complete job frame is well suited, as a proxy for person-level inequality may be misleading due to the time-varying many-to-one assignment of jobs to workers. Therefore, we use all jobs to construct person-year level annual real (deflated by the Personal Consumption Expenditures Index (PCE)) earnings files covering the period 1998-2019.

It is preferable to have both a person frame that covers a known population of interest and to have a relatively high level of confidence that the persons in that population use a consistent person identifier across all jobs. To that end, we use the U.S. Census Bureau's enhanced version of SSA's master Social Security Number (SSN) database (the Numident) to create a set of "eligible" workers each year, removing annual earnings records for ineligible workers. The first eligibility condition is that a worker have an SSN that appears on the Numident; we call such SSNs "active." Each year an "eligible" worker must meet an additional set of conditions: age (varies by sample), not reported dead, and the SSN is active. If the worker has reported earnings in a given year, that worker must also not have more than 12 reported employers during the year; otherwise we assume the SSN is being used by multiple persons and the annual earnings report is discarded.

We use the subset of "eligible" workers described above to construct two samples. The first sample is used for the cross-country comparisons, while the second sample is used to examine long-term average earnings within the U.S. The first sample contains approximately 2 billion person-year earnings records while the second sample contains approximately 1.3 billion person-year records. The two analysis sample sizes differ due to:

1. Time Period / State Entry and Exit: Sample 1 includes all years from 1998 to 2019, while sample 2 includes only the complete data period (2004-2015). To minimize the impact of firm non-reports (false zeros) on estimates of long-term average earnings, we restrict sample 2 to the complete data period.[4] However, when constructing sample 1, we include the incomplete data periods and restrict state entry and exit to two years, 2004 and 2015.[5]

2. Annual Earnings Restrictions: For much of the analysis, sample 1 imposes an earnings floor $m_t = 260 * federal\ hourly\ minimum\ wage(t)$ (about \$1,900 in 2018) and a ceiling imposed by winsorizing earnings at the 99.999999[th] quantile. Sample 2 imposes no annual earnings restrictions and zero earnings years are included as long as the worker is active at least one quarter during the analysis period.

3. Age Restrictions: Sample 1 includes workers each year that are age 25-55, while sample 2 includes workers who are age 25 to 54 in 2004 (age 36 to 65 in 2015) and eligible to work every year between 2004 and 2015.[6] Sample 1 contains a representative cross-section of workers each year allowing worker entry and exit, while sample 2 follows the same set of workers over a 12-year period with no worker entry and exit. Workers in sample 2 may have zero earnings years, but they must be eligible to work each year (we exclude from sample 2 the small number of workers who die during the analysis period).

4. Real Earnings (PCE) Reference Year: The real earnings reference year for sample 1 is 2018, while the reference year for sample 2 is 2010.

5. Binned Earnings Data: To meet Census Bureau disclosure avoidance standards for the common code cross-country earnings comparisons, the reported earnings values in sample 1 are replaced with the earnings from at least 10 adjacent persons. The various earnings and change in earnings variables are first calculated on the not-binned data at the person level prior to binning. Each earnings variable is then binned separately. We sort the data by year, sex, year -of-birth (YOB), and an earnings variable. Sorting by year, sex, and YOB preserves exact means and sums of the earnings variable by year, sex, and age. Next, we classify each observation into a bin, take the average of the earnings variable within the bin, and in the last step attach the average value to each person record by bin id.[7]

---

[4] Zero earnings years are not directly reported in the LEHD data, a zero earnings year is inferred based on the absence of reported earnings. Zero earnings years are a core part of the long-term average earnings analysis and restricting sample 2 to the complete data period ensures each worker has a consistent small probability each year of a non-report.

[5] We have previously shown that by 1998 missing state data do not significantly affect measures of inequality and volatility (Abowd, McKinney, and Zhao 2018 and McKinney and Abowd 2020). See Table 1A and 1B for sample sizes by year and a list of states with missing data.

[6] See Appendix Table B1 for the evolution of age by year for sample 2.

For the cross-country comparisons we create multiple earnings measures. Using sample 1 we create two primary measures of earnings; one measure based on annual earnings and one measure of permanent earnings. Real annual earnings are the sum of real earnings $e_{ijt}$ over all eligible employers $j$ during year $t$ for a given person $i$ subject to the minimum earnings level $m_t$

$$y_{it} = \left( \sum_j e_{ijt} \,\middle|\, \sum_j e_{ijt} > m_t \right).$$

The primary permanent earnings measure $P3_{it}$ is defined as the average of the current and the previous two years of earnings, including zeroes and values below the minimum earnings cutoff if at least one year is above the minimum earnings cutoff[8]

$$P3_{it} = \left( \frac{\sum_j e_{ijt-2} + \sum_j e_{ijt-1} + \sum_j e_{ijt}}{3} \,\middle|\, \begin{array}{c} I\left(\sum_j e_{ijt-2} > m_t\right) + I\left(\sum_j e_{ijt-1} > m_t\right) + \\ I\left(\sum_j e_{ijt} > m_t\right) \geq 1 \end{array} \right).$$

To control for earnings differences due to observable characteristics such as sex, age, and education we also calculate several measures based on residual earnings. $\varepsilon_{it}$ is the residual from a regression of log $y_{it}$ on a set of age indicator variables by sex and year. $\delta_{it}$ is the residual from a regression of log $y_{it}$ on a set of age and education (LTHS, HS, Some College, BA+) indicator

---

variables by sex and year. We also create a residual permanent log earnings measure $P_{it}$, by first calculating the average earnings of $y_{it}$ from $t-2$ to $t$ for workers with at least two non-missing values of $y_{it}$. We regress the log of the average earnings measure is on a set of age indicator variables by sex and year. The residuals from this regression are used to create $P_{it}$.

The annual change in earnings, commonly known as earnings volatility, is an important focus of our cross-country comparisons and to facilitate our analysis we create two measures of the change in residual earnings; the one-year change in residual earnings ($z = 1$) and the five-year change in residual earnings ($z = 5$)

$$g_{it}^z = \left( \varepsilon_{it+z} - \varepsilon_{it} \middle| \left( \sum_j e_{ijt} > m_t \right) \wedge \left( \sum_j e_{ijt+z} > \frac{m_{t+z}}{3} \right) \right).$$

For the cross-country comparisons, we use sample 1 to create a sample of persons representative of the active worker population each year. The base sample (BS) each year (workers ages 25-55, not reported dead, and $y_{it} > 0$) contains the set of earnings records upon which we construct seven sub-samples. The sub-samples arise from the removal of low earning workers combined with the differing availability of the various multi-year earnings measures. Table 1A provides detailed information on the construction of the various sub-samples and Table 1B shows the resulting analysis sample sizes by year. Briefly, the cross-section (CS) sample drops workers with relatively low earnings (about \$1,900 in 2018), the longitudinal (LX) samples require a minimum level of earnings in two specific years ($t, t + 5$), the heterogeneity (H) samples are a subset of the LX samples with available residual permanent log earnings ($P_{it}$) in $t - 1$, and the permanent earnings (PA) samples require non-zero permanent earnings ($P3_{it}$) for both years of a specific year pair ($[t, t + 5] \vee [t, t + 10]$). The removal of low-earning workers and the construction of multiple-year measures of earnings results in a substantially different number of workers each year across the various analysis samples. For example, in Table 1B the BS sample has approximately 97 million workers in 2005, while the PA_10 sample has about 48 million workers. Generally, the available sample size falls as more years of non-zero earnings are required for a particular earnings measure.[9]

---

[9] The differing work history requirements for each earnings measure affects the composition of the workers both over time within samples and across samples at the same point in time. Appendix Figures A1-A4 show the sample size, percent male, average age, and median real annual earnings by year for each analysis sample.

For the long-term average earnings analysis, we follow a single cohort of 109 million eligible workers for 12 years, examining the long-term earnings outcomes of eligible workers, including the impact of periods of inactivity. For the analysis using sample 2, we create a single earnings measure, average real annual earnings over all years and all employers

$$w_i = \frac{1}{12} \sum_{t=2004}^{2015} \sum_j e_{ijt}.$$

We use $w_i$ to explore the differences in long-term average earnings across 20 demographic categories based on sex, race, ethnicity, and place of birth. Specifically, we define these categories as the interaction of place of birth (native-born, foreign-born), sex (male, female) and race/ethnicity. The race/ethnicity variable is constructed from the following categories: Asian Non-Hispanic, Black Non-Hispanic, White Hispanic, White Non-Hispanic, and All Other race/ethnicity groups.

Hours of work and education are two potentially important predictors of average annual earnings. Although information on hours of work and education are not available for the entire population, we assume the data are missing at random in the sense of Little and Rubin (2002) and impute the missing observations conditional on all observed data. Hours are imputed using information from the small subset of states (WA, OR, RI, and MN) for whom hours data are reported. We estimate a least-squares regression model of log annual work hours at a given job as a function of a log earnings quartic, age quartic, race indicators, a foreign-born indicator, and NAICS 2017 industry sector indictors. If the worker has multiple jobs during the year, then earnings at all other jobs is included as an additional covariate. The imputation regression model is estimated separately for workers with different quarterly work patterns, dominant jobs, coincident jobs, and sex.[10]

Education is observed for respondents in the 2000 and 2010 Census Decennial and all available years of the American Community Survey. Missing education is imputed (~80% missing) using information about a person's observed characteristics (sex, place of birth, age, race, and ethnicity) as well as the characteristics of a person's job history such as the average earnings, modal industry, and characteristics of a person's co-workers and co-residents. The characteristics are used to form homogeneous cells of a minimum size within which the distribution of observed education values is used to impute missing education values.[11]

---

[10] A similar hours imputation methodology is found in Hahn, Hyatt, and Janicki (2021). That paper contains a detailed evaluation of the impute and a comparison to hours statistics found in other data sources.

Although education provides important information about worker skill, a much broader estimate of worker skill can be formed using the level and pattern of worker earnings over time. For example, workers with higher education levels should have relatively high earnings at all their employers compared with similar workers with less education. We estimate an AKM (Abowd, Kramarz, Margolis 1999) style earnings regression to recover the fixed person effect and the average firm effect for all workers in our analysis sample. The AKM estimation to recover these fixed firm and worker effects uses all 4.4 billion job-year earnings observations in the 1990 to 2015 LEHD infrastructure files. The long time-period used in the estimation allows us to observe and control for the impact of all observed co-workers when estimating our analysis sample fixed person and person average fixed firm effects.

# III. Cross-Country Comparisons

## III.a. Inequality

In this section we present results for the U.S. estimated using a common set of programs provided to each of the participating countries. The goal here is to produce a standard set of estimates, thus facilitating cross-country comparisons.[12] We start by using the CS sample to estimate the change in cross-sectional earnings inequality for log real annual earnings ($\log{(y_{it})}$) over time as shown in Figure 1. The *y*-axis shows the difference in log real annual earnings for a given percentile between the current year and the base year (1998). For example, using Figure 1 we see that real earnings growth for the 90th percentile (p90) from 1998 to 2019 was approximately 22 percent for males and 30 percent for females. Real annual earnings growth for the other percentiles was also positive except for the 25th percentile for males from 2009-2014 and the 10th percentile for males from 2002 to 2014. In contrast to males, female workers benefitted from higher log real annual earnings growth across almost the entire earnings distribution. Workers at the 90th percentile and above generally received consistent earnings increases over the analysis period, however the experience of males and females differ somewhat. The trends in the increase in male earnings at the very top (p90 and above) are relatively homogeneous, however earnings at the extreme top (p99.9 and p99.99) are much more variable than for male workers between p90 and p99. Overall, female earnings at the top of the distribution exhibit smoother growth over time; however female workers at the extreme top have lower earnings growth than female workers between p90 and p99. All workers in the middle (p25 to p75) of the earnings distribution had relatively low earnings growth from 2001 to 2013. Male workers below the median faced a roller-coaster ride. Male workers at the very bottom (p10) saw real earnings decline from 2001 to 2013 with a recovery to 2001 levels by 2014. Females fared somewhat better than males, but earnings inequality for both males and females increased over the period.

To more clearly see the trends in earnings inequality, Figure 2 plots the dispersion of log real annual earnings over time. Inequality increased for both males and females over the period, with a larger increase for males than females. The standard deviation based measure ($2.56 * \sigma$) of earnings inequality follows a similar trend to the non-parametric ($P90 - P10$) measure although the extended right tail of the male earnings distribution is reflected in the divergence between the two measures for males, while for females the two measures produce similar results. Overall, male earnings inequality reached a peak in 2009 (female earnings peak in 2007) with an extended decline after the Great Recession for males. The picture for females is more complicated, inequality declined substantially during the Great Recession (mostly due to compression at the bottom of the earnings distribution) but was declining using the non-parameteric measure ($P90 - P10$) and increasing using the standard deviation based measure ($2.56 * \sigma$). In summary, the changes in inequality observed over the period are a mix of a consistent increase in dispersion for the top half of the earnings distribution, while the bottom half of the earnings distribution has both increased and decreased. The combined effect of the dispersion at the bottom and top determines whether earnings inequality overall is increasing or decreasing. Up to the Great Recession the two effects worked in the same direction, increasing earnings inequality. Post Great Recession the effect of the compression at the bottom of the earnings distribution generally dominated, reducing overall earnings inequality.

Figures 3 and 4 provide insights into how earnings inequality varies by age. The earnings distribution for younger workers is generally more compressed. As workers gain experience in the labor market dispersion increases. Figure 3 shows the dispersion of log real earnings for workers at age 25. The trends for these younger workers are similar to the overall trends for both males and females shown in Figure 2; however, the compression post Great Recession at the bottom of the earnings distribution is larger and the increase in dispersion at the top half of the earnings distribution isn't as sustained.

Figure 4 shows the earnings dispersion for four entering cohorts of workers over time. For the 1998 and 2000 cohorts, earnings inequality for both males and females increases as workers gain experience (up to around age 35), with especially strong initial inequality growth for females. However, beginning with the 2005 cohort the pattern changes, especially for males. Both the 2005 and 2009 male cohorts enter the labor market with relatively high earnings inequality that persists over time. This result suggests economic conditions at the time of entry to the labor market may have relatively long-term effects for males. For females, the 2005 and 2009 entering cohorts have moderately higher initial earnings inequality, but also have substantial growth in earnings inequality over time (similar to the 1998 and 2000 cohorts).[13]

## III.b. Volatility, Skewness, and Kurtosis

In contrast to measuring the dispersion in earnings, as we did in the previous section, volatility measures the dispersion in the change in earnings. The dispersion of the change in earnings captures the extent to which workers face similar year-to-year earnings shocks. Figure 5 shows the dispersion in the earnings residuals ($g_{it}^1$) between subsequent and current years. Previous research using more standard measures of volatility across several different datasets, using either the variance of the difference in log earnings or the variance of the arc-percentage change, show similar results (Moffitt 2020). Dispersion is generally falling over the analysis period, except during recessions. Workers in 2019 generally have less dispersion in the change in earnings than workers in 1998, a result consistent with previous research showing more recent workers have fewer job changes (Davis and Haltiwanger 2014).

---

[13] Additional figures showing various features of the evolution of the U.S. earnings distribution are available in the appendix (Figures A5-A11).

Although the dispersion of the change in residual log earnings is declining over time, Figure 6 provides additional information on the composition of the changes. The pattern of Kelley skewness shown in panel (a) prior to the Great Recession highlights the relatively equal contribution of both the bottom and the top (zero Kelley skewness) except during recessions where the top of the residual log earnings change distribution compresses and the bottom expands due to a reduction in positive earnings shocks and a large increase in negative earnings shocks, respectively. Post Great Recession there is a relatively long period of consistent positive Kelley skewness due to an increase in relatively large positive earnings shocks and a reduction in relatively large negative shocks. Panel (b) looks beyond the P90 and P10 deciles of the earnings change distribution, highlighting the increase in the prevalence of large earnings shocks as the level of Crow-Siddiqi kurtosis increased over the period.

Figure 7 shows the dispersion in volatility for workers at different points in the permanent residual log earnings distribution. First, each year the support of the permanent residual log earnings distribution is divided into 41 consecutive non-overlapping bins, with each bin representing approximately 2.4% of the earnings observations. The y-axis shows the average $P90 - P10$ differential of $g_{it}^1$ across all years (2001 to 2014) for the worker year-pair observations in each bin, separated into three different age categories. Figure 7 shows a large decline in earnings volatility, a relatively small decline in Kelley skewness, and a relatively large decline in Crow-Siddiqi kurtosis as we move up the residual permanent earnings distribution, except at the very top. When constructing measures of volatility using log earnings, volatility is generally greater for workers at the bottom of the earnings distribution, a similar pattern emerges here for residual log earnings. Large percentage changes in earnings are more likely when the level of earnings is low (someone earnings $10,000 dollars per year can more easily double their earnings than someone earning $100,000 per year). However, comparing across age groups for workers at similar points in the earnings distribution we see that, except for at the very top of the earnings distribution, younger workers generally have more volatility than older workers. One caveat of this analysis is that measures using log earnings do not capture transitions in and out of active status. Many workers have significant periods of inactivity, and a large part of total earnings volatility is due to worker entry/exit (McKinney and Abowd 2020), a feature we show later in the paper is an extremely important component of long-term earnings outcomes.[14]

## III.c. Mobility

In Figures 8 and 9 we show estimates of long-term intragenerational earnings mobility, comparing a worker's permanent earnings ($P_{3it}$) rank in year $t$ with their rank in year $t + 10$ for different age groups and over time.[15]  Given the relatively stable earnings distribution over the period, both figures imply permanent earnings converge to the mean. Workers with relatively high permanent earnings ranks in the first period tend to have lower permanent earnings ranks ten years later, while workers with a relatively low permanent earnings rank in the first period tends to have a higher permanent earnings rank ten years later. Younger workers have higher earnings mobility than older workers and earnings mobility appears to be declining slightly over the period, with higher earnings mobility in 2000 compared with 2009. Overall, the results are almost identical for both males and females.

Comparing our results with rank-rank *intergenerational* regression estimates of earnings mobility provides a worthwhile benchmark.  Regression estimates from Chetty et al. (2014) using IRS tax data find a rank-rank slope of 0.34 while Mazumder (2015) using the PSID finds a slightly larger estimate of 0.4. Our intragenerational rank-rank regression estimates are about 0.67, implying there is significantly less intragenerational mobility than similar measures of intergenerational mobility.

The permanent earnings measure ($P_{3it}$) used here is designed to capture workers with at least some formal labor market activity over a three-year period, while excluding workers that left the labor market. With a well-functioning labor market this would be a reasonable compromise, however many post Great Recession workers prematurely exited the labor force permanently or for an extended period of time (McKinney, Abowd, and Zhao 2018). Incorporating eligible zero permanent earnings workers into the mobility analysis and comparing

---

[14] Additional earnings volatility figures are available in the appendix (Figures A12-A16 and A19-A22).
[15] When interpreting the results keep in mind that the "mobility" shown in Figures 8 and 9 is not necessarily the result of a change in permanent real earnings.  A worker's rank may change because of changes in the worker's permanent real earnings and/or changes in the real permanent earnings of other comparable workers

the impact of their inclusion at different points in time would be a useful extension to the results presented here.[16]

# IV. Long-Term Average Earnings

The main goal of this section is to study long-term average earnings differentials across demographic groups. For this analysis we use sample 2 to follow a cohort of prime age workers, who are 25-54 years old in 2004. We monitor these workers for 12 years, observing earnings during periods of UI-covered formal labor market activity.[17] Labor force attachment varies significantly across prime age workers, although earnings differences persist even when we control for hours of work and years of inactivity. A key aspect of sample 2 is that it contains zero- and low-earnings years compared to much of the analysis conducted in Section III, which includes only years with earnings above a time-varying minimum earnings floor. Including periods of inactivity allows us to capture earnings observations in sample 2 that result from changes in labor supply along both the intensive and the extensive margin. In Figure 10 we plot the share of workers active in the labor market for three different age groups.  The age groups are defined as follows: age group 1 workers have ages 25-34 in 2004; age group 2 workers are 35-44 in 2004; and age group 3 contains workers with ages 45-54 in 2004. At the beginning of our time series the vast majority (82-85 percent) of workers in all three age groups are active; however, at the onset of the 2007-2009 recession labor market activity decreases substantially to 77-78 percent uniformly across all the age groups. During the recovery from the Great Recession, we begin to see heterogeneity emerge. Strikingly, during the recovery, neither of the two younger age cohorts begin to approach the levels of labor market activity observed before the Great Recession. Activity increases slightly for the youngest age cohort, while activity continues to decline for the middle cohort. As expected, labor market activity for older workers continues to decline although the slope of the decline post Great Recession is likely due to both the differential effects of the recovery on older workers and retirement decisions.

---

[16] Shorter duration five-year mobility figures are available in the appendix (Figures A17 and A18).

[17] Although inactivity plays an important role in this paper, like most administrative earnings datasets the LEHD data does not contain a direct report of inactivity.  Our periods of inactivity are defined by not observing UI-covered activity.  Although LEHD coverage of the formal labor market is exceptionally broad, informal labor earnings, self-employment, and federal workers are not covered, and activity in these sectors may appear as periods of inactivity in our analysis dataset.

Using only active earners produces the log earnings profiles by age group shown in Figure 11. All three age cohorts have earnings growth before the 2007-2009 recession with the steepest growth observed by the youngest group, although the 2007-2009 recession brought small declines in average earnings across all age groups. Workers in the oldest age group had the largest decreases in earnings with slow and persistent earnings declines that continued in the subsequent economic recovery. Workers in the bottom two age groups had earnings growth starting at the beginning of the post-recession recovery, with the steepest growth observed for the youngest age group.

## IV.a. Characteristics of Long-term Earnings

In the previous section we documented the changes in labor market activity and earnings for workers in sample 2. In this section, we focus on average real (2010 PCE) annual earnings $w_i$ which summarizes the impact of changes in labor supply and earnings over the twelve-year period from 2004 to 2015. Our focus here is on the distribution of $w_i$ both within and across twenty sex, race, ethnicity, and place of birth demographic groups. For each group, Tables 2A and 2B show the 10th, 25th, 50th, 75th and 90th average real earnings percentiles. For average real earnings we show the actual percentile while for the other earnings and activity measures, we show averages for workers with earnings in the neighborhood of the reported percentile. We sorted all workers by the value of their average annual earnings. The amount shown in the column "Average Annual Earnings" is the percentile of this distribution. We then used this sort order to compute average values of the other variables for workers at the indicated percentile. These averages use a window of the percentile plus or minus one percentile point. For example, the "Share of Active Each 4 Year Period" for Asian Non-Hispanic Foreign-born Females shown as 0.06 in the table is the average value for all such women whose average annual earnings are between the 9th and the 11th percentile in the average annual earnings distribution for Asian Non-Hispanic foreign-born females.

In Table 3A and 3B we expand the set of characteristics to include geography (Census division), industry, age, and education.[18] Tables 2 and 3 are both grouped into a part A and part B, with part A containing statistics for the foreign-born and part B containing statistics for the native-born. Figure 12 illustrates the relative average annual earnings differences between each demographic group and our reference group (native-born White Non-Hispanic males) at each of the reported own-group percentiles.

---

[18] See Appendix Table B2 for the definitions of the geography and industry variables. Figure B1 provides a map of the Census divisions.

Before we discuss the differences in average annual earnings across demographic groups, we would like to emphasize a key point of Table 2A and 2B. Our analysis of average annual earnings compactly captures much of the earnings dynamics and variation in labor market activity across percentiles. That is, we can learn much about the earnings history of workers by looking at their percentiles in the average annual earnings distribution. To illustrate this idea, we define labor market activity by dividing our 12-year analysis period into three consecutive non-overlapping 4-year sub-periods. A worker is considered long-term active if they have at least one quarter of positive earnings in each 4-year period. Even using this weak measure of labor market attachment, average annual earnings capture much of the variation in labor market activity across percentiles. If we look at average annual earnings growth between the first and the last 4-year sub-period for workers active in each 4-year sub-period, we see a strong positive correlation between earnings growth and average earnings; workers at the top of the earnings distribution have noticeably more earnings growth than workers at the bottom. Workers at the top of the earnings distribution also have lower earnings volatility, more hours paid, and fewer years of inactivity. Simply knowing a worker's long-term average annual earnings conveys a large amount of information about a worker's earning dynamics and work history.

The reference group for our comparative analysis of earnings differences is native-born White Non-Hispanic males, making it natural to start our discussion of the tables with this group. In Table 2B, native White Non-Hispanic males have reported average earnings of $3,469 at the 10th percentile. These numbers increase steadily to $38,960 at the median and to $110,400 at the 90th percentile. In comparison, native Black Non-Hispanic males have substantially lower earnings at all percentiles. For example, at the 10th percentile, we observe annual earnings of only $617. This represents only 18 percent of the earnings found for a similarly located worker in the reference group. Figure 12 facilitates these types of comparisons, showing the ratio of average real annual earnings for all groups relative to native-born White Non-Hispanic males.

Alternatively, for groups with large earnings differences, it is useful to compare average annual earnings across percentiles. For example, the 25th percentile of the Black average earnings distribution is comparable to the 10th percentile of the White distribution with Black workers earnings $3,927 (compared to $3,469 for White workers) with similar results for hours paid with 820 hours paid (compared to 887 hours paid for White workers). Median average earnings for Black workers are $16,780, which represents 43 percent of White median earnings. At the 90th percentile, average long-term earnings are $59,180 for Black workers. That is, at the 90th percentile of the earnings distribution Black workers earnings are less than the 75th percentile of the White distribution with more hours paid than White workers at the 90th percentile. In contrast, native-born Asian Non-Hispanic males earn more than White Non-Hispanic males at all percentiles of the earnings distribution. The relative earnings of Hispanic and the All Other race/ethnicity group fits between Black and White workers with White Hispanic workers having higher earnings than the All Other race/ethnicity group at every percentile.

Foreign-born males have more mixed outcomes by race and ethnicity. Figure 12 panel (b) shows that while foreign-born Black Non-Hispanic males at the bottom of the distribution have large average earnings differentials compared to the native-born White Non-Hispanic male reference group, these differentials are smaller than those observed for foreign-born White (both Hispanic and Non-Hispanic) workers and the All Other race/ethnicity group. However, at higher percentiles, average earnings for foreign-born White Non-Hispanic males and foreign-born Asian Non-Hispanic males exceed those of native-born White Non-Hispanic males with an earnings differential of 15-20 percent at the 90th percentile.

By sex, Table 2B shows that native-born White Non-Hispanic females earn $1,727 at the 10th percentile, increasing to $23,790 at the median, and $69,010 at the 90th percentile. The earnings among females of this group are lower than comparable percentile calculations for males as seen in Figure 12 panel (c). The earnings differences are even more stark among the Black and All Other groups of native-born females at the 10th and 25th percentiles. Native-born Black females earn $1,208 and $6,807 and the All Other group females earn $843 and $4,946 at the 10th and 25th percentiles, respectively. Asian Non-Hispanic females earn slightly less than White Non-Hispanic *males* at each percentile, except at the 75th percentile, where they slightly exceed male earnings.

Among foreign-born Black Non-Hispanic females, we find smaller earnings differentials relative to native-born White Non-Hispanic males at lower earnings percentiles than those recorded for native Black Non-Hispanic females, as described in the previous paragraph. Earnings among Black Non-Hispanic females are $2,585 at the 10th percentile and $10,920 at the 25th percentile. However, these represent only 75 and 67 percent of the respective earnings for the reference group of native-born White Non-Hispanic males as seen in Figure 12 panel (a). Earnings among foreign-born Black Non-Hispanic females exceed the earnings of the All Other race/ethnicity group at the 10th, 25th, and 50th percentiles. Earnings of foreign-born Asian and White females exceed those of Black females only at the 75th and 90th percentiles.

In Tables 3A and 3B we show the variation among demographic groups by age, education, geography, and industry across percentiles. Unlike most of the work history measures in Table 2 and many of the measures in Table 3, average age does not increase monotonically with the average earnings percentiles. Although average earnings increase monotonically by age for the 25th percentile and above native-born White Non-Hispanic males, many of the groups have different patterns (i.e., the oldest workers are found in both the bottom and the top of the native-born and foreign-born Asian Non-Hispanic female average earnings distribution). Low earners in each demographic group tend to be less educated than higher earning workers. The share of workers with less than a high school degree is highest among workers in the 10th percentile and lowest among workers in the 90th percentile. The converse is true for workers with at least a BA degree. There is also substantial variation in education across these groups. The share of Asian workers (of any gender and place-of-birth) with a BA degree or higher at the 90th percentile exceeds 70 percent. In contrast, only 14 percent of foreign-born White Hispanic workers have a BA degree or higher. There are differences in industry composition across percentiles for each demographic group. For example, workers at the 10th percentile are usually employed in industry sectors: construction (D), retail trade (G), administrative and support (N), and manufacturing (E). At the 90th percentile, only manufacturing is found in common with the workers at the 10th percentile. Workers at the 90th percentile are most often found in professional, technical, and scientific services (L), wholesale trade (F), and finance and insurance (J). At the 90th percentile, these industries account for 51 percent of employment. Geography varies as well with low earners at the 10th percentile found in the South Atlantic and East North Central Census Divisions and high earners at the 90th percentile found in the East North Central and Middle Atlantic Census Divisions.[19]

Education differences are only one measure of skill differentials. We can also use AKM-style fixed person and firm effects to provide an alternative description of the types of workers at each percentile in terms of their portable earnings component and the type of firms with which they match. In Tables 2A and 2B we detail the average fixed person and firm effects for each demographic group. We generally find higher person-effect workers correspond to higher earnings percentiles, although the pattern is not monotone at the lower percentiles. For example, workers at the 10th percentile often have a larger person effect than those found at the 25th percentile. Workers with higher earnings are often found in high-paying firms. This is true at higher earnings percentiles for all groups. However, foreign-born Asian workers at the 10th percentile of the average annual earnings distribution match with slightly better firms than those found at the 25th percentile. Differences in observables for other demographic groups not specifically discussed in the text are detailed in Table 2A and 2B. In the next section, we explore the role of these factors in explaining average earnings across demographic groups.

---

[19] See appendix Table B2 for the definition of Census geography divisions.

## IV.b. Least Squares Adjusted Average Earnings Differentials

Although the unadjusted average earnings differentials across groups are large, observable characteristics associated with each worker may account for most of the differences. To control for differences in observable characteristics our first approach is to estimate an OLS regression. We estimate the following pooled earnings model:

$$\log (w_i) = \gamma_g + x_i\delta + \varepsilon_{i.}$$

We regress our real average annual earnings measure $w_i$ (defined as an average over 12 years of individual $i$ earnings) on $\gamma_g$, an indicator variable for each of our 20 demographic groups of interest ($g = 0, \ldots, 19$), and $x_i$ a vector of covariates including an average annual hours-worked quartic, years of inactivity, years of partial activity, division indicators, industry indicators, initial age, education indicators, fixed person effects, and average (over all employers for $i$) fixed firm effects. The last two covariates were obtained from an AKM regression as described in the data section. We begin with a minimal specification and add additional explanatory variables with each successive model. The results are presented in Table 4.

Model 1 in Table 4 shows the unconditional log average earnings differentials using native-born White Non-Hispanic males as the reference group ($g = 0$). The coefficients for model 1 are the same as the unadjusted average earnings differentials from Table 2A and Table 2B except that the differences are now shown in log points not dollars $\left(\log(w_g) - \log(w_0)\right)$. Relative to the reference group, native-born Black Non-Hispanic males have average earnings lower by just over 1 log point, which is equal to approximately $16,700. In contrast, native-born White Non-Hispanic females have earnings that are lower by 0.54 log points, almost half as

small as for native-born Black Non-Hispanic males. The most important covariate in any of the models is hours worked (first included in model 2). Simply controlling for hours worked, the proportion of the total variation in average earnings explained increases from 0.04 in model 1 to 0.87 in model 2. Although hours worked is expected to have a large effect given that in most cases average earnings are a positive linear function of average hours worked, however hours worked differences across the groups do not explain all the differences across demographic groups. As we add additional covariates, the differences decrease substantially but do not completely disappear. For example, in Model 5 the average earnings of native-born Black Non-Hispanics males is 0.32 log points lower than for the reference White males, a substantial reduction compared with the 1.02 log points in Model 1, but no reduction compared with Model 2. The adjusted earnings differences for females are also smaller with richer specifications, but once again do not completely disappear. For example, the full model specification finds that earnings for native-born White Non-Hispanic females is -0.14 log points lower than for males compared with -0.54 log points in Model 1.

The addition of AKM human capital variables has an interesting effect on earnings differences for many groups. For example, the addition of the AKM measures *increases* earnings differentials for Asian and Black Non-Hispanic groups relative to the reference group. That is, the indicator variable for these groups becomes more negative when comparing Model 4 and Model 5. Recall, that the AKM measures capture person-specific skills and the quality of the employer that is separate from what can be captured by the educational attainment variable alone. This persistent differential captures some characteristics of the labor market that point to the possibility of additional labor market frictions (through job matching or race discrimination). Further analysis is beyond the scope of this paper, but considerable additional research is needed to formalize the mechanisms behind these differentials.

## IV.c. Quantile Regression Adjusted Average Earnings Differentials

The richness of our data allows us to go beyond an analysis of real average earnings differentials across groups at the mean. In this section, we investigate the magnitude of earnings differentials between demographic groups at different percentiles of the average earnings distribution. Although we showed in the previous section that observable characteristics holding the return to those characteristics constant across groups do not completely explain mean average earnings differences, we now relax this restriction, allowing the impact of observable characteristics to differ across the group earnings distributions.

### IV.c.i. Estimation Methodology

We define the regression estimate of quantile $\theta$ for each demographic group $g$ as $Q_\theta \left( \ln (w) \big| x(g) \right) = x(g)'\beta_\theta(g)$ where $w$ represents real average annual earnings, $x(g)$ represents a vector of covariates for group $g$ and $\beta_\theta(g)$ represents coefficients at the estimation quantile $\theta$ for workers in group $g$. For each demographic group $g$ we estimate quantile regressions including the set of regressors in Model 4 of Table 4. Similar to our least-squares estimates, we conduct our analysis relative to the native-born White Non-Hispanic male reference group, which is indexed by $g = 0$.

Our goal is to estimate the conditional real annual earnings distribution for each group of interest and our reference group and then use the estimated coefficients to decompose earnings differences into components due to coefficients, covariates, and a residual following the methodology outlined in Machado and Mata (2005) and Albrecht et al. (2003). First, we define the observed density of log real average annual earnings corresponding to each of our groups $g$ by $f\left(\ln\left(w(g)\right)\right)$ and the simulated average earnings density for group $g$ as $f_w^*\left(\beta(g); x(g)\right)$. To simulate the conditional average earnings distribution for group $g$ we start by estimating 99 separate quantile regressions, one for each quantile $\theta = 1, \dots, 99$. Next, we take one draw from a uniform $(0,1)$ distribution for each person in group $g$ and assign each of them a $\theta_i$ based on dividing the support of the uniform distribution into 99 equal size bins. Using the $\theta_i$ values from the previous step we calculate the predicted average earnings $\ln\left(w_i(g)\right) = x_i(g)'\hat{\beta}_{\theta_i}(g)$ for each person in group $g$. The resulting simulated earnings values can then be used to estimate quantiles or any other statistic of the log average earnings distribution $f_w^*\left(\beta(g); x(g)\right)$. As we show below, the power of this approach is its ability to easily simulate counterfactual average earnings distributions by replacing for example $\hat{\beta}(g)$ with $\hat{\beta}(0)$.

We define the difference in observed log average earnings for group $g$ and our reference group at a specific quantile as $\Theta\left(f\left(\ln\left(w(g)\right)\right)\right)$- $\Theta\left(f\left(\ln\left(w(0)\right)\right)\right)$. This earnings difference can be decomposed into three components. The first component is defined as earnings differentials that arise to due differences in covariates while holding the coefficients constant at common values. The second component is the earnings difference due to changes in coefficients holding covariates fixed at common values. The third component is the residual. More formally, we define the decomposition using the following equation:

$$\Theta\Big(f\big(\ln\big(w(g)\big)\big)\Big) - \Theta\Big(f\big(\ln(w(0))\big)\Big) =$$

$$\underbrace{\Theta\Big(f_w^*\big(\beta(g); x(g)\big)\Big) - \Theta\Big(f_w^*\big(\beta(g); x(0)\big)\Big)}_{Covariates} +$$

$$\underbrace{\Theta\Big(f_w^*\big(\beta(g); x(0)\big)\Big) - \Theta\Big(f_w^*\big(\beta(0); x(0)\big)\Big)}_{Coefficients} +$$

$$Residual.$$

In this form of the decomposition the counterfactual distribution $f_w^*\big(\beta(g); x(0)\big)$ estimates the conditional earnings distribution using the covariates of the reference group 0 combined with the estimated coefficients of group $g$.[20] For example, using this approach we could estimate the annual earnings distribution for native-born White Non-Hispanic males using the returns to the observables of native-born Black Non-Hispanic females. Comparing this earnings distribution with the predicted earnings distribution for native-born White Non-Hispanic males reveals the change in earnings if White workers received the same returns to their observable characteristics as Black workers. The estimates of the individual components for both forms of the decomposition are shown in Table 5. The decompositions themselves are shown in Table 6.

### IV.c.ii. Results

As shown in the OLS results, log real average earnings differentials exist across all groups. As we show below, the role of the coefficient and covariate components varies across both groups and percentiles of the earnings distribution within each group as seen in Table 6. We plot the components of the decomposition for each demographic group in Figures 13-17.

To focus the discussion, we first present results from the decomposition for native-born male Black-White earnings in Figure 16 panel (b). Figure 16 panel (b) presents a visualization of the results of the earnings decomposition between Black and White native-born Non-Hispanic males from Table 5B. As shown previously, there are striking differences in earnings between Black and White workers throughout the distribution. The decomposition illustrated in Figure 16 panel (b) shows that much of the differential at the lower percentiles of the earnings distribution is due to the covariate component rather than the coefficient component. For example, more than 90 percent of the earnings differentials predicted by our model between Black and White Non-Hispanic males below the median can be accounted for by difference in observables that make up the covariate component. As we move up the earnings distribution, the earnings differential decreases (in absolute value) and the relative contribution of the covariate component also decreases. At the 75th percentile, the contribution of the coefficient component begins to exceed that of the covariate component and continues to increase among workers in the higher percentiles of the earnings distribution.

---

[20] Alternatively, we can express the decomposition as $\Theta\Big(f\big(\ln\big(w(g)\big)\big)\Big) - \Theta\Big(f\big(\ln(w(0))\big)\Big) =$ $\Theta\Big(f_w^*\big(\beta(g); x(g)\big)\Big) - \Theta\Big(f_w^*\big(\beta(0); x(g)\big)\Big) + \Theta\Big(f_w^*\big(\beta(0); x(g)\big)\Big) - \Theta\Big(f_w^*\big(\beta(0); x(0)\big)\Big) +$ $Residual.$

Although the differences are smaller and more uniform across the earnings distribution, a similar pattern holds for native-born White Hispanic males in Figure 16 panel (c). Covariate differences also account for most of the earnings discrepancy between White and Asian males (Figure 16 panel (a)) at the bottom of the earnings distribution, although native-born Asian Non-Hispanic males have higher earnings than workers in the reference group. Note that because our results are relative to the reference group, Figure 16 panel (d) shows no difference relative to itself. We should also note that our quantile regressions generally fit the data well with the residual component in Figures 13-17 generally very close to zero.

For ease of interpretation, we construct shares of the total predicted earnings differential attributable to the differences in coefficients and covariates. These are presented in Table 6A and 6B. The share of earnings differentials accounted for by differences in model coefficients increases as we move up the earnings distribution. However, the rate of substitution between these two components varies depending on the demographic group.

For example, using Decomposition 1 the share of the total earnings differential between native-born Hispanic and Non-Hispanic White males accounted for by the coefficient component increases from 12 percent at the 10th percentile of the earnings distribution to 67 percent at the median to a 97 percent at the 90th percentile. The share of the covariate component follows the opposite pattern, consistent with the small residual component in our regression analysis. The earnings discrepancy accounted for by the coefficient component is particularly large among high-earning Hispanic workers, with smaller levels found for other demographic categories.

At the bottom of the earnings distribution, differences in covariates play a strong role in explaining real average earnings differences. As a demographic group increases hours paid, , finds employment in higher paying industries, and/or acquires more education, the earnings gap relative to the reference group decreases dramatically.[21] However, as we move up the earnings distribution, differences in the returns to these observables play a dominant role. This increased role of the coefficient component corresponds to a difference across groups in the return to observables such as education, hours paid, etc. For workers above the median, the path to greater real average earnings is less clear. Even if higher earning workers are employed in the same industries and have similar observable education levels, they will be faced with a significant earnings gap relative to the reference group because of the differences in their coefficients—the implicit labor market "returns" to their characteristics. Are workers in certain groups not employed in similar occupations within high earning industries? Is there workplace discrimination? Disentangling the determinants of the differences in the return to observables across groups is a worthwhile area of future research.

### IV.c.iii. Counterfactual Earnings Differentials

Finally, we use the estimated counterfactual earnings distributions to create two figures similar in spirit to the unadjusted earnings differentials shown in Figure 12. We use the counterfactual earnings distributions $f_w^*\big(\beta(g); x(0)\big)$ and $f_w^*\big(\beta(0); x(g)\big)$ to set or adjust each group's characteristics or coefficients to the reference group, respectively. The first counterfactual is the predicted earnings distribution of group $g$ when observable characteristics are those of the reference group, that is, the earnings distribution of group $g$ when we control for differences in covariates (such as education, industry, division, age, etc.). The second counterfactual is the predicted earnings distribution of group $g$ when the "returns" to observables are those of the reference group. For both counterfactuals, the comparison group is the predicted real average earnings of the reference group, $f_w^*\big(\beta(0); x(0)\big)$. We present the counterfactual earnings differentials at each percentile of interest with reference group characteristics in Figure 18, where each point is expressed as a share of the reference group $\exp\left(\Theta(f_w^*(\beta(g); x(0)))\right)/\exp\left(\Theta(f_w^*(\beta(0); x(0)))\right)$. Figure 19 contains counterfactual earnings differentials with reference group coefficients expressed as $\exp\left(\Theta(f_w^*(\beta(0); x(g)))\right)/\exp\left(\Theta(f_w^*(\beta(0); x(0)))\right)$. The elements underlying these figures are found in Table 5.

---

[21] Underlying our figures and tables are 1,980 separately estimated quantile regression models (99 for each of the 20 demographic groups). The large number of estimated coefficients precludes including them in the paper, however the signs of the coefficients do not differ from prior expectations. For example, the return to additional hours worked and the return to additional formal education is always positive. Thus, for all demographic groups, increasing hours worked and/or education will increase average earnings, reducing the earnings gap with the reference group (one exception is native-Born Asian Non-Hispanic males who have higher average earnings than the reference group).

Figure 18 isolates the role of different coefficients holding characteristics at the reference group level. At the lower percentiles, earnings differentials decrease and compress for all groups when we control for differences in observables as seen in Figure 18. This implies, for example, that *most* of the earnings differences we observe between native-born White and Black Non-Hispanic males are due to characteristics such as education, industry, division of employment, and age as shown in Figure 18 panel (d) at the lower percentiles. At the 10[th] percentile of the Black Non-Hispanic average earnings distribution, the earnings differential controlling for observable characteristics is less than 10 percent when compared to the reference group of native-born White Non-Hispanic males (the actual average earnings difference at the 10[th] percentile is over 80 percent). Much of the earnings premium we observe between Asian and White Non-Hispanic males is also due to observable differences at the lower percentiles. In contrast, earnings differentials between races at higher percentiles vary little when we control for differences in covariates. For example, at the 90[th] percentile of the Black Non-Hispanic earnings distribution the estimated average earnings differential controlling for observable characteristics is 37% while the actual average earnings differential is 46%.

Figure 19 isolates the role of different observables holding the coefficients at the reference group levels. In this counterfactual, the average earnings differentials among low-earning workers are close to their actual values in Figure 12. In contrast to the results shown in Figure 18, the earnings differentials in Figure 19 are due to observable characteristics. What about workers with high average earnings? Earnings disparities decrease among workers with high average earnings for the Black, Hispanic, and All Other race/ethnicity groups in Figure 19. For high-earning workers, this finding implies that the returns to observable differences are generally larger for Non-Hispanic White workers than for other race and ethnicity groups. It is important to note that earnings differences do not disappear among high earners even when we control for differences in coefficients or returns to observables. The starkest contrast is that of native-born Black Non-Hispanic males where earners at the 90[th] percentile have earnings that are 76 percent of those of the reference group of native-born White Non-Hispanic males an improvement compared with the actual value of 54 percent from Figure 12, but still a relatively large gap.

Echoing the results in previous sections, the counterfactuals from the quantile regression approach suggests much of the earnings differences observed at the lower percentiles of the earnings distribution can be attributed to differences in observable characteristics, such as hours, education, industry etc.[22] Earnings differentials at the higher percentiles are more difficult to interpret since they primarily reflect differences in the return to the observable characteristics, not differences in those characteristics. These returns could be interpreted as prices, but they could also take the form of skills, quality of job matches, or discrimination. We believe these results serve as a strong motivation for future work.

## V. Conclusion

From 1998 to 2019 earnings inequality in the U.S. increased while volatility decreased. Although long-term mobility over a worker's lifetime is moderate with some regression to the mean, the U.S. also has persistent differences in long-term average earnings both within and across sex, race, ethnicity, and place of birth demographic groups. Going beyond the standard OLS log earnings regression, we show that the structure of earnings differentials relative to native-born White Non-Hispanic males differs throughout the earnings distribution. At the bottom of the earnings distribution differences in earnings across groups are largely due to observable characteristics suggesting that workers at the bottom of the earnings distribution may have the clearest path to improving their position. Increasing hours paid, changing employers, and attaining additional education, while difficult in many cases, is one of the standard pathways to higher real earnings. For workers above the median, differences in the return to characteristics is the dominant component. The pathway to reducing differences in the returns to observable characteristics across demographic groups is less clear. Future research towards a better understanding of the differences in the returns to observable characteristics would be a worthwhile endeavor.

---

[22] Due to space constraints and Census Bureau disclosure limitations we do not further explore the separate role of each observable characteristic in detail. The regression results in secion IV.b. suggests that hours account for at least 75 percent of the variation in log earnings at the annual level. Karahan *et al.* (2019) and Gregory et al. (2021) also find a large role for hours among low-income workers.

## Table 1A: Analysis Sample Descriptions

| Sample Name | Description |
|---|---|
| Base (BS) | Workers age 25-55, not reported dead, and $(y_{it}) > 0$ |
| Cross-Section (CS) | BS sample with annual earnings $(y_{it}) > m(t)$ |
| Longitudinal 1 (LX_1) | CS sample and annual earnings $(y_{it+1}) > 1/3*m(t+1)$ |
| Longitudinal 5 (LX_5) | CS sample and annual earnings $(y_{it+5}) > 1/3*m(t+5)$ |
| Heterogeneity 1 (H_1) | LX_1 sample and residual permanent log earnings $(P_{it-1})$ not missing |
| Heterogeneity 5 (H_5) | LX_5 sample and residual permanent log earnings $(P_{it-1})$ not missing |
| Permanent Earnings 5 (PA_5) | BS sample and permanent earnings $(P3_{it})$ not missing in t and t+5 |
| Permanent Earnings 10 (PA_10) | BS sample and permanent earnings $(P3_{it})$ not missing in t and t+10 |

Notes: Analysis samples use 1998-2019 LEHD person-year earnings data. All states plus DC report earnings from 2004-2015. Prior to 2004 DE, IA, NW, UT, OK, VT, AL, MA, DC, AR, NH, MS have non-reporting quarters and are not included. After 2015 AK, AR, and MS have non-reporting quarters and are not included. All analysis samples are constructed from the Base (BS) sample. m(t) is equal to 260*federal minimum wage in year t. Detailed descriptions of each earnings measure are included in the text.

## Table 1B: Number of Observations by Analysis Sample

| Year | BS | CS | LX_1 | LX_5 | H_1 | H_5 | PA_5 | PA_10 |
|---|---|---|---|---|---|---|---|---|
| 1998 | 86,210,000 | 82,120,000 | 76,040,000 | 60,700,000 | | | | |
| 1999 | 87,250,000 | 83,330,000 | 77,140,000 | 61,930,000 | | | | |
| 2000 | 88,450,000 | 84,680,000 | 77,790,000 | 62,460,000 | | | 66,260,000 | 49,630,000 |
| 2001 | 88,390,000 | 84,700,000 | 76,940,000 | 62,500,000 | 66,630,000 | 53,850,000 | 66,330,000 | 48,680,000 |
| 2002 | 87,220,000 | 83,520,000 | 76,000,000 | 61,860,000 | 66,300,000 | 53,610,000 | 66,090,000 | 47,950,000 |
| 2003 | 86,550,000 | 82,860,000 | 76,100,000 | 61,080,000 | 66,360,000 | 53,010,000 | 65,650,000 | 47,480,000 |
| 2004 | 96,770,000 | 92,740,000 | 85,290,000 | 66,590,000 | 66,720,000 | 51,920,000 | 65,770,000 | 47,810,000 |
| 2005 | 97,270,000 | 93,330,000 | 85,850,000 | 65,990,000 | 66,990,000 | 51,420,000 | 65,110,000 | 47,770,000 |
| 2006 | 97,880,000 | 94,040,000 | 86,440,000 | 66,420,000 | 74,330,000 | 57,080,000 | 69,900,000 | 51,010,000 |
| 2007 | 98,200,000 | 94,460,000 | 86,390,000 | 66,880,000 | 74,660,000 | 57,650,000 | 69,540,000 | 51,100,000 |
| 2008 | 97,610,000 | 93,650,000 | 84,130,000 | 66,750,000 | 73,290,000 | 57,770,000 | 69,500,000 | 51,080,000 |
| 2009 | 94,560,000 | 90,130,000 | 81,410,000 | 65,140,000 | 71,540,000 | 56,800,000 | 69,080,000 | 50,760,000 |
| 2010 | 93,420,000 | 88,750,000 | 80,910,000 | 64,610,000 | 70,930,000 | 56,140,000 | 68,620,000 | |
| 2011 | 93,710,000 | 89,160,000 | 81,580,000 | 63,810,000 | 70,540,000 | 54,660,000 | 67,050,000 | |
| 2012 | 94,370,000 | 89,940,000 | 82,550,000 | 64,510,000 | 70,480,000 | 54,530,000 | 67,090,000 | |
| 2013 | 95,230,000 | 90,910,000 | 83,640,000 | 65,430,000 | 71,070,000 | 55,010,000 | 67,410,000 | |
| 2014 | 96,370,000 | 92,180,000 | 84,970,000 | 66,620,000 | 71,990,000 | 55,850,000 | 68,120,000 | |
| 2015 | 97,550,000 | 93,530,000 | 84,590,000 | | 71,570,000 | | | |
| 2016 | 96,660,000 | 92,790,000 | 85,490,000 | | 72,540,000 | | | |
| 2017 | 97,450,000 | 93,690,000 | 86,440,000 | | 73,670,000 | | | |
| 2018 | 98,310,000 | 94,720,000 | 87,440,000 | | 74,690,000 | | | |
| 2019 | 99,000,000 | 95,560,000 | | | | | | |

Notes: Analysis samples use 1998-2019 LEHD person-year earnings data. All states plus DC report earnings from 2004-2015. Prior to 2004 DE, IA, NW, UT, OK, VT, AL, MA, DC, AR, NH, MS have non-reporting quarters and are not included. After 2015 AK, AR, and MS have non-reporting quarters and are not included. See Table 1A for analysis sample descriptions.

# Table 2A: Foreign-Born Earnings and Activity Measures

| Race/Ethnicity | N | Percentile | Average Annual Earnings | Share Active Each 4 Year Period | Earnings Growth (Active Each 4 Year Period) | Earnings Volatility (Arc Pct Change) | Average Annual Hours Worked | Years Partially Active | Years Inactive | HC Fixed Person Effect | HC Average Firm Effect |
|---|---|---|---|---|---|---|---|---|---|---|---|
| | | | | | Foreign-Born Females | | | | | | |
| Asian Non-Hispanic | 2,416,000 | 10 | $1,062 | 0.06 | -0.08 | 3.24 | 598 | 1.96 | 9.45 | 0.01 | -0.04 |
| | | 25 | $6,283 | 0.46 | 0.06 | 1.64 | 1,029 | 2.64 | 5.28 | -0.05 | -0.05 |
| | | 50 | $21,870 | 0.87 | 0.07 | 0.62 | 1,617 | 1.56 | 1.53 | -0.03 | 0.01 |
| | | 75 | $49,200 | 0.96 | 0.17 | 0.30 | 1,988 | 0.91 | 0.63 | 0.12 | 0.17 |
| | | 90 | $88,620 | 0.99 | 0.20 | 0.17 | 2,059 | 0.64 | 0.27 | 0.45 | 0.28 |
| Black Non-Hispanic | 797,000 | 10 | $2,585 | 0.22 | -0.24 | 2.56 | 836 | 2.78 | 7.80 | 0.01 | -0.04 |
| | | 25 | $10,920 | 0.71 | -0.07 | 1.30 | 1,238 | 2.89 | 3.25 | 0.00 | -0.02 |
| | | 50 | $24,950 | 0.95 | 0.04 | 0.48 | 1,728 | 1.43 | 0.80 | 0.04 | 0.04 |
| | | 75 | $42,950 | 0.98 | 0.13 | 0.24 | 2,064 | 0.83 | 0.33 | 0.12 | 0.14 |
| | | 90 | $68,270 | 0.99 | 0.16 | 0.17 | 2,193 | 0.62 | 0.20 | 0.33 | 0.22 |
| White Hispanic | 3,213,000 | 10 | $646 | 0.06 | -0.15 | 3.69 | 429 | 2.02 | 9.72 | 0.00 | -0.08 |
| | | 25 | $3,884 | 0.35 | -0.06 | 2.08 | 844 | 3.13 | 6.31 | -0.04 | -0.07 |
| | | 50 | $12,880 | 0.85 | 0.05 | 0.85 | 1,285 | 2.49 | 1.91 | -0.04 | -0.06 |
| | | 75 | $25,170 | 0.97 | 0.08 | 0.33 | 1,771 | 1.05 | 0.54 | 0.00 | 0.00 |
| | | 90 | $41,470 | 0.98 | 0.11 | 0.21 | 2,048 | 0.70 | 0.33 | 0.08 | 0.10 |
| White Non-Hispanic | 1,935,000 | 10 | $680 | 0.05 | -0.18 | 3.60 | 442 | 1.85 | 9.78 | 0.06 | -0.04 |
| | | 25 | $4,937 | 0.34 | 0.01 | 1.86 | 919 | 2.56 | 6.25 | -0.01 | -0.03 |
| | | 50 | $19,330 | 0.83 | 0.05 | 0.74 | 1,454 | 1.77 | 1.92 | 0.01 | 0.01 |
| | | 75 | $42,400 | 0.95 | 0.12 | 0.32 | 1,852 | 0.95 | 0.68 | 0.15 | 0.11 |
| | | 90 | $74,350 | 0.97 | 0.17 | 0.21 | 1,943 | 0.68 | 0.39 | 0.43 | 0.22 |
| All Other | 533,000 | 10 | $666 | 0.06 | -0.15 | 3.72 | 471 | 1.99 | 9.74 | -0.03 | -0.07 |
| | | 25 | $4,315 | 0.37 | -0.11 | 2.03 | 902 | 3.03 | 6.17 | -0.08 | -0.06 |
| | | 50 | $14,840 | 0.87 | 0.02 | 0.78 | 1,412 | 2.15 | 1.71 | -0.08 | -0.04 |
| | | 75 | $29,890 | 0.97 | 0.09 | 0.32 | 1,906 | 1.02 | 0.54 | -0.05 | 0.04 |
| | | 90 | $50,650 | 0.98 | 0.13 | 0.21 | 2,264 | 0.67 | 0.40 | 0.05 | 0.17 |
| | | | | | Foreign-Born Males | | | | | | |
| Asian Non-Hispanic | 2,423,000 | 10 | $1,646 | 0.07 | -0.17 | 3.01 | 741 | 2.02 | 9.34 | -0.07 | -0.02 |
| | | 25 | $9,923 | 0.55 | -0.07 | 1.44 | 1,240 | 2.73 | 4.54 | -0.12 | -0.04 |
| | | 50 | $32,940 | 0.88 | 0.04 | 0.54 | 1,872 | 1.36 | 1.37 | -0.07 | 0.06 |
| | | 75 | $75,270 | 0.95 | 0.20 | 0.30 | 2,079 | 0.89 | 0.68 | 0.24 | 0.24 |
| | | 90 | $125,700 | 0.98 | 0.23 | 0.17 | 2,097 | 0.58 | 0.32 | 0.53 | 0.38 |
| Black Non-Hispanic | 794,000 | 10 | $1,509 | 0.10 | -0.30 | 3.03 | 749 | 2.25 | 9.20 | -0.05 | -0.03 |
| | | 25 | $8,709 | 0.46 | -0.29 | 1.74 | 1,232 | 3.08 | 5.26 | -0.07 | -0.02 |
| | | 50 | $25,230 | 0.91 | -0.05 | 0.62 | 1,758 | 1.76 | 1.20 | -0.06 | 0.01 |
| | | 75 | $45,790 | 0.97 | 0.11 | 0.27 | 2,151 | 0.92 | 0.43 | 0.03 | 0.10 |
| | | 90 | $72,960 | 0.98 | 0.17 | 0.19 | 2,226 | 0.67 | 0.29 | 0.21 | 0.23 |
| White Hispanic | 3,976,000 | 10 | $1,197 | 0.07 | -0.29 | 3.42 | 614 | 2.16 | 9.52 | -0.15 | -0.02 |
| | | 25 | $7,361 | 0.44 | -0.25 | 1.86 | 1,125 | 3.44 | 5.55 | -0.16 | -0.02 |
| | | 50 | $21,470 | 0.90 | -0.08 | 0.68 | 1,641 | 2.28 | 1.35 | -0.15 | -0.01 |
| | | 75 | $37,940 | 0.98 | 0.03 | 0.27 | 2,067 | 1.02 | 0.41 | -0.09 | 0.05 |
| | | 90 | $57,890 | 0.98 | 0.08 | 0.17 | 2,258 | 0.63 | 0.25 | 0.04 | 0.12 |
| White Non-Hispanic | 2,188,000 | 10 | $949 | 0.04 | -0.26 | 3.60 | 554 | 1.81 | 9.89 | -0.01 | -0.01 |
| | | 25 | $8,021 | 0.37 | -0.14 | 1.80 | 1,165 | 2.71 | 6.03 | -0.04 | 0.01 |
| | | 50 | $32,190 | 0.85 | -0.01 | 0.65 | 1,764 | 1.68 | 1.69 | 0.00 | 0.06 |
| | | 75 | $72,230 | 0.94 | 0.11 | 0.31 | 2,019 | 0.92 | 0.71 | 0.27 | 0.19 |
| | | 90 | $132,100 | 0.96 | 0.17 | 0.24 | 2,027 | 0.70 | 0.48 | 0.67 | 0.32 |
| All Other | 610,000 | 10 | $890 | 0.05 | -0.24 | 3.72 | 562 | 1.95 | 9.82 | -0.16 | -0.03 |
| | | 25 | $6,139 | 0.36 | -0.23 | 2.00 | 1,105 | 3.15 | 6.21 | -0.19 | -0.02 |
| | | 50 | $20,660 | 0.87 | -0.08 | 0.76 | 1,640 | 2.31 | 1.66 | -0.18 | -0.01 |
| | | 75 | $39,670 | 0.96 | 0.04 | 0.30 | 2,107 | 1.03 | 0.51 | -0.12 | 0.06 |
| | | 90 | $65,050 | 0.97 | 0.13 | 0.22 | 2,264 | 0.67 | 0.40 | 0.05 | 0.17 |

Notes: Estimates are created using the 108,800,000 worker sample 2. Average annual earnings show the percentile of worker average earnings at all jobs over the 12 year sample period. All measures except average annual earnings are calculated using the 2% of workers with earnings greater than the p-1 and less than the p+1 percentile. A worker is active each 4 year period if they have at least one quarter of positive earnings in each consecutive 4 year period. Earnings growth shows the percentage increase in average earnings from the first 4 year period to the last 4 year period. Earnings volatility is the variance of the year-to-year change in average annual earnings. Every year a worker is either full year active (earnings in all 4 quarters), partial year active (earnings in at least 1 quarter), or inactive (earnings in 0 quarters). HC fixed person effects and HC average firm effects are estimated using an AKM style earnings regression.

# Table 2B: Native-Born Earnings and Activity Measures

| Race/Ethnicity | N | Percentile | Average Annual Earnings | Share Active Each 4 Year Period | Earnings Growth (Active Each 4 Year Period) | Earnings Volatility (Arc Pct Change) | Average Annual Hours Worked | Years Partially Active | Years Inactive | HC Fixed Person Effect | HC Average Firm Effect |
|---|---|---|---|---|---|---|---|---|---|---|---|
| | | | | | Native-Born Females | | | | | | |
| Asian Non-Hispanic | 321,000 | 10 | $2,696 | 0.24 | -0.12 | 2.31 | 938 | 2.60 | 7.48 | -0.03 | 0.01 |
| | | 25 | $14,390 | 0.67 | -0.09 | 1.19 | 1,326 | 2.33 | 3.42 | -0.05 | 0.05 |
| | | 50 | $38,350 | 0.92 | 0.07 | 0.44 | 1,786 | 1.18 | 0.94 | 0.03 | 0.14 |
| | | 75 | $69,530 | 0.97 | 0.17 | 0.23 | 1,920 | 0.72 | 0.41 | 0.26 | 0.25 |
| | | 90 | $108,900 | 0.98 | 0.24 | 0.18 | 1,952 | 0.63 | 0.29 | 0.51 | 0.33 |
| Black Non-Hispanic | 6,311,000 | 10 | $1,208 | 0.23 | -0.23 | 2.99 | 673 | 3.12 | 8.03 | 0.01 | -0.07 |
| | | 25 | $6,807 | 0.65 | -0.13 | 1.59 | 1,009 | 3.49 | 3.89 | -0.03 | -0.05 |
| | | 50 | $19,320 | 0.93 | -0.09 | 0.58 | 1,473 | 1.73 | 1.01 | 0.00 | 0.01 |
| | | 75 | $35,090 | 0.97 | 0.00 | 0.25 | 1,861 | 0.87 | 0.38 | 0.08 | 0.10 |
| | | 90 | $54,370 | 0.99 | 0.05 | 0.17 | 1,987 | 0.62 | 0.23 | 0.23 | 0.19 |
| White Hispanic | 2,600,000 | 10 | $1,658 | 0.24 | -0.15 | 2.76 | 662 | 2.99 | 7.82 | -0.09 | -0.06 |
| | | 25 | $7,922 | 0.65 | -0.03 | 1.49 | 1,023 | 3.16 | 3.81 | -0.13 | -0.03 |
| | | 50 | $21,590 | 0.93 | 0.00 | 0.56 | 1,535 | 1.64 | 1.02 | -0.10 | 0.03 |
| | | 75 | $38,970 | 0.98 | 0.07 | 0.23 | 1,890 | 0.79 | 0.35 | -0.01 | 0.11 |
| | | 90 | $59,630 | 0.99 | 0.12 | 0.15 | 1,999 | 0.51 | 0.20 | 0.16 | 0.19 |
| White Non-Hispanic | 34,340,000 | 10 | $1,727 | 0.22 | -0.14 | 2.51 | 607 | 2.73 | 7.77 | -0.06 | -0.07 |
| | | 25 | $8,553 | 0.63 | -0.02 | 1.32 | 1,006 | 2.67 | 3.80 | -0.11 | -0.04 |
| | | 50 | $23,790 | 0.91 | -0.02 | 0.48 | 1,515 | 1.33 | 1.05 | -0.06 | 0.01 |
| | | 75 | $44,180 | 0.97 | 0.05 | 0.22 | 1,791 | 0.71 | 0.41 | 0.11 | 0.10 |
| | | 90 | $69,010 | 0.99 | 0.10 | 0.15 | 1,869 | 0.51 | 0.24 | 0.33 | 0.17 |
| All Other | 1,348,000 | 10 | $843 | 0.15 | -0.29 | 3.33 | 582 | 2.69 | 8.79 | -0.07 | -0.07 |
| | | 25 | $4,946 | 0.51 | -0.14 | 1.97 | 921 | 3.58 | 5.20 | -0.12 | -0.06 |
| | | 50 | $16,970 | 0.88 | -0.04 | 0.81 | 1,374 | 2.23 | 1.62 | -0.13 | -0.01 |
| | | 75 | $34,450 | 0.97 | 0.06 | 0.31 | 1,817 | 1.01 | 0.47 | -0.06 | 0.08 |
| | | 90 | $55,400 | 0.99 | 0.13 | 0.19 | 1,965 | 0.66 | 0.27 | 0.10 | 0.17 |
| | | | | | Native-Born Males | | | | | | |
| Asian Non-Hispanic | 352,000 | 10 | $3,905 | 0.29 | -0.27 | 2.42 | 1,035 | 2.89 | 7.21 | -0.20 | 0.01 |
| | | 25 | $19,900 | 0.76 | -0.05 | 1.05 | 1,506 | 2.32 | 2.69 | -0.20 | 0.03 |
| | | 50 | $49,030 | 0.94 | 0.11 | 0.36 | 1,949 | 1.02 | 0.74 | -0.08 | 0.14 |
| | | 75 | $89,870 | 0.98 | 0.18 | 0.20 | 2,035 | 0.64 | 0.38 | 0.18 | 0.29 |
| | | 90 | $146,400 | 0.98 | 0.28 | 0.19 | 2,018 | 0.62 | 0.29 | 0.49 | 0.38 |
| Black Non-Hispanic | 5,757,000 | 10 | $617 | 0.14 | -0.22 | 3.77 | 434 | 2.59 | 9.13 | -0.01 | -0.10 |
| | | 25 | $3,927 | 0.45 | -0.16 | 2.37 | 820 | 3.84 | 6.03 | -0.06 | -0.06 |
| | | 50 | $16,780 | 0.84 | -0.09 | 0.98 | 1,392 | 2.66 | 1.99 | -0.07 | -0.01 |
| | | 75 | $36,410 | 0.96 | -0.02 | 0.32 | 1,906 | 1.07 | 0.51 | 0.07 | 0.07 |
| | | 90 | $59,180 | 0.98 | 0.04 | 0.19 | 2,093 | 0.66 | 0.28 | 0.17 | 0.16 |
| White Hispanic | 2,556,000 | 10 | $2,065 | 0.25 | -0.31 | 2.93 | 762 | 3.20 | 7.87 | -0.20 | -0.02 |
| | | 25 | $10,550 | 0.65 | -0.17 | 1.60 | 1,201 | 3.49 | 3.93 | -0.22 | 0.00 |
| | | 50 | $29,160 | 0.94 | -0.01 | 0.54 | 1,760 | 1.68 | 0.94 | -0.19 | 0.05 |
| | | 75 | $52,060 | 0.98 | 0.08 | 0.22 | 2,071 | 0.77 | 0.33 | -0.06 | 0.13 |
| | | 90 | $80,540 | 0.99 | 0.13 | 0.14 | 2,156 | 0.49 | 0.20 | 0.13 | 0.22 |
| White Non-Hispanic | 35,000,000 | 10 | $3,469 | 0.28 | -0.32 | 2.44 | 887 | 3.06 | 7.29 | -0.21 | -0.01 |
| | | 25 | $16,370 | 0.71 | -0.20 | 1.15 | 1,345 | 2.73 | 3.07 | -0.22 | 0.00 |
| | | 50 | $38,960 | 0.94 | -0.04 | 0.36 | 1,823 | 1.16 | 0.73 | -0.13 | 0.05 |
| | | 75 | $67,890 | 0.98 | 0.05 | 0.19 | 1,988 | 0.65 | 0.34 | 0.09 | 0.15 |
| | | 90 | $110,400 | 0.98 | 0.11 | 0.16 | 2,002 | 0.52 | 0.27 | 0.39 | 0.24 |
| All Other | 1,284,000 | 10 | $980 | 0.16 | -0.34 | 3.51 | 569 | 2.78 | 8.80 | -0.21 | -0.05 |
| | | 25 | $6,121 | 0.51 | -0.26 | 2.07 | 978 | 3.82 | 5.31 | -0.26 | -0.02 |
| | | 50 | $22,140 | 0.88 | -0.08 | 0.81 | 1,538 | 2.37 | 1.60 | -0.25 | 0.02 |
| | | 75 | $45,210 | 0.97 | 0.05 | 0.29 | 1,957 | 1.03 | 0.48 | -0.14 | 0.10 |
| | | 90 | $73,890 | 0.99 | 0.12 | 0.18 | 2,074 | 0.64 | 0.27 | 0.05 | 0.21 |

Notes: Estimates are created using the 108,800,000 worker sample 2. Average annual earnings show the percentile of worker average earnings at all jobs over the 12 year sample period. All measures except average annual earnings are calculated using the 2% of workers with earnings greater than the p-1 and less than the p+1 percentile. A worker is active each 4 year period if they have at least one quarter of positive earnings in each consecutive 4 year period. Earnings growth shows the percentage increase in average earnings from the first 4 year period to the last 4 year period. Earnings volatility is the variance of the year-to-year change in average annual earnings. Every year a worker is either full year active (earnings in all 4 quarters), partial year active (earnings in at least 1 quarter), or inactive (earnings in 0 quarters). HC fixed person effects and HC average firm effects are estimated using an AKM style earnings regression.

# Table 3A: Foreign-Born Job and Worker Characteristics

| Race/Ethnicity | N | Percentile | Division (Top 2) First | Second | Share | Industry (Top 4) First | Second | Third | Fourth | Share | Age (2004) | Education <HS | BA+ |
|---|---|---|---|---|---|---|---|---|---|---|---|---|---|
| | | | | | | | Foreign Born Females | | | | | | |
| Asian Non-Hispanic | 2,416,000 | 10 | 9 | 2 | 0.55 | R | S | G | P | 0.56 | 39.10 | 0.24 | 0.34 |
| | | 25 | 9 | 2 | 0.58 | R | S | G | P | 0.58 | 39.09 | 0.24 | 0.33 |
| | | 50 | 9 | 2 | 0.55 | P | E | G | R | 0.58 | 39.30 | 0.19 | 0.36 |
| | | 75 | 9 | 2 | 0.61 | P | E | O | L | 0.56 | 37.75 | 0.07 | 0.59 |
| | | 90 | 9 | 2 | 0.61 | P | L | E | J | 0.68 | 37.31 | 0.02 | 0.78 |
| Black Non-Hispanic | 797,000 | 10 | 5 | 2 | 0.66 | P | N | R | G | 0.66 | 38.13 | 0.25 | 0.17 |
| | | 25 | 5 | 2 | 0.70 | P | R | G | N | 0.68 | 37.83 | 0.23 | 0.17 |
| | | 50 | 5 | 2 | 0.71 | P | G | O | R | 0.71 | 38.18 | 0.18 | 0.20 |
| | | 75 | 2 | 5 | 0.72 | P | O | J | T | 0.72 | 38.69 | 0.10 | 0.32 |
| | | 90 | 2 | 5 | 0.72 | P | O | J | L | 0.78 | 39.09 | 0.06 | 0.47 |
| White Hispanic | 3,213,000 | 10 | 9 | 5 | 0.53 | N | R | P | G | 0.55 | 39.17 | 0.58 | 0.08 |
| | | 25 | 9 | 5 | 0.53 | N | P | R | E | 0.52 | 38.77 | 0.57 | 0.08 |
| | | 50 | 9 | 5 | 0.51 | P | E | R | G | 0.55 | 38.61 | 0.54 | 0.08 |
| | | 75 | 9 | 5 | 0.50 | P | E | G | R | 0.58 | 38.33 | 0.44 | 0.11 |
| | | 90 | 9 | 5 | 0.55 | P | O | E | J | 0.54 | 37.95 | 0.26 | 0.22 |
| White Non-Hispanic | 1,935,000 | 10 | 9 | 5 | 0.42 | G | P | R | O | 0.51 | 39.14 | 0.18 | 0.30 |
| | | 25 | 9 | 5 | 0.41 | P | G | O | R | 0.54 | 38.90 | 0.15 | 0.31 |
| | | 50 | 9 | 2 | 0.38 | P | G | O | E | 0.56 | 39.80 | 0.13 | 0.31 |
| | | 75 | 9 | 2 | 0.40 | P | O | L | J | 0.56 | 39.60 | 0.06 | 0.46 |
| | | 90 | 9 | 2 | 0.47 | P | O | L | J | 0.66 | 39.52 | 0.03 | 0.64 |
| All Other | 533,000 | 10 | 9 | 2 | 0.48 | N | R | P | G | 0.56 | 38.67 | 0.43 | 0.16 |
| | | 25 | 9 | 2 | 0.51 | P | N | R | G | 0.56 | 38.15 | 0.43 | 0.15 |
| | | 50 | 2 | 9 | 0.50 | P | E | R | G | 0.59 | 38.03 | 0.41 | 0.14 |
| | | 75 | 9 | 2 | 0.50 | P | E | O | R | 0.57 | 37.95 | 0.28 | 0.21 |
| | | 90 | 9 | 2 | 0.57 | P | O | J | E | 0.57 | 37.82 | 0.15 | 0.38 |
| | | | | | | | Foreign Born Males | | | | | | |
| Asian Non-Hispanic | 2,423,000 | 10 | 9 | 2 | 0.57 | R | G | S | N | 0.50 | 39.16 | 0.23 | 0.36 |
| | | 25 | 9 | 2 | 0.56 | R | G | E | L | 0.59 | 39.00 | 0.23 | 0.35 |
| | | 50 | 9 | 2 | 0.54 | E | G | R | L | 0.54 | 38.54 | 0.16 | 0.40 |
| | | 75 | 9 | 2 | 0.54 | L | E | P | O | 0.58 | 36.87 | 0.05 | 0.68 |
| | | 90 | 9 | 2 | 0.58 | L | E | J | I | 0.66 | 36.73 | 0.02 | 0.85 |
| Black Non-Hispanic | 794,000 | 10 | 5 | 2 | 0.62 | N | G | R | D | 0.52 | 38.69 | 0.24 | 0.20 |
| | | 25 | 5 | 2 | 0.65 | N | G | P | R | 0.48 | 38.14 | 0.24 | 0.20 |
| | | 50 | 5 | 2 | 0.67 | P | G | E | N | 0.47 | 38.31 | 0.22 | 0.20 |
| | | 75 | 2 | 5 | 0.66 | P | E | H | O | 0.46 | 38.67 | 0.14 | 0.29 |
| | | 90 | 2 | 5 | 0.67 | O | P | H | E | 0.50 | 39.51 | 0.09 | 0.42 |
| White Hispanic | 3,976,000 | 10 | 9 | 5 | 0.50 | D | N | R | E | 0.58 | 38.94 | 0.60 | 0.07 |
| | | 25 | 9 | 5 | 0.53 | D | N | E | R | 0.57 | 38.96 | 0.61 | 0.07 |
| | | 50 | 9 | 5 | 0.51 | E | D | N | R | 0.56 | 38.99 | 0.60 | 0.07 |
| | | 75 | 9 | 5 | 0.50 | E | D | G | F | 0.55 | 38.06 | 0.52 | 0.09 |
| | | 90 | 9 | 7 | 0.52 | E | D | F | H | 0.52 | 37.98 | 0.42 | 0.14 |
| White Non-Hispanic | 2,188,000 | 10 | 9 | 2 | 0.42 | R | D | G | N | 0.50 | 39.26 | 0.21 | 0.29 |
| | | 25 | 9 | 2 | 0.43 | G | D | R | E | 0.47 | 38.87 | 0.18 | 0.31 |
| | | 50 | 2 | 5 | 0.37 | E | G | D | R | 0.46 | 39.50 | 0.13 | 0.34 |
| | | 75 | 2 | 9 | 0.42 | E | L | O | D | 0.49 | 39.69 | 0.06 | 0.52 |
| | | 90 | 9 | 2 | 0.47 | L | E | F | J | 0.57 | 39.72 | 0.03 | 0.73 |
| All Other | 610,000 | 10 | 9 | 2 | 0.46 | N | D | R | G | 0.55 | 38.83 | 0.45 | 0.15 |
| | | 25 | 9 | 2 | 0.48 | N | D | E | G | 0.53 | 38.32 | 0.46 | 0.13 |
| | | 50 | 9 | 2 | 0.48 | E | D | N | R | 0.52 | 38.35 | 0.46 | 0.13 |
| | | 75 | 9 | 2 | 0.49 | E | D | G | R | 0.48 | 37.81 | 0.38 | 0.18 |
| | | 90 | 9 | 2 | 0.56 | E | D | O | H | 0.42 | 37.87 | 0.24 | 0.31 |

Notes: Estimates are created using the 108,800,000 worker sample 2. All measures are calculated using the 2% of workers with earnings greater than the p-1 and less than the p+1 percentile. Please see Appendix Table B2 for definitions of the division and industry codes. The Share shows the percent of workers in the top 2 divisions or the top 4 industries.

# Table 3B: Native-Born Job and Worker Characteristics

| Race/Ethnicity | N | Percentile | Division (Top 2) | | | Industry (Top 4) | | | | | Age (2004) | Education | |
|---|---|---|---|---|---|---|---|---|---|---|---|---|---|
| | | | First | Second | Share | First | Second | Third | Fourth | Share | | <HS | BA+ |
| Native-Born Females | | | | | | | | | | | | | |
| Asian Non-Hispanic | 321,000 | 10 | 9 | 5 | 0.65 | P | G | O | R | 0.52 | 36.77 | 0.13 | 0.33 |
| | | 25 | 9 | 5 | 0.65 | P | O | G | L | 0.54 | 36.30 | 0.10 | 0.35 |
| | | 50 | 9 | 2 | 0.70 | P | O | L | J | 0.55 | 35.91 | 0.05 | 0.43 |
| | | 75 | 9 | 2 | 0.74 | P | O | L | J | 0.62 | 35.87 | 0.02 | 0.62 |
| | | 90 | 9 | 2 | 0.74 | P | L | J | E | 0.60 | 35.24 | 0.02 | 0.74 |
| Black Non-Hispanic | 6,311,000 | 10 | 5 | 3 | 0.49 | P | N | R | G | 0.65 | 38.73 | 0.26 | 0.10 |
| | | 25 | 5 | 3 | 0.50 | P | R | N | G | 0.64 | 37.98 | 0.22 | 0.10 |
| | | 50 | 5 | 7 | 0.52 | P | O | G | E | 0.63 | 38.34 | 0.15 | 0.12 |
| | | 75 | 5 | 3 | 0.50 | P | O | T | J | 0.62 | 38.57 | 0.08 | 0.20 |
| | | 90 | 5 | 2 | 0.49 | O | P | T | J | 0.64 | 39.00 | 0.04 | 0.37 |
| White Hispanic | 2,600,000 | 10 | 7 | 9 | 0.56 | P | G | R | N | 0.58 | 36.70 | 0.26 | 0.12 |
| | | 25 | 7 | 9 | 0.56 | P | G | O | R | 0.59 | 36.22 | 0.23 | 0.12 |
| | | 50 | 7 | 9 | 0.56 | P | O | G | J | 0.58 | 36.43 | 0.16 | 0.14 |
| | | 75 | 9 | 7 | 0.56 | P | O | T | J | 0.59 | 36.32 | 0.09 | 0.23 |
| | | 90 | 9 | 7 | 0.58 | O | P | T | J | 0.61 | 36.94 | 0.04 | 0.38 |
| White Non-Hispanic | 34,340,000 | 10 | 5 | 3 | 0.37 | G | P | O | R | 0.57 | 39.38 | 0.12 | 0.22 |
| | | 25 | 3 | 5 | 0.38 | P | G | O | R | 0.59 | 39.24 | 0.10 | 0.21 |
| | | 50 | 3 | 5 | 0.38 | P | O | G | E | 0.59 | 39.97 | 0.06 | 0.23 |
| | | 75 | 5 | 3 | 0.36 | O | P | J | E | 0.61 | 39.79 | 0.02 | 0.40 |
| | | 90 | 3 | 2 | 0.34 | P | O | J | L | 0.67 | 40.39 | 0.01 | 0.58 |
| All Other | 1,348,000 | 10 | 9 | 8 | 0.42 | P | G | R | N | 0.57 | 37.66 | 0.23 | 0.13 |
| | | 25 | 9 | 8 | 0.42 | P | G | R | N | 0.57 | 36.78 | 0.20 | 0.13 |
| | | 50 | 9 | 8 | 0.42 | P | G | O | R | 0.55 | 36.78 | 0.14 | 0.15 |
| | | 75 | 9 | 7 | 0.44 | P | O | T | J | 0.56 | 36.86 | 0.08 | 0.22 |
| | | 90 | 9 | 2 | 0.49 | O | P | T | J | 0.59 | 37.34 | 0.04 | 0.38 |
| Native-Born Males | | | | | | | | | | | | | |
| Asian Non-Hispanic | 352,000 | 10 | 9 | 5 | 0.66 | G | N | R | D | 0.44 | 36.15 | 0.15 | 0.28 |
| | | 25 | 9 | 5 | 0.65 | G | R | L | P | 0.41 | 35.52 | 0.11 | 0.29 |
| | | 50 | 9 | 2 | 0.70 | G | L | E | O | 0.39 | 35.88 | 0.06 | 0.38 |
| | | 75 | 9 | 2 | 0.74 | L | E | T | P | 0.50 | 36.15 | 0.02 | 0.61 |
| | | 90 | 9 | 2 | 0.73 | L | E | P | J | 0.63 | 36.52 | 0.01 | 0.76 |
| Black Non-Hispanic | 5,757,000 | 10 | 5 | 3 | 0.48 | N | R | D | G | 0.59 | 37.98 | 0.28 | 0.09 |
| | | 25 | 5 | 3 | 0.48 | N | R | E | D | 0.55 | 37.55 | 0.26 | 0.09 |
| | | 50 | 5 | 7 | 0.50 | E | N | G | H | 0.48 | 37.92 | 0.21 | 0.09 |
| | | 75 | 5 | 7 | 0.51 | E | T | H | P | 0.48 | 38.55 | 0.14 | 0.14 |
| | | 90 | 5 | 7 | 0.47 | E | T | H | O | 0.54 | 39.20 | 0.08 | 0.22 |
| White Hispanic | 2,556,000 | 10 | 9 | 7 | 0.53 | D | N | G | R | 0.54 | 36.42 | 0.29 | 0.11 |
| | | 25 | 9 | 7 | 0.54 | D | G | E | N | 0.51 | 35.76 | 0.25 | 0.11 |
| | | 50 | 7 | 9 | 0.55 | E | G | D | F | 0.46 | 35.91 | 0.20 | 0.12 |
| | | 75 | 9 | 7 | 0.58 | E | D | T | O | 0.45 | 36.33 | 0.13 | 0.19 |
| | | 90 | 9 | 7 | 0.58 | T | E | D | O | 0.45 | 37.05 | 0.07 | 0.32 |
| White Non-Hispanic | 35,000,000 | 10 | 5 | 3 | 0.36 | D | G | N | E | 0.52 | 39.30 | 0.17 | 0.18 |
| | | 25 | 5 | 3 | 0.38 | D | E | G | N | 0.50 | 39.12 | 0.14 | 0.18 |
| | | 50 | 3 | 5 | 0.38 | E | D | G | F | 0.50 | 39.21 | 0.09 | 0.21 |
| | | 75 | 3 | 5 | 0.36 | E | D | O | L | 0.47 | 39.62 | 0.04 | 0.36 |
| | | 90 | 3 | 2 | 0.33 | E | L | F | J | 0.51 | 40.39 | 0.02 | 0.60 |
| All Other | 1,284,000 | 10 | 9 | 8 | 0.42 | N | D | R | G | 0.55 | 37.63 | 0.25 | 0.11 |
| | | 25 | 9 | 8 | 0.43 | D | N | G | E | 0.50 | 36.64 | 0.22 | 0.11 |
| | | 50 | 9 | 8 | 0.42 | E | D | G | T | 0.45 | 36.57 | 0.16 | 0.13 |
| | | 75 | 9 | 7 | 0.47 | E | D | T | G | 0.45 | 37.05 | 0.10 | 0.20 |
| | | 90 | 9 | 7 | 0.51 | E | T | D | L | 0.45 | 37.61 | 0.05 | 0.33 |

Notes: Estimates are created using the 108,800,000 worker sample 2. All measures are calculated using the 2% of workers with earnings greater than the p-1 and less than the p+1 percentile. Please see Appendix Table B2 for definitions of the division and industry codes. The Share shows the percent of workers in the top 2 divisions or the top 4 industries.

# Table 4: Average Annual Earnings OLS Regression Estimates

| Parameter | Model 1 | Model 2 | Model 3 | Model 4 | Model 5 |
|---|---|---|---|---|---|
| Intercept | 10.17 | 6.683 | 6.52 | 6.465 | 6.836 |
| **Foreign-Born Females** | | | | | |
| Asian Non Hispanic | -0.64 | -0.18 | -0.20 | -0.23 | -0.30 |
| Black Non-Hispanic | -0.43 | -0.40 | -0.44 | -0.41 | -0.49 |
| White Hispanic | -1.21 | -0.42 | -0.39 | -0.33 | -0.40 |
| White Non-Hispanic | -0.84 | -0.10 | -0.15 | -0.16 | -0.28 |
| All Other | -1.08 | -0.43 | -0.43 | -0.38 | -0.39 |
| **Foreign-Born Males** | | | | | |
| Asian Non Hispanic | -0.23 | -0.08 | -0.10 | -0.13 | -0.20 |
| Black Non-Hispanic | -0.53 | -0.39 | -0.39 | -0.37 | -0.38 |
| White Hispanic | -0.71 | -0.46 | -0.41 | -0.35 | -0.31 |
| White Non-Hispanic | -0.35 | 0.03 | 0.01 | -0.02 | -0.18 |
| All Other | -0.78 | -0.43 | -0.40 | -0.36 | -0.27 |
| **Native-Born Females** | | | | | |
| Asian Non Hispanic | -0.08 | 0.04 | -0.05 | -0.05 | -0.18 |
| Black Non-Hispanic | -0.80 | -0.32 | -0.33 | -0.29 | -0.41 |
| White Hispanic | -0.64 | -0.27 | -0.30 | -0.25 | -0.26 |
| White Non-Hispanic | -0.54 | -0.05 | -0.09 | -0.09 | -0.14 |
| All Other | -0.94 | -0.26 | -0.27 | -0.23 | -0.22 |
| **Native-Born Males** | | | | | |
| Asian Non Hispanic | 0.22 | 0.08 | 0.03 | 0.02 | -0.01 |
| Black Non-Hispanic | -1.02 | -0.32 | -0.28 | -0.24 | -0.32 |
| White Hispanic | -0.36 | -0.27 | -0.27 | -0.22 | -0.15 |
| White Non-Hispanic | 0.00 | 0.00 | 0.00 | 0.00 | 0.00 |
| All Other | -0.70 | -0.22 | -0.21 | -0.17 | -0.07 |
| **Covariates** | | | | | |
| Hours | No | Yes | Yes | Yes | Yes |
| Division/Industry | No | No | Yes | Yes | Yes |
| Age and Education | No | No | No | Yes | No |
| Age, HC Fixed Person, and Firm Effects | No | No | No | No | Yes |
| **Summary Statistics** | | | | | |
| R2 | 0.04 | 0.87 | 0.89 | 0.89 | 0.93 |
| Observations | 108,800,000 | 108,800,000 | 108,800,000 | 108,800,000 | 108,800,000 |

Notes: Estimates are created using the 108,800,000 worker sample 2. The hours covariates include a quartic in hours worked, paritial years worked, and inactive years worked. The region/industry covariates include indicator variables for 9 Census Divisions and 21 NAICS 2017 industry sectors. Please see Appendix Table B2 for definitions of the division and industry codes. The age and education covariates include age in 2004 and education indicator variables for less than HS, HS grad, some college, and BA+. The Age, HC Fixed Person and HC Average Firm effects include age in 2004 and the fixed effects from an AKM style earnings regression. Due to the large sample size, standard errors are not reported.

## Table 5A: Foreign-Born Earnings Simulation

| Race/Ethnicity | Percentile θ | Q(θ)\|g - Q(θ)\|0 | Q(θ)\|β(g),x(g) | Q(θ)\|β(0),x(0) | Q(θ)\|β(0),x(g) | Q(θ)\|β(g),x(0) | Predicted Diff Log Earn | Residual Diff Log Earn |
|---|---|---|---|---|---|---|---|---|
| | | | **Foreign-Born Females** | | | | | |
| Asian Non-Hispanic | 10 | -1.18 | 6.77 | 7.91 | 6.94 | 7.75 | -1.14 | -0.04 |
| | 25 | -0.96 | 8.62 | 9.71 | 8.78 | 9.53 | -1.09 | 0.13 |
| | 50 | -0.58 | 10.11 | 10.64 | 10.32 | 10.41 | -0.53 | -0.05 |
| | 75 | -0.32 | 10.77 | 11.08 | 11.05 | 10.82 | -0.31 | -0.01 |
| | 90 | -0.22 | 11.26 | 11.56 | 11.67 | 11.23 | -0.30 | 0.08 |
| Black Non-Hispanic | 10 | -0.29 | 7.63 | 7.91 | 7.77 | 7.75 | -0.28 | -0.02 |
| | 25 | -0.40 | 9.25 | 9.71 | 9.48 | 9.43 | -0.46 | 0.06 |
| | 50 | -0.45 | 10.20 | 10.64 | 10.53 | 10.28 | -0.44 | -0.01 |
| | 75 | -0.46 | 10.64 | 11.08 | 11.14 | 10.61 | -0.44 | -0.02 |
| | 90 | -0.48 | 11.03 | 11.56 | 11.77 | 10.96 | -0.53 | 0.05 |
| White Hispanic | 10 | -1.68 | 6.37 | 7.91 | 6.48 | 7.72 | -1.54 | -0.14 |
| | 25 | -1.44 | 8.06 | 9.71 | 8.22 | 9.41 | -1.65 | 0.21 |
| | 50 | -1.11 | 9.49 | 10.64 | 9.73 | 10.22 | -1.16 | 0.05 |
| | 75 | -0.99 | 10.20 | 11.08 | 10.60 | 10.54 | -0.88 | -0.11 |
| | 90 | -0.98 | 10.59 | 11.56 | 11.16 | 10.83 | -0.97 | -0.01 |
| White Non-Hispanic | 10 | -1.63 | 6.35 | 7.91 | 6.56 | 7.80 | -1.56 | -0.07 |
| | 25 | -1.20 | 8.34 | 9.71 | 8.43 | 9.63 | -1.37 | 0.17 |
| | 50 | -0.70 | 9.96 | 10.64 | 10.06 | 10.47 | -0.68 | -0.02 |
| | 75 | -0.47 | 10.66 | 11.08 | 10.85 | 10.88 | -0.42 | -0.05 |
| | 90 | -0.40 | 11.12 | 11.56 | 11.43 | 11.32 | -0.44 | 0.04 |
| All Other | 10 | -1.65 | 6.38 | 7.91 | 6.55 | 7.69 | -1.53 | -0.12 |
| | 25 | -1.33 | 8.18 | 9.71 | 8.40 | 9.39 | -1.53 | 0.20 |
| | 50 | -0.97 | 9.66 | 10.64 | 9.97 | 10.22 | -0.98 | 0.02 |
| | 75 | -0.82 | 10.36 | 11.08 | 10.81 | 10.56 | -0.72 | -0.10 |
| | 90 | -0.78 | 10.76 | 11.56 | 11.39 | 10.90 | -0.80 | 0.02 |
| | | | **Foreign-Born Males** | | | | | |
| Asian Non-Hispanic | 10 | -0.75 | 7.17 | 7.91 | 7.33 | 7.78 | -0.74 | 0.00 |
| | 25 | -0.50 | 9.17 | 9.71 | 9.32 | 9.60 | -0.54 | 0.04 |
| | 50 | -0.17 | 10.50 | 10.64 | 10.66 | 10.50 | -0.14 | -0.03 |
| | 75 | 0.10 | 11.14 | 11.08 | 11.26 | 10.97 | 0.06 | 0.04 |
| | 90 | 0.13 | 11.69 | 11.56 | 11.85 | 11.45 | 0.13 | 0.00 |
| Black Non-Hispanic | 10 | -0.83 | 7.07 | 7.91 | 7.25 | 7.75 | -0.84 | 0.00 |
| | 25 | -0.63 | 9.00 | 9.71 | 9.24 | 9.49 | -0.72 | 0.08 |
| | 50 | -0.43 | 10.22 | 10.64 | 10.53 | 10.34 | -0.42 | -0.01 |
| | 75 | -0.39 | 10.71 | 11.08 | 11.14 | 10.69 | -0.37 | -0.02 |
| | 90 | -0.41 | 11.12 | 11.56 | 11.72 | 11.06 | -0.44 | 0.03 |
| White Hispanic | 10 | -1.06 | 6.87 | 7.91 | 6.99 | 7.75 | -1.05 | -0.02 |
| | 25 | -0.80 | 8.77 | 9.71 | 9.01 | 9.45 | -0.94 | 0.14 |
| | 50 | -0.60 | 10.04 | 10.64 | 10.37 | 10.32 | -0.60 | 0.00 |
| | 75 | -0.58 | 10.56 | 11.08 | 10.97 | 10.64 | -0.52 | -0.06 |
| | 90 | -0.65 | 10.91 | 11.56 | 11.45 | 11.00 | -0.65 | 0.00 |
| White Non-Hispanic | 10 | -1.30 | 6.61 | 7.91 | 6.80 | 7.81 | -1.30 | 0.00 |
| | 25 | -0.71 | 8.91 | 9.71 | 8.97 | 9.70 | -0.80 | 0.09 |
| | 50 | -0.19 | 10.48 | 10.64 | 10.52 | 10.59 | -0.16 | -0.03 |
| | 75 | 0.06 | 11.13 | 11.08 | 11.14 | 11.09 | 0.05 | 0.01 |
| | 90 | 0.18 | 11.74 | 11.56 | 11.69 | 11.63 | 0.18 | 0.00 |
| All Other | 10 | -1.36 | 6.60 | 7.91 | 6.78 | 7.73 | -1.31 | -0.05 |
| | 25 | -0.98 | 8.58 | 9.71 | 8.83 | 9.46 | -1.14 | 0.15 |
| | 50 | -0.63 | 10.01 | 10.64 | 10.34 | 10.32 | -0.63 | 0.00 |
| | 75 | -0.54 | 10.61 | 11.08 | 11.02 | 10.68 | -0.47 | -0.07 |
| | 90 | -0.53 | 11.02 | 11.56 | 11.55 | 11.08 | -0.54 | 0.01 |

Notes: Estimates are created using the 108,800,000 worker sample 2. All measures are calculated using the 2% of workers with earnings greater than the p-1 and less than the p+1 percentile. See section IV.c.i. of the paper for more details.

## Table 5B: Native-Born Earnings Simulation

| Race/Ethnicity | Percentile θ | Q(θ)\|g - Q(θ)\|0 | Q(θ)\|β(g),x(g) | Q(θ)\|β(0),x(0) | Q(θ)\|β(0),x(g) | Q(θ)\|β(g),x(0) | Predicted Diff Log Earn | Residual Diff Log Earn |
|---|---|---|---|---|---|---|---|---|
| | | | Predicted Log Earnings at Quantile Q(θ) | | | | | |
| | | | *Native-Born Females* | | | | | |
| | 10 | -0.25 | 7.67 | 7.91 | 7.75 | 7.82 | -0.24 | -0.01 |
| | 25 | -0.13 | 9.60 | 9.71 | 9.63 | 9.69 | -0.11 | -0.02 |
| Asian Non-Hispanic | 50 | -0.02 | 10.60 | 10.64 | 10.63 | 10.60 | -0.04 | 0.02 |
| | 75 | 0.02 | 11.08 | 11.08 | 11.14 | 11.07 | 0.00 | 0.02 |
| | 90 | -0.01 | 11.56 | 11.56 | 11.70 | 11.55 | 0.00 | -0.01 |
| | 10 | -1.05 | 6.89 | 7.91 | 6.94 | 7.79 | -1.02 | -0.04 |
| | 25 | -0.88 | 8.69 | 9.71 | 8.80 | 9.49 | -1.02 | 0.14 |
| Black Non-Hispanic | 50 | -0.70 | 9.94 | 10.64 | 10.15 | 10.34 | -0.70 | 0.00 |
| | 75 | -0.66 | 10.49 | 11.08 | 10.82 | 10.67 | -0.59 | -0.07 |
| | 90 | -0.71 | 10.83 | 11.56 | 11.36 | 11.00 | -0.73 | 0.02 |
| | 10 | -0.74 | 7.19 | 7.91 | 7.23 | 7.80 | -0.72 | -0.02 |
| | 25 | -0.73 | 8.87 | 9.71 | 8.99 | 9.50 | -0.84 | 0.12 |
| White Hispanic | 50 | -0.59 | 10.06 | 10.64 | 10.26 | 10.37 | -0.58 | -0.01 |
| | 75 | -0.56 | 10.57 | 11.08 | 10.87 | 10.72 | -0.51 | -0.05 |
| | 90 | -0.62 | 10.91 | 11.56 | 11.39 | 11.06 | -0.65 | 0.03 |
| | 10 | -0.70 | 7.22 | 7.91 | 7.22 | 7.88 | -0.69 | 0.00 |
| | 25 | -0.65 | 8.96 | 9.71 | 8.96 | 9.64 | -0.75 | 0.10 |
| White Non-Hispanic | 50 | -0.49 | 10.17 | 10.64 | 10.21 | 10.52 | -0.47 | -0.02 |
| | 75 | -0.43 | 10.68 | 11.08 | 10.80 | 10.91 | -0.40 | -0.03 |
| | 90 | -0.47 | 11.05 | 11.56 | 11.33 | 11.29 | -0.51 | 0.04 |
| | 10 | -1.41 | 6.59 | 7.91 | 6.64 | 7.80 | -1.32 | -0.09 |
| | 25 | -1.20 | 8.33 | 9.71 | 8.41 | 9.52 | -1.38 | 0.19 |
| All Other | 50 | -0.83 | 9.80 | 10.64 | 9.95 | 10.37 | -0.84 | 0.01 |
| | 75 | -0.68 | 10.48 | 11.08 | 10.74 | 10.74 | -0.60 | -0.08 |
| | 90 | -0.69 | 10.86 | 11.56 | 11.29 | 11.08 | -0.70 | 0.01 |
| | | | *Native-Born Males* | | | | | |
| | 10 | 0.12 | 8.06 | 7.91 | 8.11 | 7.90 | 0.15 | -0.03 |
| | 25 | 0.20 | 9.97 | 9.71 | 9.96 | 9.74 | 0.26 | -0.06 |
| Asian Non-Hispanic | 50 | 0.23 | 10.82 | 10.64 | 10.80 | 10.68 | 0.18 | 0.05 |
| | 75 | 0.28 | 11.35 | 11.08 | 11.30 | 11.19 | 0.27 | 0.01 |
| | 90 | 0.28 | 11.89 | 11.56 | 11.84 | 11.69 | 0.33 | -0.05 |
| | 10 | -1.73 | 6.30 | 7.91 | 6.37 | 7.82 | -1.61 | -0.12 |
| | 25 | -1.43 | 8.07 | 9.71 | 8.14 | 9.55 | -1.65 | 0.22 |
| Black Non-Hispanic | 50 | -0.84 | 9.78 | 10.64 | 9.93 | 10.40 | -0.86 | 0.02 |
| | 75 | -0.62 | 10.53 | 11.08 | 10.79 | 10.75 | -0.55 | -0.07 |
| | 90 | -0.62 | 10.92 | 11.56 | 11.29 | 11.09 | -0.64 | 0.02 |
| | 10 | -0.52 | 7.38 | 7.91 | 7.43 | 7.85 | -0.53 | 0.01 |
| | 25 | -0.44 | 9.21 | 9.71 | 9.35 | 9.57 | -0.51 | 0.07 |
| White Hispanic | 50 | -0.29 | 10.37 | 10.64 | 10.56 | 10.46 | -0.27 | -0.02 |
| | 75 | -0.27 | 10.83 | 11.08 | 11.07 | 10.84 | -0.25 | -0.02 |
| | 90 | -0.32 | 11.22 | 11.56 | 11.54 | 11.23 | -0.34 | 0.02 |
| | 10 | 0.00 | 7.91 | 7.91 | 7.91 | 7.91 | 0.00 | 0.00 |
| | 25 | 0.00 | 9.71 | 9.71 | 9.71 | 9.71 | 0.00 | 0.00 |
| White Non-Hispanic | 50 | 0.00 | 10.64 | 10.64 | 10.64 | 10.64 | 0.00 | 0.00 |
| | 75 | 0.00 | 11.08 | 11.08 | 11.08 | 11.08 | 0.00 | 0.00 |
| | 90 | 0.00 | 11.56 | 11.56 | 11.56 | 11.56 | 0.00 | 0.00 |
| | 10 | -1.26 | 6.69 | 7.91 | 6.73 | 7.85 | -1.22 | -0.05 |
| | 25 | -0.98 | 8.57 | 9.71 | 8.64 | 9.61 | -1.14 | 0.16 |
| All Other | 50 | -0.57 | 10.09 | 10.64 | 10.22 | 10.48 | -0.55 | -0.02 |
| | 75 | -0.41 | 10.72 | 11.08 | 10.91 | 10.87 | -0.36 | -0.05 |
| | 90 | -0.40 | 11.14 | 11.56 | 11.40 | 11.26 | -0.42 | 0.02 |

Notes: Estimates are created using the 108,800,000 worker sample 2. All measures are calculated using the 2% of workers with earnings greater than the p-1 and less than the p+1 percentile. See section IV.c.i. of the paper for more details.

# Table 6A: Foreign-Born Earnings Decompositions

| Race/Ethnicity | Percentile θ | Predicted Diff Log Earn | Decomposition 1 Q(θ)|β(g),x(g) - Q(θ)|β(g),x(0) + Q(θ)|β(g),x(0) - Q(θ)|β(0),x(0) | | | | Decomposition 2 Q(θ)|β(0),x(g) - Q(θ)|β(0),x(0) + Q(θ)|β(g),x(g) - Q(θ)|β(0),x(g) | | | |
| | | | Components | | Share of Difference | | Components | | Share of Difference | |
| | | | Covariates | Coefficients | Covariates | Coefficients | Covariates | Coefficients | Covariates | Coefficients |
| --- | --- | --- | --- | --- | --- | --- | --- | --- | --- | --- |
| | | | | | **Foreign-Born Females** | | | | | |
| Asian Non-Hispanic | 10 | -1.14 | -0.98 | -0.16 | 0.86 | 0.14 | -0.97 | -0.18 | 0.85 | 0.15 |
| | 25 | -1.09 | -0.91 | -0.19 | 0.83 | 0.17 | -0.93 | -0.16 | 0.85 | 0.15 |
| | 50 | -0.53 | -0.30 | -0.23 | 0.57 | 0.43 | -0.32 | -0.21 | 0.60 | 0.40 |
| | 75 | -0.31 | -0.05 | -0.26 | 0.16 | 0.84 | -0.03 | -0.28 | 0.10 | 0.90 |
| | 90 | -0.30 | 0.03 | -0.33 | -0.10 | 1.10 | 0.11 | -0.41 | -0.37 | 1.37 |
| Black Non-Hispanic | 10 | -0.28 | -0.11 | -0.16 | 0.41 | 0.59 | -0.14 | -0.13 | 0.52 | 0.48 |
| | 25 | -0.46 | -0.18 | -0.29 | 0.38 | 0.62 | -0.23 | -0.23 | 0.50 | 0.50 |
| | 50 | -0.44 | -0.08 | -0.36 | 0.18 | 0.82 | -0.11 | -0.33 | 0.25 | 0.75 |
| | 75 | -0.44 | 0.03 | -0.47 | -0.07 | 1.07 | 0.06 | -0.50 | -0.14 | 1.14 |
| | 90 | -0.53 | 0.07 | -0.60 | -0.13 | 1.13 | 0.21 | -0.74 | -0.40 | 1.40 |
| White Hispanic | 10 | -1.54 | -1.35 | -0.19 | 0.88 | 0.12 | -1.43 | -0.10 | 0.93 | 0.07 |
| | 25 | -1.65 | -1.35 | -0.31 | 0.81 | 0.19 | -1.49 | -0.16 | 0.90 | 0.10 |
| | 50 | -1.16 | -0.74 | -0.42 | 0.64 | 0.36 | -0.91 | -0.25 | 0.79 | 0.21 |
| | 75 | -0.88 | -0.34 | -0.54 | 0.39 | 0.61 | -0.48 | -0.40 | 0.55 | 0.45 |
| | 90 | -0.97 | -0.24 | -0.73 | 0.25 | 0.75 | -0.40 | -0.57 | 0.41 | 0.59 |
| White Non-Hispanic | 10 | -1.56 | -1.45 | -0.11 | 0.93 | 0.07 | -1.35 | -0.21 | 0.86 | 0.14 |
| | 25 | -1.37 | -1.29 | -0.09 | 0.94 | 0.06 | -1.28 | -0.09 | 0.93 | 0.07 |
| | 50 | -0.68 | -0.51 | -0.17 | 0.75 | 0.25 | -0.58 | -0.10 | 0.85 | 0.15 |
| | 75 | -0.42 | -0.22 | -0.20 | 0.52 | 0.48 | -0.23 | -0.19 | 0.55 | 0.45 |
| | 90 | -0.44 | -0.20 | -0.24 | 0.45 | 0.55 | -0.13 | -0.31 | 0.30 | 0.70 |
| All Other | 10 | -1.53 | -1.31 | -0.22 | 0.86 | 0.14 | -1.36 | -0.18 | 0.88 | 0.12 |
| | 25 | -1.53 | -1.21 | -0.32 | 0.79 | 0.21 | -1.31 | -0.22 | 0.85 | 0.15 |
| | 50 | -0.98 | -0.56 | -0.42 | 0.57 | 0.43 | -0.68 | -0.31 | 0.69 | 0.31 |
| | 75 | -0.72 | -0.20 | -0.52 | 0.28 | 0.72 | -0.27 | -0.45 | 0.37 | 0.63 |
| | 90 | -0.80 | -0.14 | -0.66 | 0.18 | 0.82 | -0.17 | -0.63 | 0.21 | 0.79 |
| | | | | | **Foreign-Born Males** | | | | | |
| Asian Non-Hispanic | 10 | -0.74 | -0.62 | -0.13 | 0.83 | 0.17 | -0.58 | -0.17 | 0.78 | 0.22 |
| | 25 | -0.54 | -0.42 | -0.12 | 0.78 | 0.22 | -0.39 | -0.15 | 0.72 | 0.28 |
| | 50 | -0.14 | 0.00 | -0.14 | 0.00 | 1.00 | 0.02 | -0.16 | -0.14 | 1.14 |
| | 75 | 0.06 | 0.17 | -0.11 | 2.83 | -1.83 | 0.18 | -0.12 | 3.00 | -2.00 |
| | 90 | 0.13 | 0.24 | -0.11 | 1.85 | -0.85 | 0.29 | -0.16 | 2.23 | -1.23 |
| Black Non-Hispanic | 10 | -0.84 | -0.68 | -0.16 | 0.81 | 0.19 | -0.66 | -0.18 | 0.79 | 0.21 |
| | 25 | -0.72 | -0.49 | -0.23 | 0.68 | 0.32 | -0.47 | -0.24 | 0.66 | 0.34 |
| | 50 | -0.42 | -0.12 | -0.30 | 0.29 | 0.71 | -0.11 | -0.31 | 0.26 | 0.74 |
| | 75 | -0.37 | 0.02 | -0.39 | -0.05 | 1.05 | 0.06 | -0.43 | -0.16 | 1.16 |
| | 90 | -0.44 | 0.06 | -0.50 | -0.14 | 1.14 | 0.16 | -0.60 | -0.36 | 1.36 |
| White Hispanic | 10 | -1.05 | -0.88 | -0.16 | 0.85 | 0.15 | -0.92 | -0.12 | 0.88 | 0.12 |
| | 25 | -0.94 | -0.68 | -0.26 | 0.72 | 0.28 | -0.70 | -0.24 | 0.75 | 0.25 |
| | 50 | -0.60 | -0.28 | -0.32 | 0.47 | 0.53 | -0.27 | -0.33 | 0.45 | 0.55 |
| | 75 | -0.52 | -0.08 | -0.44 | 0.15 | 0.85 | -0.11 | -0.41 | 0.21 | 0.79 |
| | 90 | -0.65 | -0.09 | -0.56 | 0.14 | 0.86 | -0.11 | -0.54 | 0.17 | 0.83 |
| White Non-Hispanic | 10 | -1.30 | -1.20 | -0.09 | 0.93 | 0.07 | -1.11 | -0.19 | 0.85 | 0.15 |
| | 25 | -0.80 | -0.79 | -0.01 | 0.98 | 0.02 | -0.75 | -0.05 | 0.93 | 0.07 |
| | 50 | -0.16 | -0.11 | -0.05 | 0.69 | 0.31 | -0.12 | -0.04 | 0.75 | 0.25 |
| | 75 | 0.05 | 0.04 | 0.01 | 0.80 | 0.20 | 0.06 | -0.01 | 1.20 | -0.20 |
| | 90 | 0.18 | 0.11 | 0.07 | 0.61 | 0.39 | 0.13 | 0.05 | 0.72 | 0.28 |
| All Other | 10 | -1.31 | -1.13 | -0.18 | 0.86 | 0.14 | -1.13 | -0.18 | 0.86 | 0.14 |
| | 25 | -1.14 | -0.88 | -0.25 | 0.78 | 0.22 | -0.88 | -0.26 | 0.78 | 0.22 |
| | 50 | -0.63 | -0.31 | -0.32 | 0.49 | 0.51 | -0.30 | -0.33 | 0.48 | 0.52 |
| | 75 | -0.47 | -0.07 | -0.40 | 0.15 | 0.85 | -0.06 | -0.41 | 0.13 | 0.87 |
| | 90 | -0.54 | -0.06 | -0.48 | 0.11 | 0.89 | -0.01 | -0.53 | 0.02 | 0.98 |

Notes: Estimates are created using the 108,800,000 worker sample 2. All measures are calculated using the 2% of workers with earnings greater than the p-1 and less than the p+1 percentile. See section IV.c.i. of the paper for more details.

# Table 6B: Native-Born Earnings Decompositions

| Race/Ethnicity | Percentile θ | Predicted Diff Log Earn | Decomposition 1 Q(θ)\|β(g),x(g) - Q(θ)\|β(g),x(0) + Q(θ)\|β(g),x(0) - Q(θ)\|β(0),x(0) | | | | Decomposition 2 Q(θ)\|β(0),x(g) - Q(θ)\|β(0),x(0) + Q(θ)\|β(g),x(g) - Q(θ)\|β(0),x(g) | | | |
|---|---|---|---|---|---|---|---|---|---|---|
| | | | Components | | Share of Difference | | Components | | Share of Difference | |
| | | | Covariates | Coefficients | Covariates | Coefficients | Covariates | Coefficients | Covariates | Coefficients |
| **Native-Born Females** | | | | | | | | | | |
| Asian Non-Hispanic | 10 | -0.24 | -0.16 | -0.09 | 0.65 | 0.35 | -0.16 | -0.09 | 0.64 | 0.36 |
| | 25 | -0.11 | -0.08 | -0.03 | 0.75 | 0.25 | -0.08 | -0.03 | 0.75 | 0.25 |
| | 50 | -0.04 | 0.00 | -0.04 | 0.00 | 1.00 | -0.01 | -0.03 | 0.25 | 0.75 |
| | 75 | 0.00 | 0.01 | -0.01 | | | 0.06 | -0.06 | | |
| | 90 | 0.00 | 0.01 | -0.01 | | | 0.14 | -0.14 | | |
| Black Non-Hispanic | 10 | -1.02 | -0.89 | -0.12 | 0.88 | 0.12 | -0.97 | -0.05 | 0.95 | 0.05 |
| | 25 | -1.02 | -0.80 | -0.22 | 0.78 | 0.22 | -0.91 | -0.11 | 0.89 | 0.11 |
| | 50 | -0.70 | -0.40 | -0.30 | 0.57 | 0.43 | -0.49 | -0.21 | 0.70 | 0.30 |
| | 75 | -0.59 | -0.18 | -0.41 | 0.31 | 0.69 | -0.26 | -0.33 | 0.44 | 0.56 |
| | 90 | -0.73 | -0.17 | -0.56 | 0.23 | 0.77 | -0.20 | -0.53 | 0.27 | 0.73 |
| White Hispanic | 10 | -0.72 | -0.60 | -0.11 | 0.84 | 0.16 | -0.68 | -0.04 | 0.94 | 0.06 |
| | 25 | -0.84 | -0.63 | -0.22 | 0.74 | 0.26 | -0.72 | -0.12 | 0.85 | 0.15 |
| | 50 | -0.58 | -0.31 | -0.27 | 0.53 | 0.47 | -0.38 | -0.20 | 0.66 | 0.34 |
| | 75 | -0.51 | -0.15 | -0.36 | 0.29 | 0.71 | -0.21 | -0.30 | 0.41 | 0.59 |
| | 90 | -0.65 | -0.15 | -0.50 | 0.23 | 0.77 | -0.17 | -0.48 | 0.26 | 0.74 |
| White Non-Hispanic | 10 | -0.69 | -0.67 | -0.03 | 0.96 | 0.04 | -0.70 | 0.00 | 1.00 | 0.00 |
| | 25 | -0.75 | -0.68 | -0.07 | 0.90 | 0.10 | -0.76 | 0.01 | 1.01 | -0.01 |
| | 50 | -0.47 | -0.35 | -0.12 | 0.74 | 0.26 | -0.43 | -0.04 | 0.91 | 0.09 |
| | 75 | -0.40 | -0.23 | -0.17 | 0.58 | 0.42 | -0.28 | -0.12 | 0.70 | 0.30 |
| | 90 | -0.51 | -0.24 | -0.27 | 0.47 | 0.53 | -0.23 | -0.28 | 0.45 | 0.55 |
| All Other | 10 | -1.32 | -1.21 | -0.11 | 0.91 | 0.09 | -1.27 | -0.05 | 0.96 | 0.04 |
| | 25 | -1.38 | -1.19 | -0.19 | 0.86 | 0.14 | -1.30 | -0.08 | 0.94 | 0.06 |
| | 50 | -0.84 | -0.57 | -0.27 | 0.68 | 0.32 | -0.69 | -0.15 | 0.82 | 0.18 |
| | 75 | -0.60 | -0.26 | -0.34 | 0.43 | 0.57 | -0.34 | -0.26 | 0.57 | 0.43 |
| | 90 | -0.70 | -0.22 | -0.48 | 0.31 | 0.69 | -0.27 | -0.43 | 0.39 | 0.61 |
| **Native-Born Males** | | | | | | | | | | |
| Asian Non-Hispanic | 10 | 0.15 | 0.16 | -0.01 | 1.08 | -0.08 | 0.20 | -0.05 | 1.32 | -0.32 |
| | 25 | 0.26 | 0.23 | 0.03 | 0.89 | 0.11 | 0.25 | 0.00 | 0.98 | 0.02 |
| | 50 | 0.18 | 0.14 | 0.04 | 0.78 | 0.22 | 0.16 | 0.02 | 0.89 | 0.11 |
| | 75 | 0.27 | 0.16 | 0.11 | 0.59 | 0.41 | 0.22 | 0.05 | 0.81 | 0.19 |
| | 90 | 0.33 | 0.20 | 0.13 | 0.61 | 0.39 | 0.28 | 0.05 | 0.85 | 0.15 |
| Black Non-Hispanic | 10 | -1.61 | -1.52 | -0.09 | 0.94 | 0.06 | -1.54 | -0.07 | 0.96 | 0.04 |
| | 25 | -1.65 | -1.49 | -0.16 | 0.90 | 0.10 | -1.57 | -0.07 | 0.96 | 0.04 |
| | 50 | -0.86 | -0.62 | -0.24 | 0.72 | 0.28 | -0.71 | -0.15 | 0.82 | 0.18 |
| | 75 | -0.55 | -0.22 | -0.33 | 0.40 | 0.60 | -0.29 | -0.26 | 0.53 | 0.47 |
| | 90 | -0.64 | -0.17 | -0.47 | 0.27 | 0.73 | -0.27 | -0.37 | 0.42 | 0.58 |
| White Hispanic | 10 | -0.53 | -0.47 | -0.06 | 0.88 | 0.12 | -0.48 | -0.05 | 0.91 | 0.09 |
| | 25 | -0.51 | -0.36 | -0.14 | 0.72 | 0.28 | -0.36 | -0.14 | 0.72 | 0.28 |
| | 50 | -0.27 | -0.09 | -0.18 | 0.33 | 0.67 | -0.08 | -0.19 | 0.30 | 0.70 |
| | 75 | -0.25 | -0.01 | -0.24 | 0.04 | 0.96 | -0.01 | -0.24 | 0.04 | 0.96 |
| | 90 | -0.34 | -0.01 | -0.33 | 0.03 | 0.97 | -0.02 | -0.32 | 0.06 | 0.94 |
| White Non-Hispanic | 10 | 0.00 | 0.00 | 0.00 | | | 0.00 | 0.00 | | |
| | 25 | 0.00 | 0.00 | 0.00 | | | 0.00 | 0.00 | | |
| | 50 | 0.00 | 0.00 | 0.00 | | | 0.00 | 0.00 | | |
| | 75 | 0.00 | 0.00 | 0.00 | | | 0.00 | 0.00 | | |
| | 90 | 0.00 | 0.00 | 0.00 | | | 0.00 | 0.00 | | |
| All Other | 10 | -1.22 | -1.16 | -0.06 | 0.95 | 0.05 | -1.18 | -0.03 | 0.97 | 0.03 |
| | 25 | -1.14 | -1.04 | -0.10 | 0.91 | 0.09 | -1.07 | -0.07 | 0.94 | 0.06 |
| | 50 | -0.55 | -0.39 | -0.16 | 0.71 | 0.29 | -0.42 | -0.13 | 0.76 | 0.24 |
| | 75 | -0.36 | -0.15 | -0.21 | 0.42 | 0.58 | -0.17 | -0.19 | 0.47 | 0.53 |
| | 90 | -0.42 | -0.12 | -0.30 | 0.29 | 0.71 | -0.16 | -0.26 | 0.38 | 0.62 |

Notes: Estimates are created using the 108,800,000 worker sample 2. All measures are calculated using the 2% of workers with earnings greater than the p-1 and less than the p+1 percentile. See section IV.c.i. of the paper for more details.

Figure 1: Change in the Percentiles of Log Real Annual Earnings

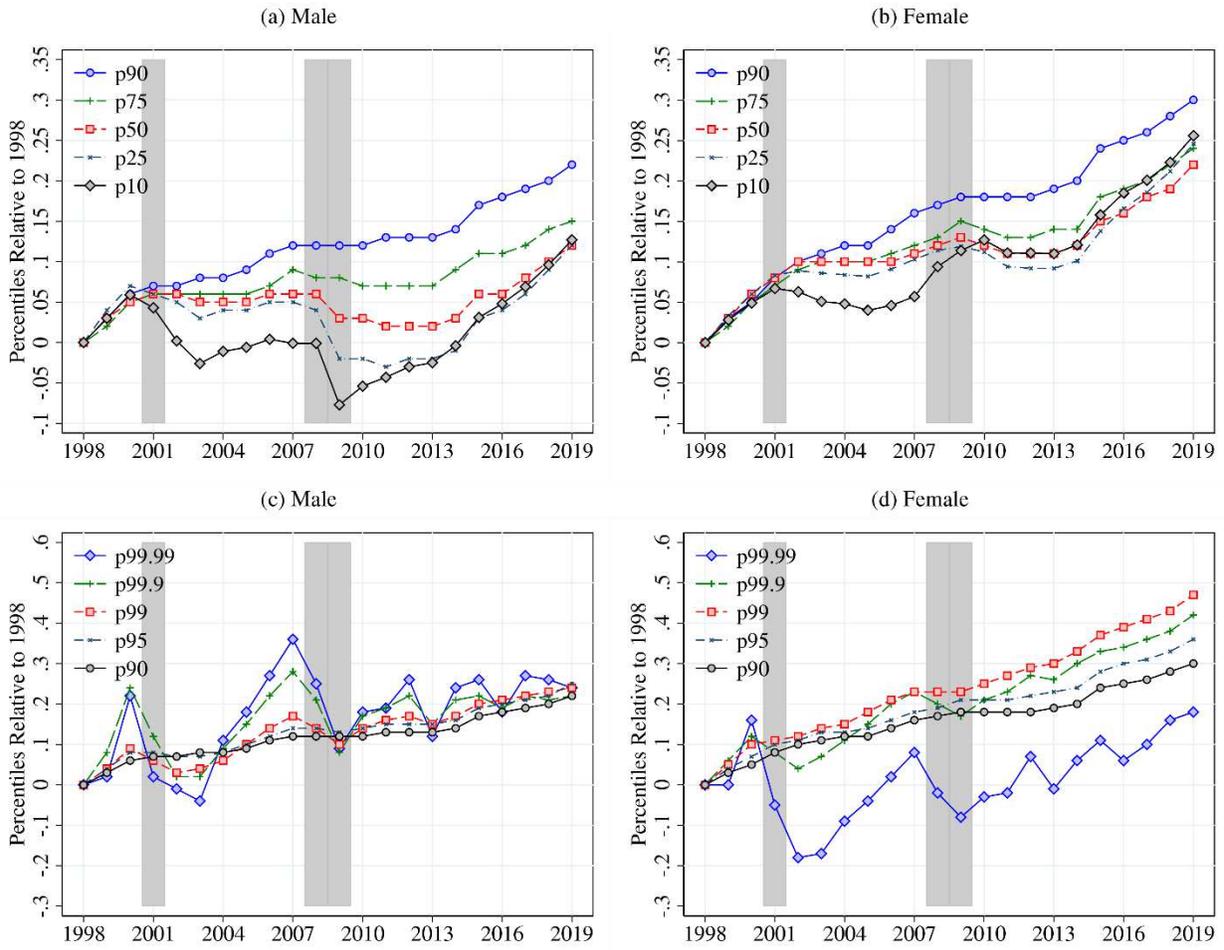

Notes: LEHD CS sample. Shaded areas are recessions. The analysis variable is the log of $y_{it}$. $y_{it}$ must be greater than 260*federal hourly minimum wage. The percentiles relative to 1998 are calculated as $pX_t - pX_{1998}$, where $X$ is the percentile.

## Figure 2: Dispersion of Log Real Annual Earnings

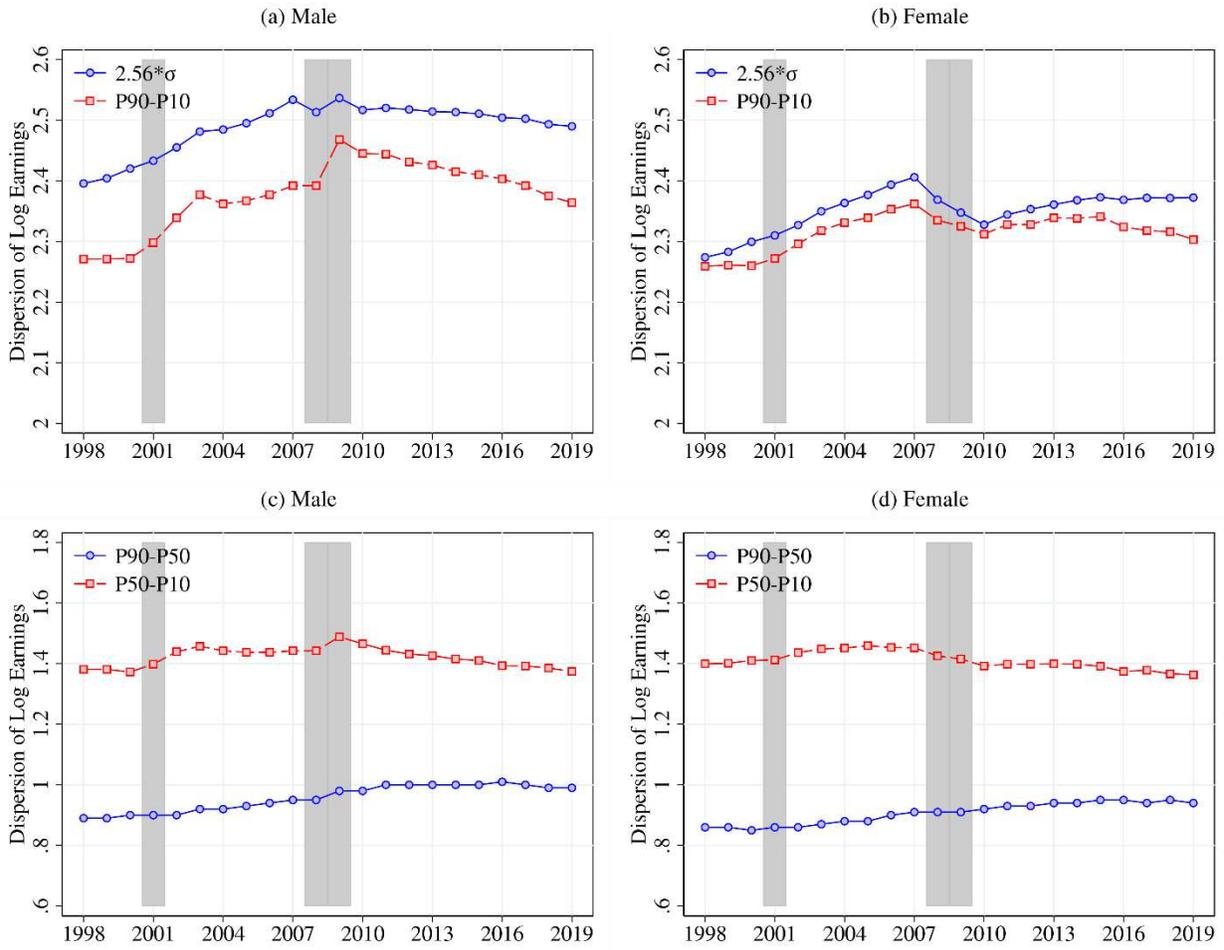

Notes: LEHD CS sample. Shaded areas are recessions. The analysis variable is the log of $y_{it}$. $y_{it}$ must be greater than 260*federal hourly minimum wage. $2.56 * \sigma$ corresponds to $P90 - P10$ for the normal distribution.

## Figure 3: Dispersion of Log Real Earnings (Age 25)

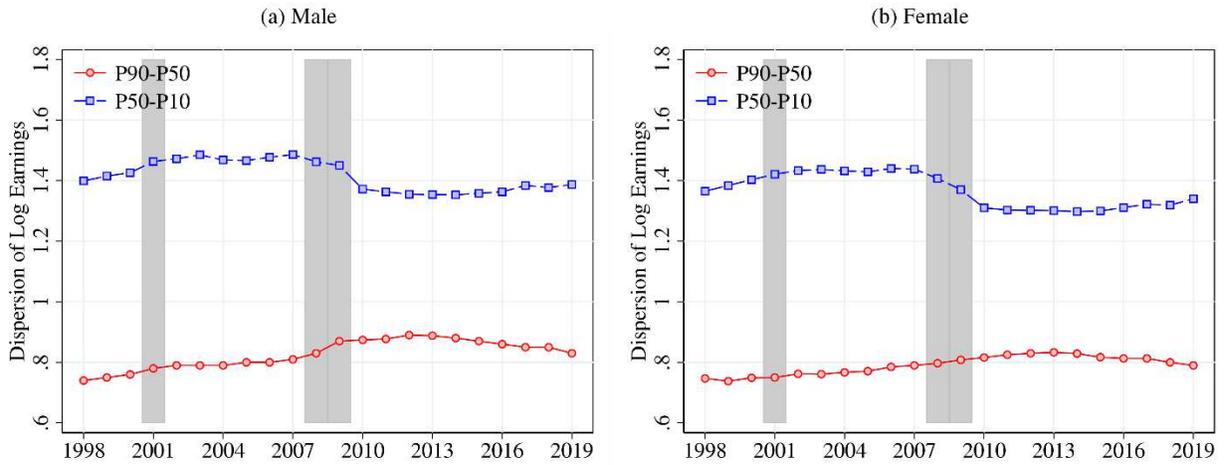

Notes: LEHD CS sample for workers age 25. Shaded areas are recessions. The analysis variable is the log of $y_{it}$. $y_{it}$ must be greater than 260*federal hourly minimum wage.

## Figure 4: Life-Cycle Earnings Inequality by Age Cohort

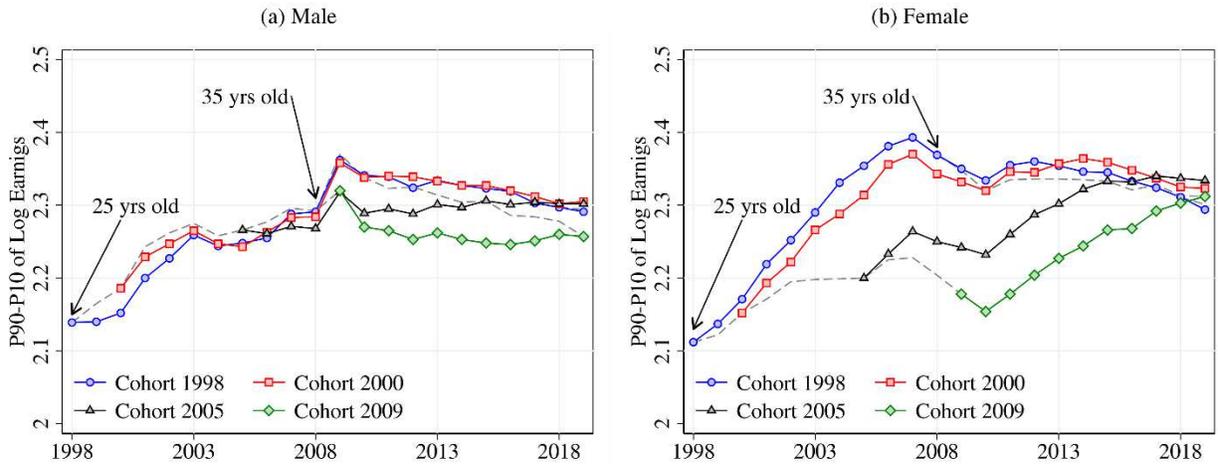

Notes: LEHD CS sample. The analysis variable is the log of $y_{it}$. $y_{it}$ must be greater than 260*federal hourly minimum wage. Each colored line represents $P90 - P10$ for the set of workers who were age 25 in 1998, 2000, 2005, and 2009. The grey dashed lines show the $P90 - P10$ for workers at age 25 and 35.

## Figure 5: Dispersion of One-Year Residual Log Earnings Changes

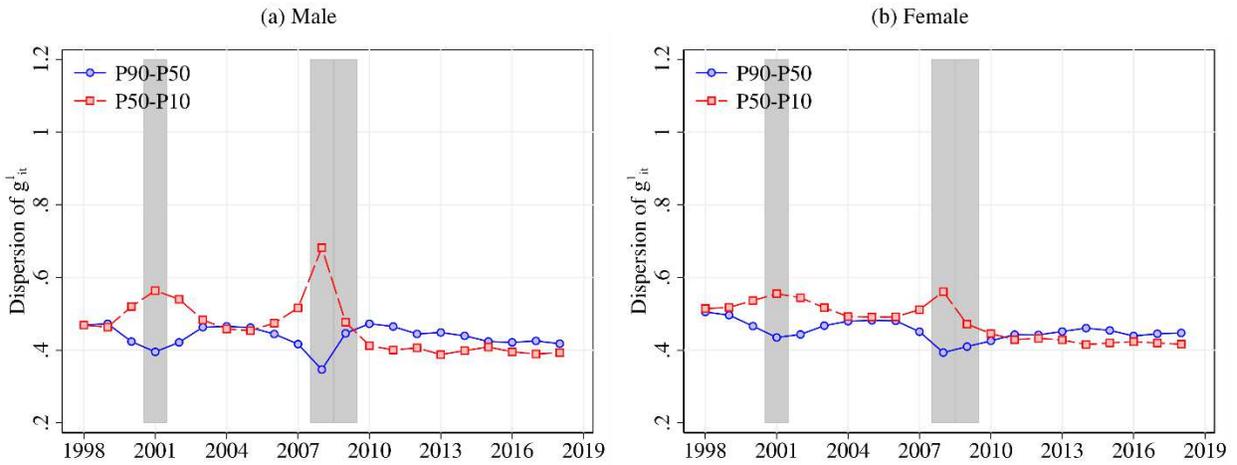

Notes: LEHD LX_1 sample. Shaded areas are recessions. Residual log earnings $g_{it}^1 = \varepsilon_{it+1} - \varepsilon_{it}$. $\varepsilon_{it}$ is the residual from a regression of log $y_{it}$ on a set of age indicator variables by sex and year. $y_{it}$ must be greater than 260*federal hourly minimum wage in $t$ and greater than 1/3*260*federal hourly minimum wage in $t + 1$.

## Figure 6: Skewness and Kurtosis of One-Year Residual Log Earnings Changes

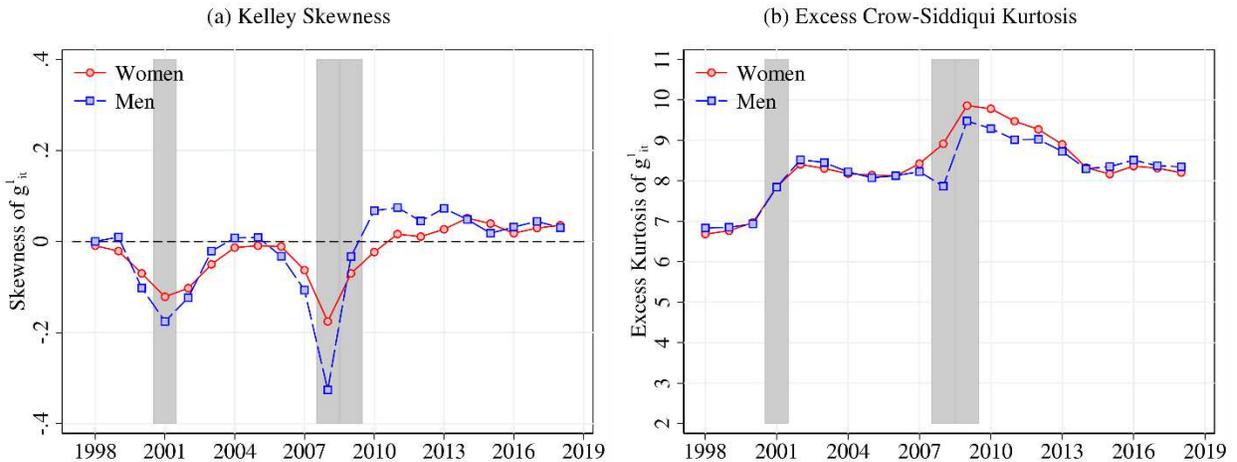

Notes: LEHD LX_1 sample. Shaded areas are recessions. Residual Log Earnings $g_{it}^1 = \varepsilon_{it+1} - \varepsilon_{it}$. $\varepsilon_{it}$ is the residual from a regression of log $y_{it}$ on a set of age indicator variables by sex and year. $y_{it}$ must be greater than 260*federal hourly minimum wage in $t$ and greater than 1/3*260*federal hourly minimum wage in $t + 1$. Kelley skewness is $\frac{(P90-P50)-(P50-P10)}{(P90-P10)}$. Excess Crow-Siddiqui kurtosis is $\frac{(P97.5-P2.5)}{(P75-P25)} - 2.91$. 2.91 is the kurtosis value for the normal distribution.

Figure 7: Dispersion, Skewness, and Kurtosis of One-Year Residual Log Earnings Changes

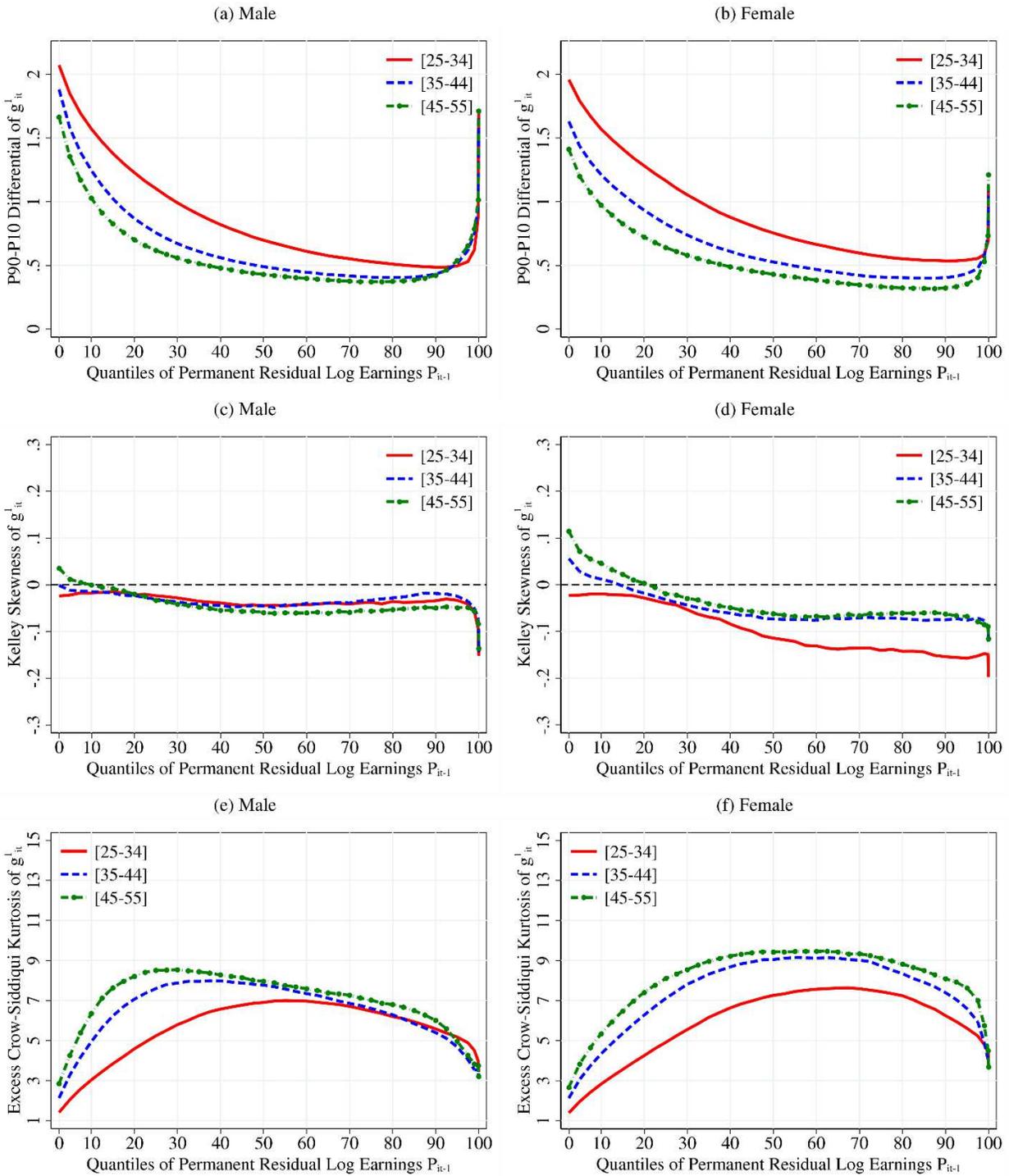

Notes: LEHD H_1 sample (2001-2014). Residual Log Earnings $g^1_{it} = \varepsilon_{it+1} - \varepsilon_{it}$. $\varepsilon_{it}$ is the residual from a regression of log $y_{it}$ on a set of age indicator variables by sex and year. $y_{it}$ must be greater than 260*federal hourly minimum wage in $t$ and greater than $1/3$*260*federal hourly minimum wage in $t + 1$. Kelley skewness is $\frac{(P90-P50)-(P50-P10)}{(P90-P10)}$. Excess Crow-Siddiqui kurtosis is $\frac{(P97.5-P2.5)}{(P75-P25)} - 2.91$. 2.91 is the kurtosis value for the normal distribution. $P_{it-1}$ is a three-year measure of permanent residual log earnings (see the text for more details).

## Figure 8: Ten-Year Earnings Mobility by Age

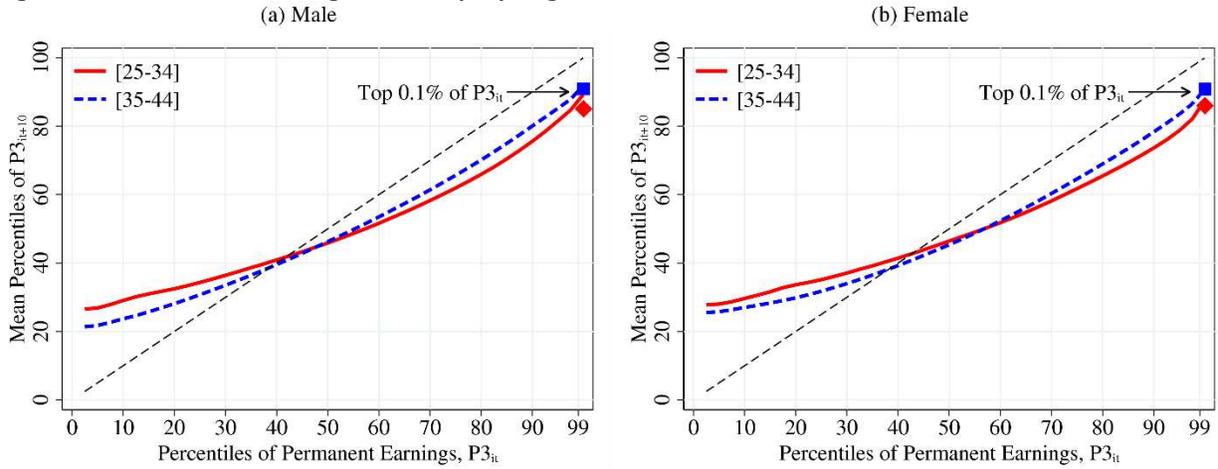

Notes: LEHD PA_10 sample. Permanent earnings $P3_{it} = \frac{(\sum_j e_{ijt-2} + \sum_j e_{ijt-1} + \sum_j e_{ijt})}{3}$. $P3_{it}$ is missing unless $\sum_j e_{ijt} > 260*$federal hourly minimum wage in at least one of the three years. In both $t$ and $t+10$ permanent earnings are ranked (0,100] separately by sex and age. The vertical axis shows the average rank in $t+10$ for workers of a given rank (percentile) in $t$.

## Figure 9: Ten-Year Earnings Mobility by Selected Years

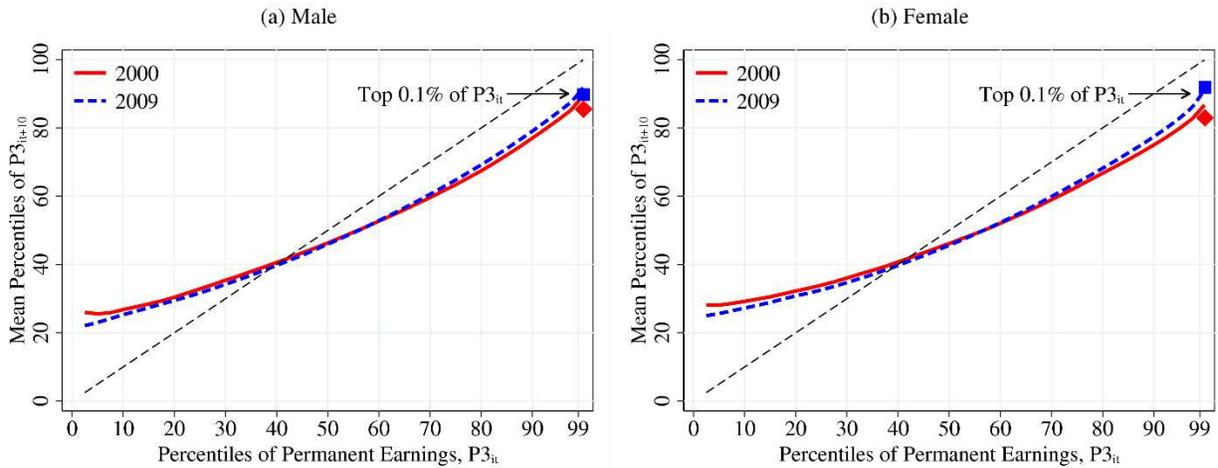

Notes: LEHD PA_10 sample. Permanent earnings $P3_{it} = \frac{(\sum_j e_{ijt-2} + \sum_j e_{ijt-1} + \sum_j e_{ijt})}{3}$. $P3_{it}$ is missing unless $\sum_j e_{ijt} > 260*$federal hourly minimum wage in at least one of the three years. In both $t$ and $t+10$ permanent earnings are ranked (0,100] separately by sex and age. The vertical axis shows the average rank in $t+10$ for workers of a given rank (percentile) in $t$.

Figure 10: Percent Active in Sample 2 by Age Group and Year

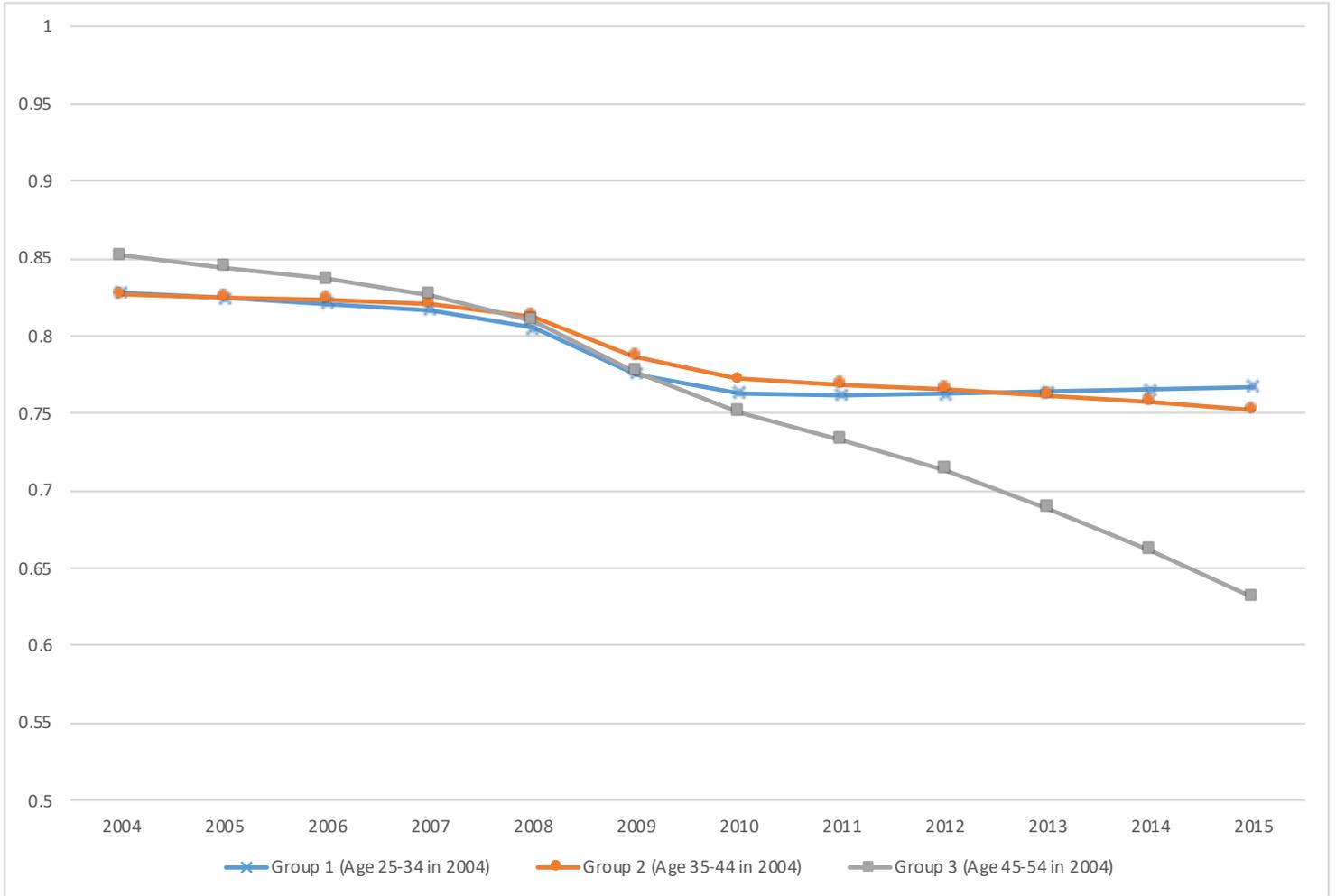

Notes: Calculations are based on the 1.3 billion person-year records in sample 2. A worker is active if they have positive earnings in at least 1 quarter during the year.

Figure 11: Mean Log Real Annual Earnings by Age Group and Year

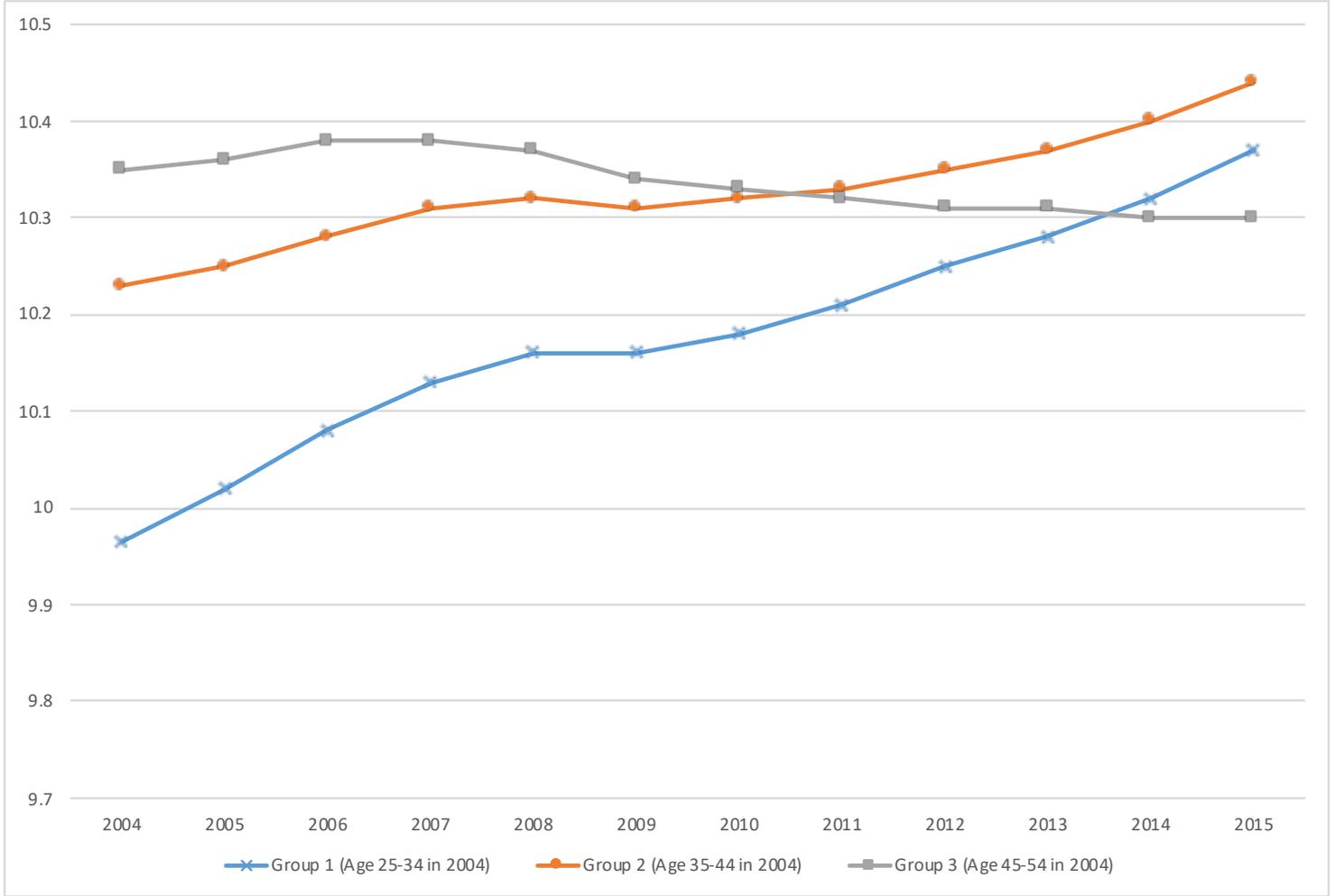

Notes: Calculations are based on the 1.3 billion person -year records in sample 2. Log real (2010 PCE) annual earnings at all jobs. To be included in a given year's estimates, the worker must have at least 1 quarter of positive earnings.

Figure 12: Log Real Average Annual Earnings as a Share of the Reference Group

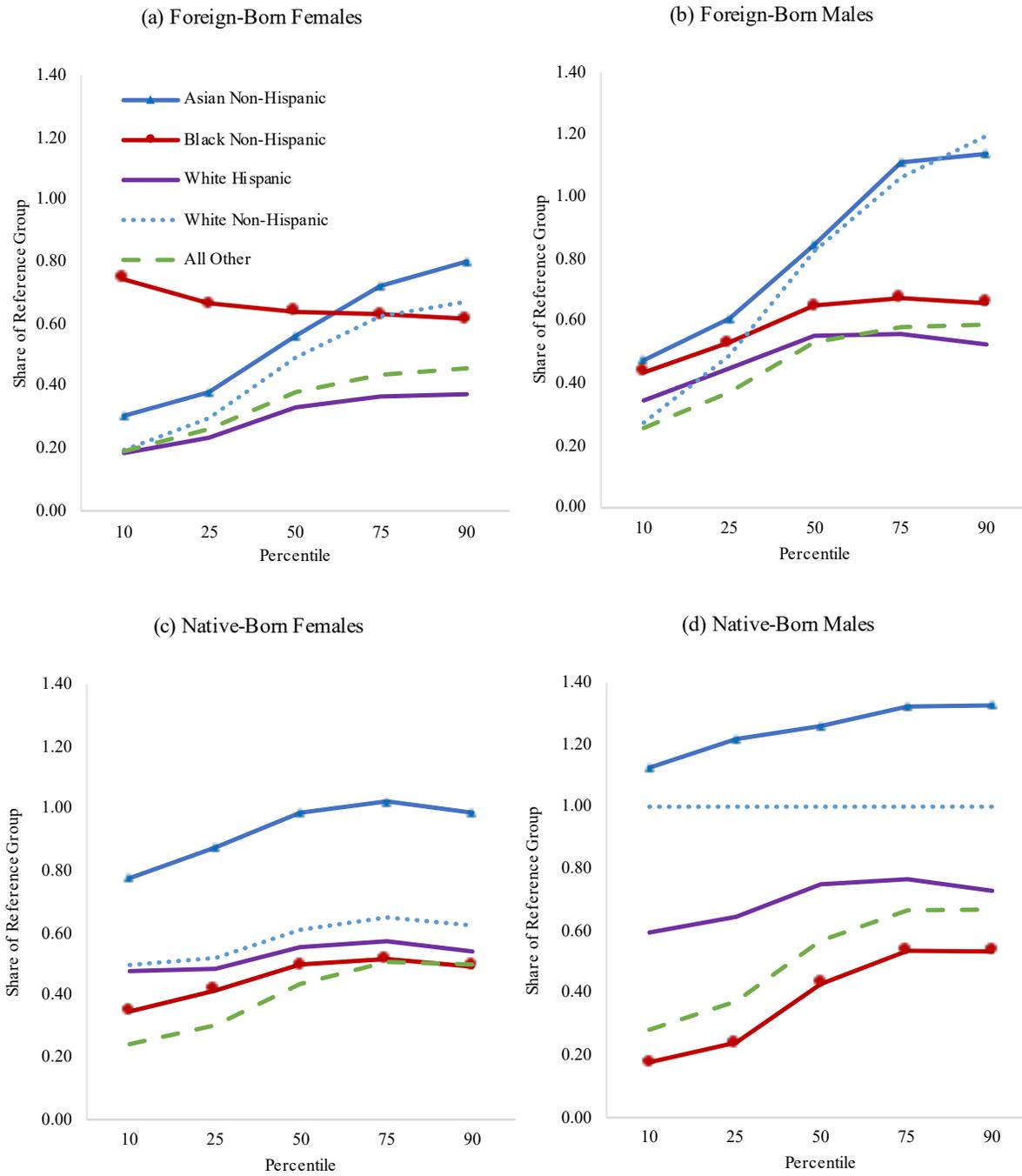

Notes: Calculations are based on the 108,800,000 worker records in sample 2. Log real (2010 PCE) average annual earnings by demographic group expressed as a share of reference group earnings (native-born White Non-Hispanic males). See text for data and estimation details.

Figure 13: Earnings Decomposition by Demographic Group (Foreign-Born Females)

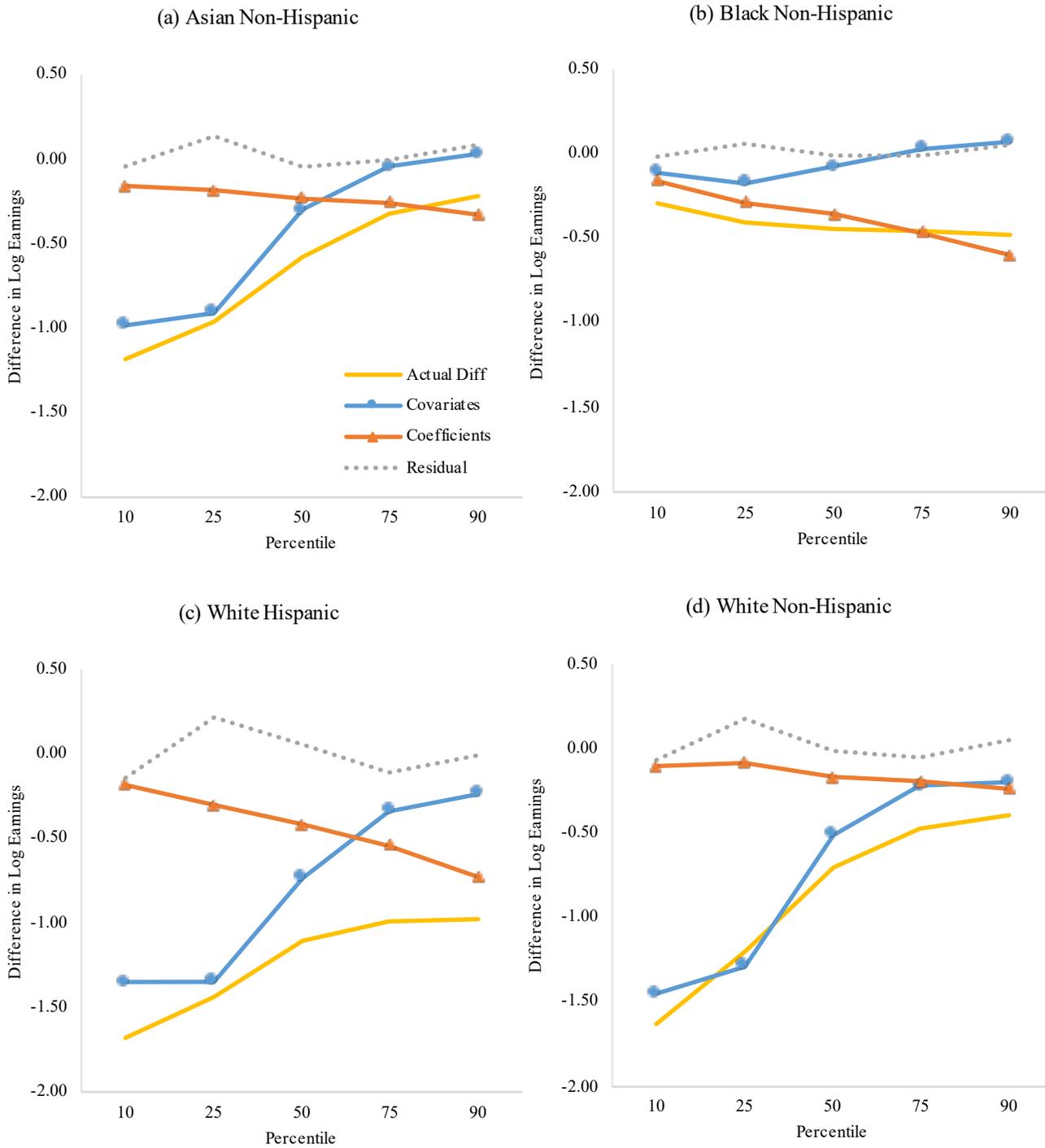

Notes: Calculations are based on the 108,800,000 worker records in sample 2. Each panel shows the actual difference in log real (2010 PCE) average annual earnings and the components for decomposition method #1 by demographic group. The reference group is native-born White Non-Hispanic males. Actual Diff = Covariates + Coefficients + Residual. See text for data and estimation details. The all other race group is in Figure 17.

Figure 14: Earnings Decomposition by Demographic Group (Foreign-Born Males)

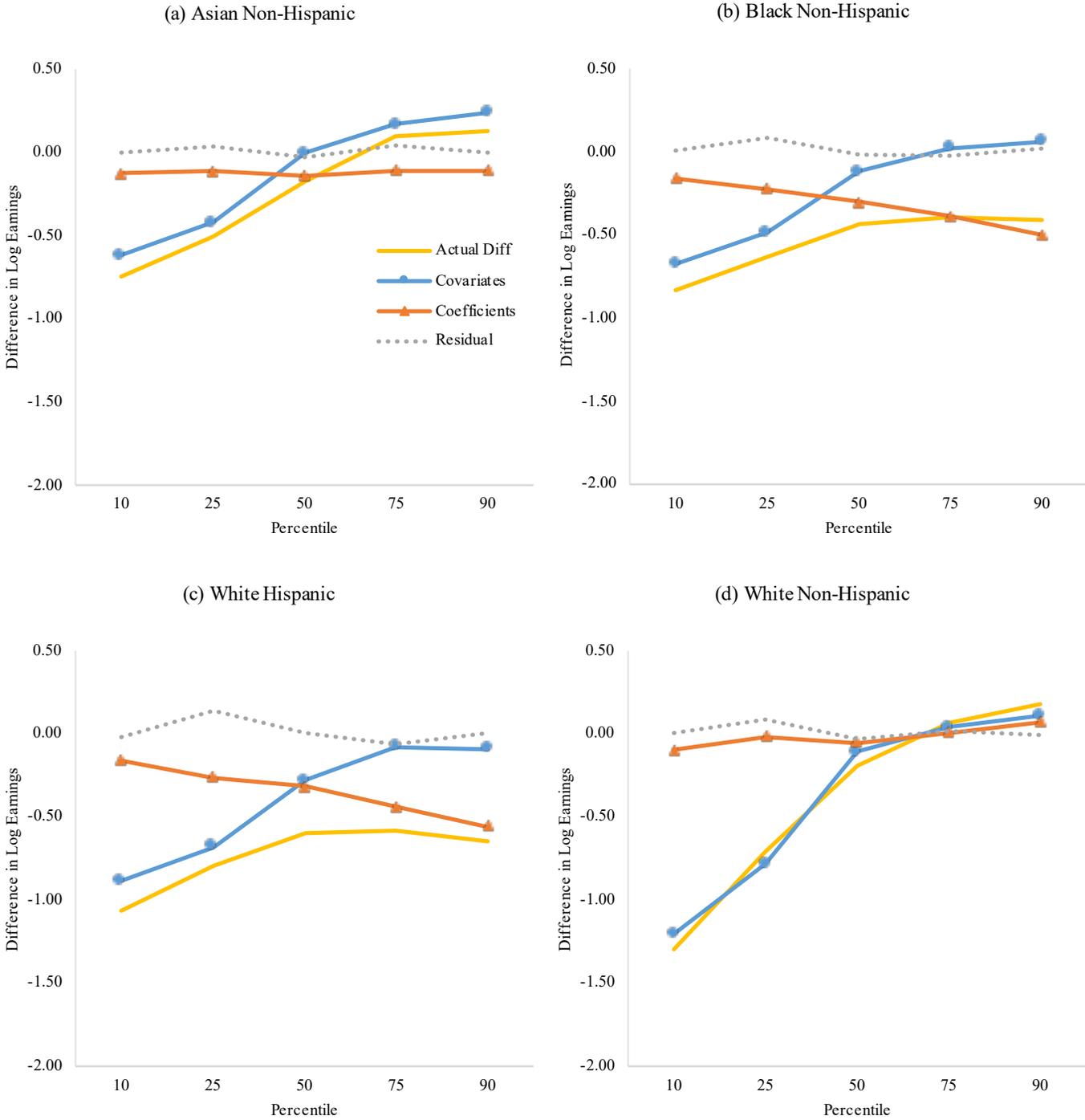

Notes: Calculations are based on the 108,800,000 worker records in sample 2. Each panel shows the actual difference in log real (2010 PCE) average annual earnings and the components for decomposition method #1 by demographic group. The reference group is native-born White Non-Hispanic males. Actual Diff = Covariates + Coefficients + Residual. See text for data and estimation details. The all other race group is in Figure 17.
.

Figure 15: Earnings Decomposition by Demographic Group (Native-Born Females)

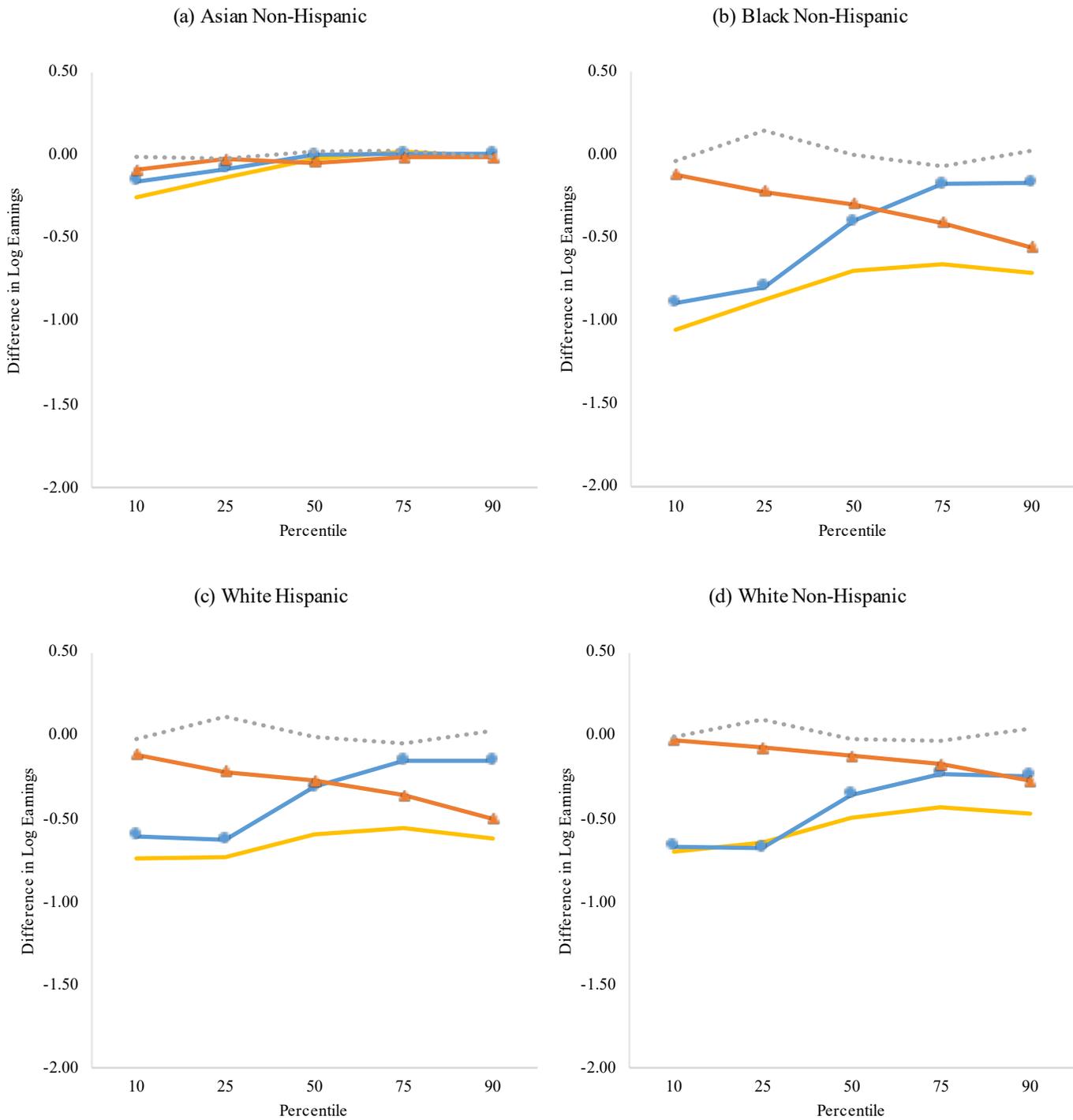

Notes: Calculations are based on the 108,800,000 worker records in sample 2. Each panel shows the actual difference in log real (2010 PCE) average annual earnings and the components for decomposition method #1 by demographic group. The reference group is White Non-Hispanic males. Actual Diff = Covariates + Coefficients + Residual. See text for data and estimation details. The all other race group is in Figure 17.
.

Figure 16: Earnings Decomposition by Demographic Group (Native-Born Males)

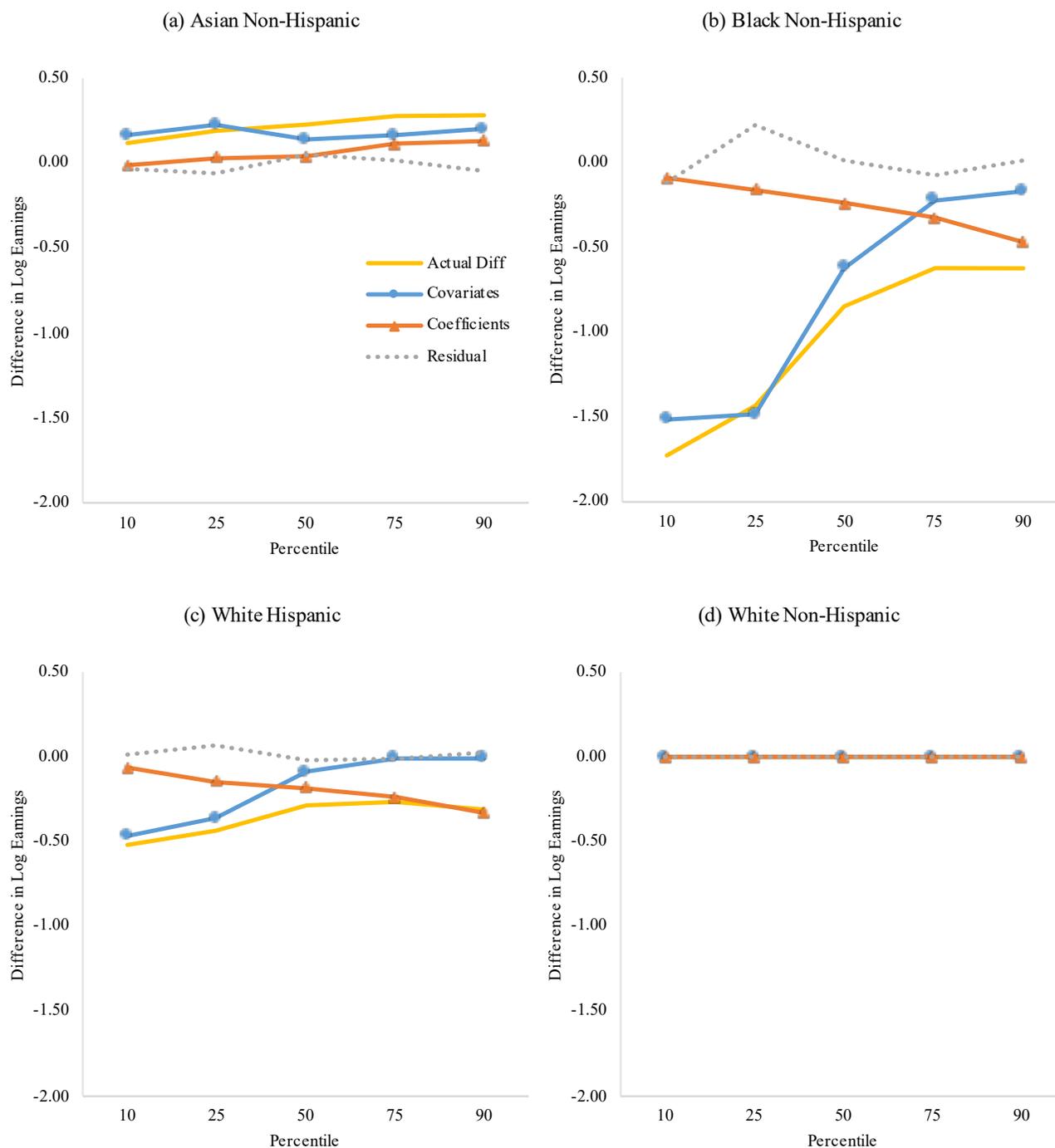

Notes: Calculations are based on the 108,800,000 worker records in sample 2. Each panel shows the actual difference in log real (2010 PCE) average annual earnings and the components for decomposition method #1 by demographic group. The reference group is native-born White Non-Hispanic males. Actual Diff = Covariates + Coefficients + Residual. See text for data and estimation details. The all other race group is in Figure 17.

Figure 17: Earnings Decomposition for All Other Races

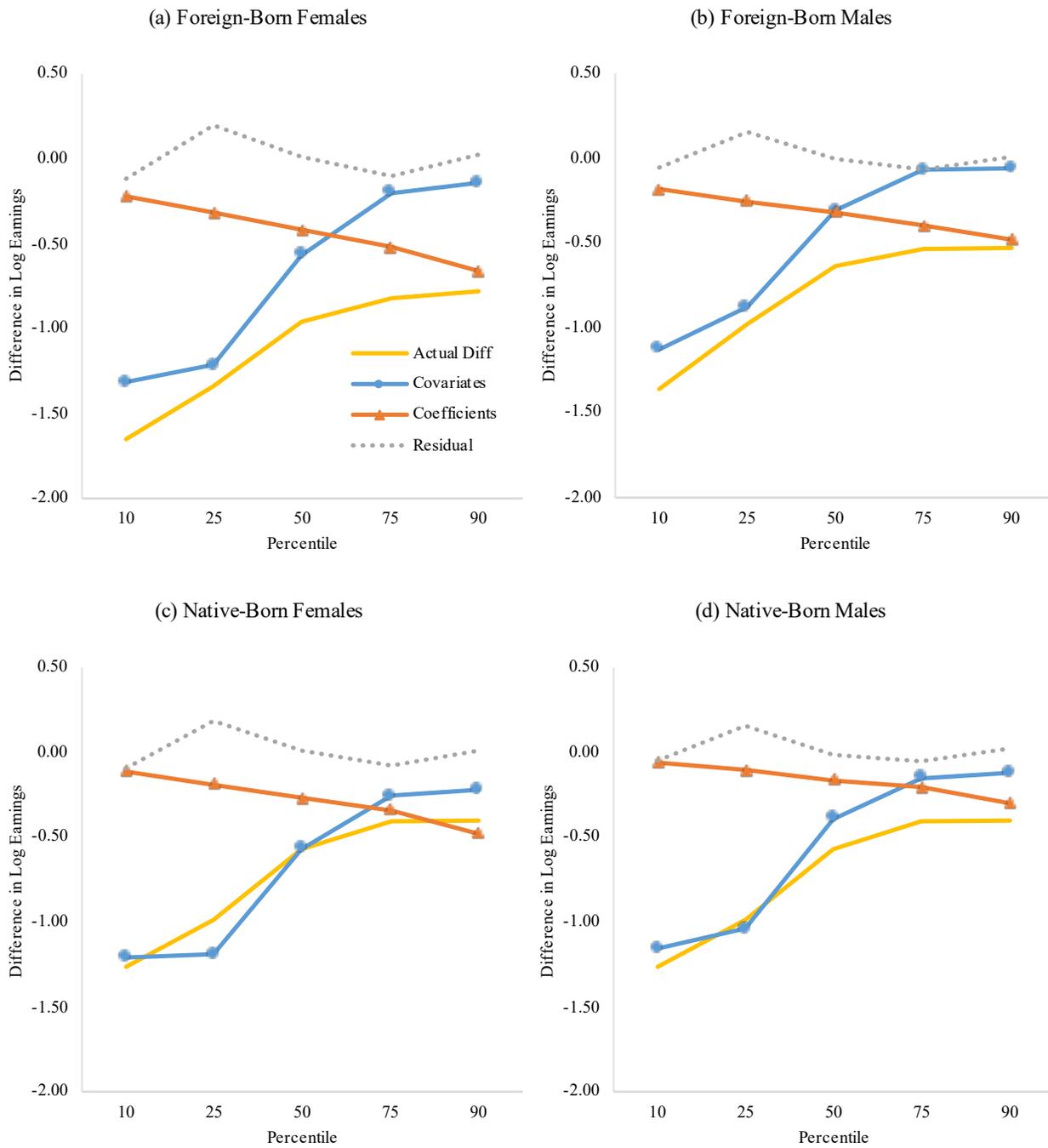

Notes: Calculations are based on the 108,800,000 worker records in sample 2. Each panel shows the actual difference in log real (2010 PCE) average annual earnings and the components for decomposition method #1 by demographic group. The reference group is native-born White Non-Hispanic males. Actual Diff = Covariates + Coefficients + Residual. See text for data and estimation details.

Figure 18: Counterfactual Earnings Differentials with Reference Group Characteristics

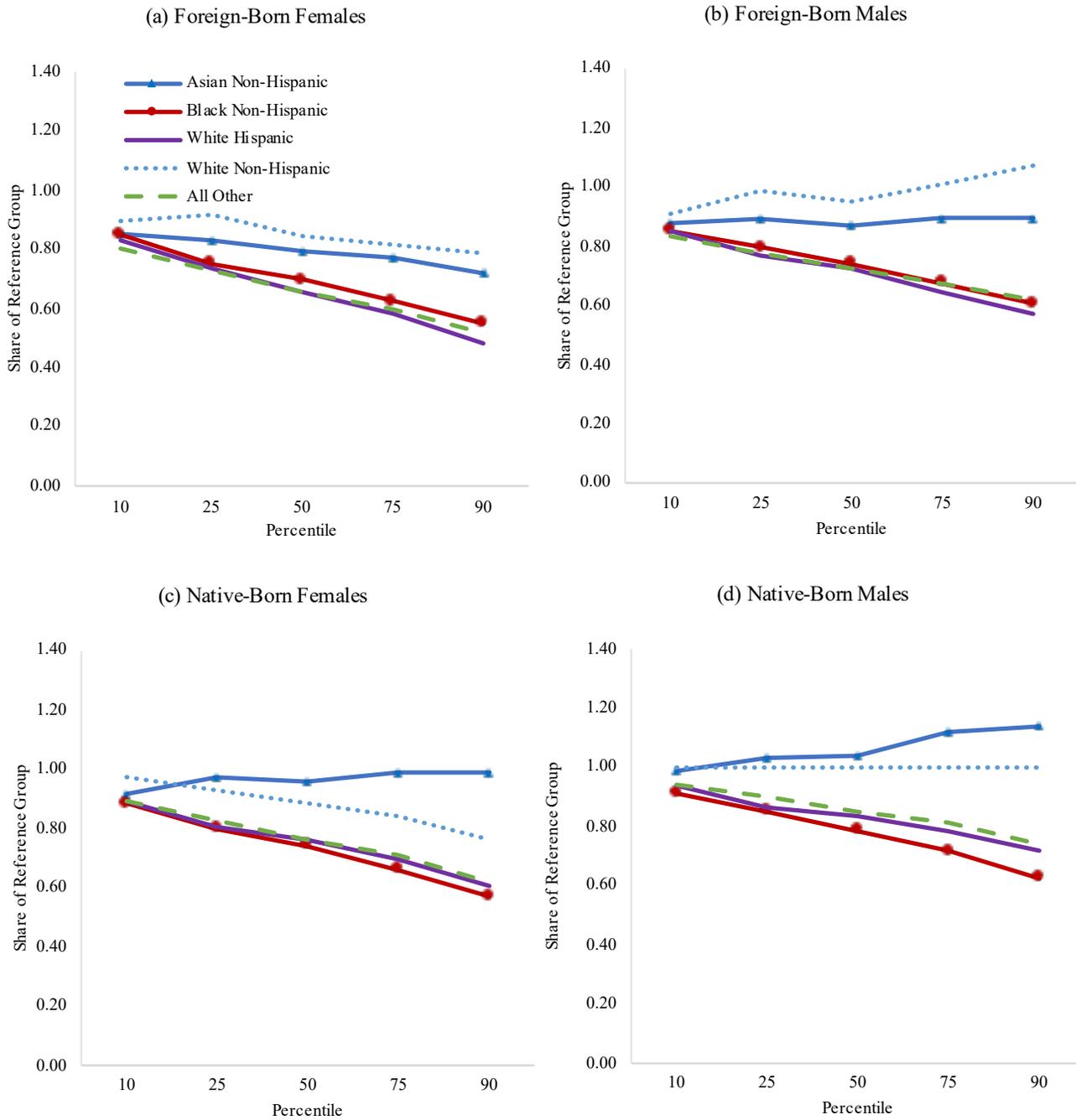

Notes: Calculations are based on the 108,800,000 worker records in sample 2. Each panel is based on the counterfactual log real (2010 PCE) average earnings of demographic group g with the characteristics of the reference group (native-born White Non-Hispanic males). The y-axis shows the share of reference group earnings. The actual share of reference groups earnings is shown in Figure 12. See text for estimation details.

Figure 19: Counterfactual Earnings Differentials with Reference Group Coefficients

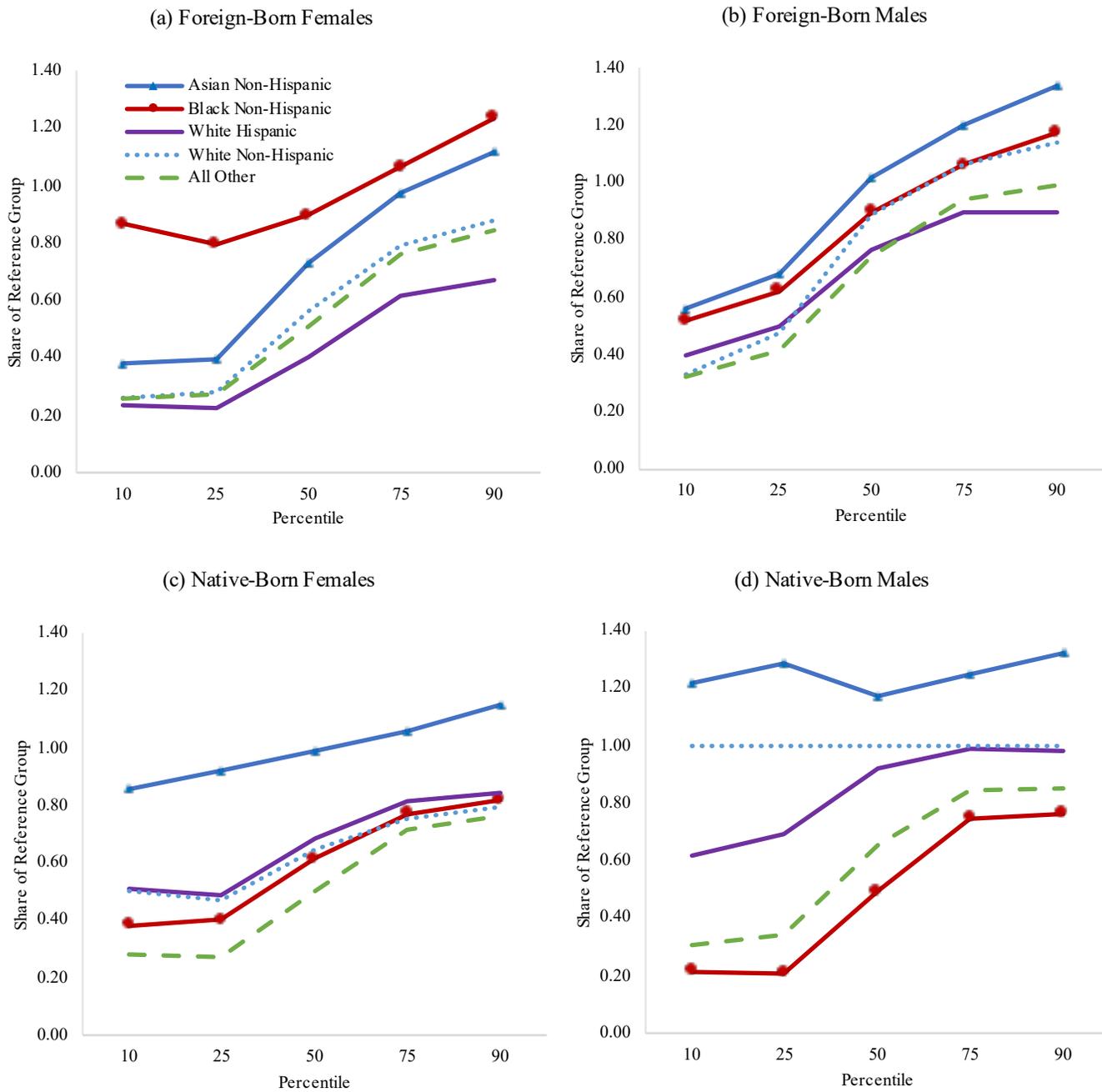

Notes: Calculations are based on the 108,800,000 worker records in sample 2. Each panel is based on the counterfactual log real (2010 PCE) average earnings of demographic group g with the coefficients of the reference group (native-born White Non-Hispanic males). The y-axis shows the share of reference group earnings. The actual share of reference groups earnings is shown in Figure 12. See text for estimation details.

Appendix A: Supplemental Results for the Cross-Country Comparisons

Figure A1: Number of Observations by Sample

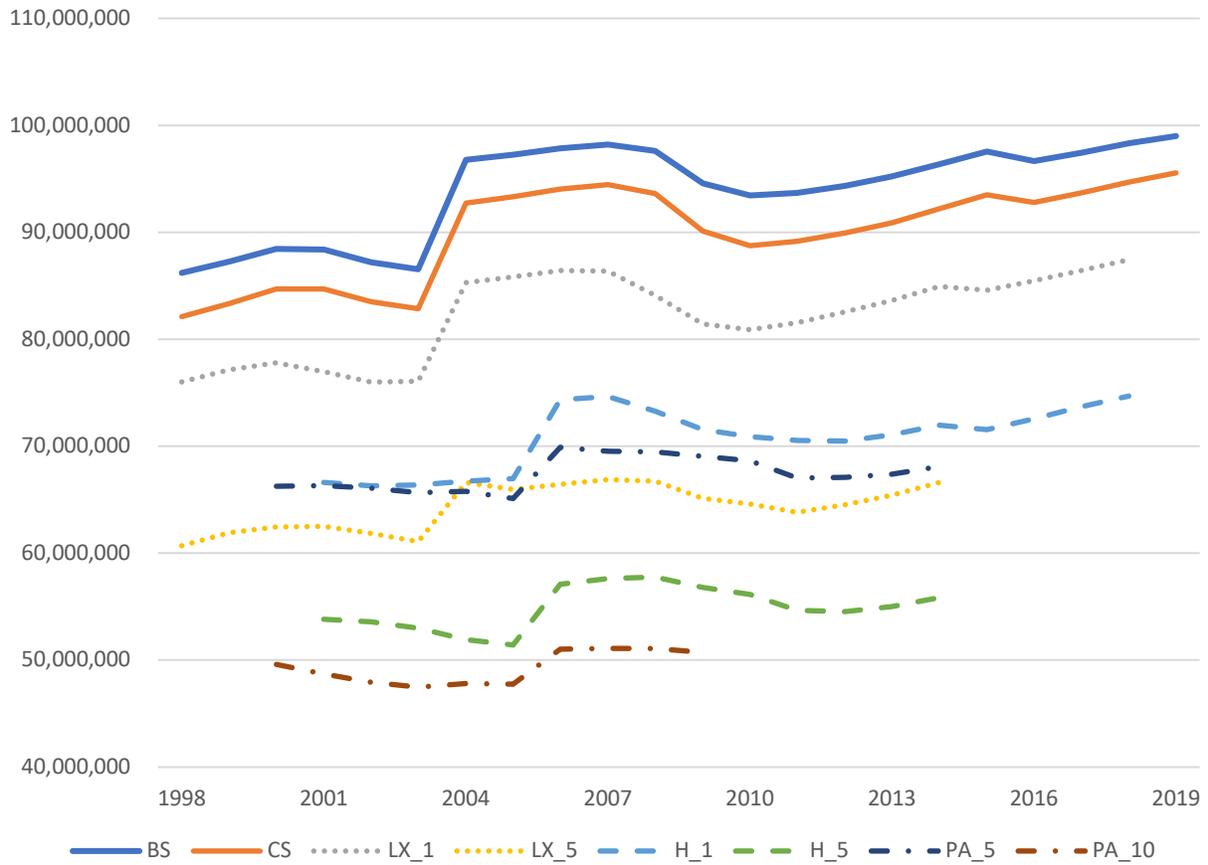

Notes: LEHD person-year annual earnings extract. All analysis samples are constructed from the Base (BS) sample. See table 1A for sample descriptions.

Figure A2: Percent Male by Sample

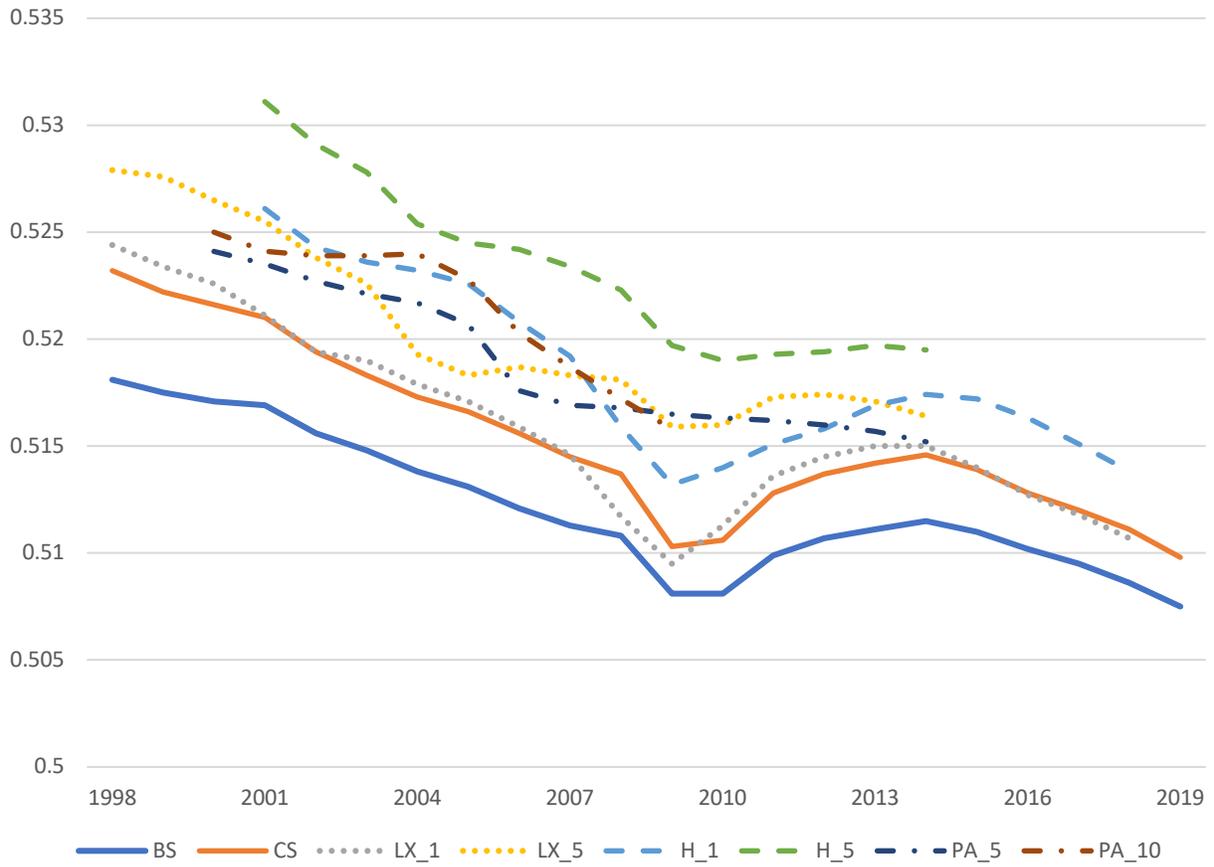

Notes: LEHD person-year annual earnings extract. All analysis samples are constructed from the Base (BS) sample. See table 1A for sample descriptions.

Figure A3: Average Age by Sample

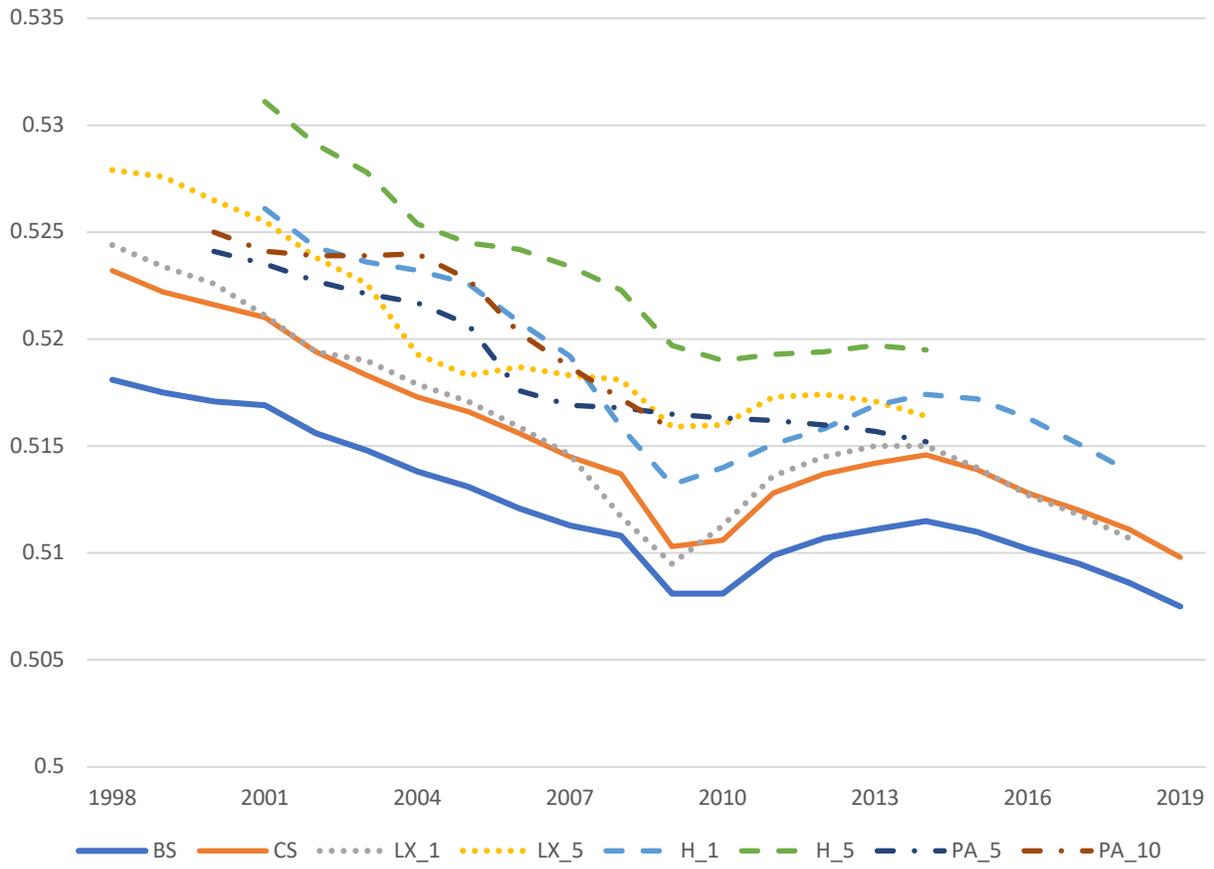

Notes: LEHD person-year annual earnings extract. All analysis samples are constructed from the Base (BS) sample. See table 1A for sample descriptions.

Figure A4: Median Earnings by Sample

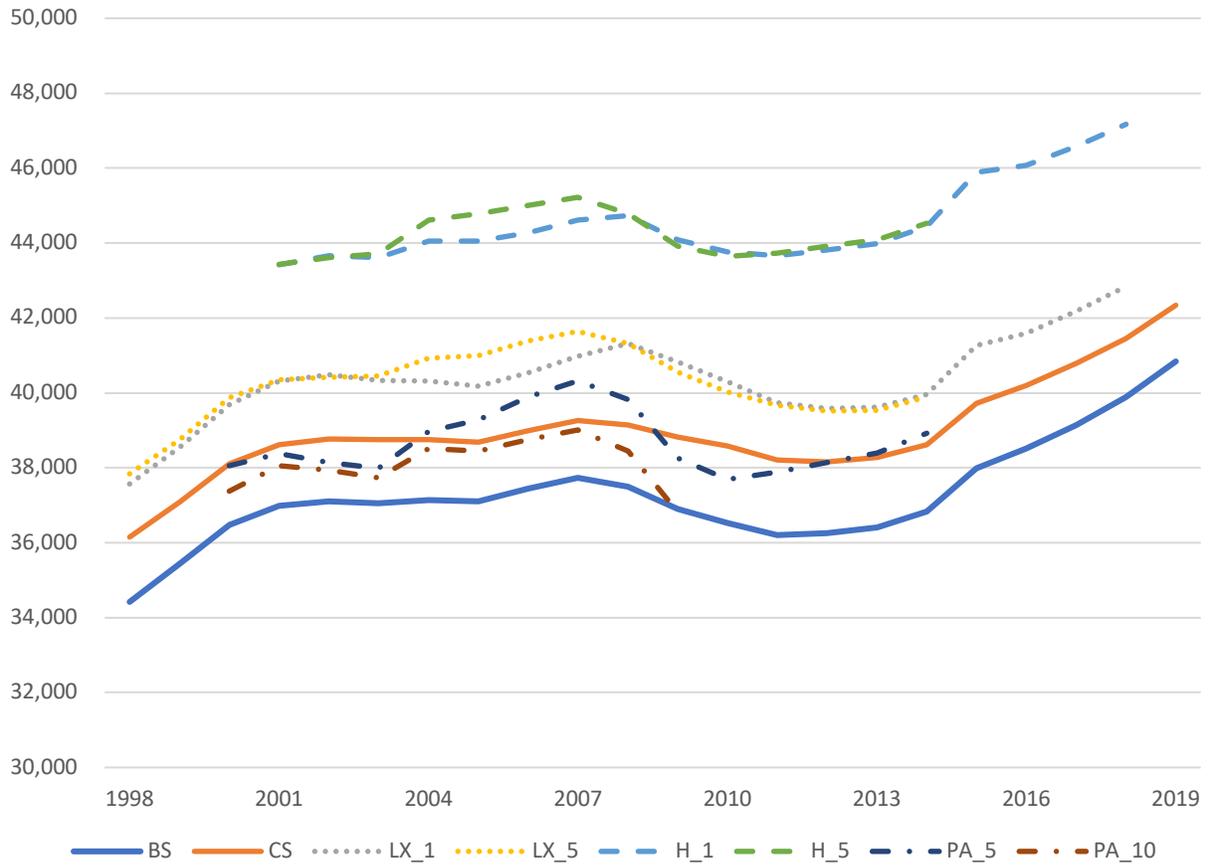

Notes: LEHD workers age 25-55 annual earnings extract. All analysis samples are constructed from the Base (BS) sample. See table 1A for sample descriptions. Median real (2018) annual earnings from all jobs

Figure A5: Distribtion of Earnings (All Workers)

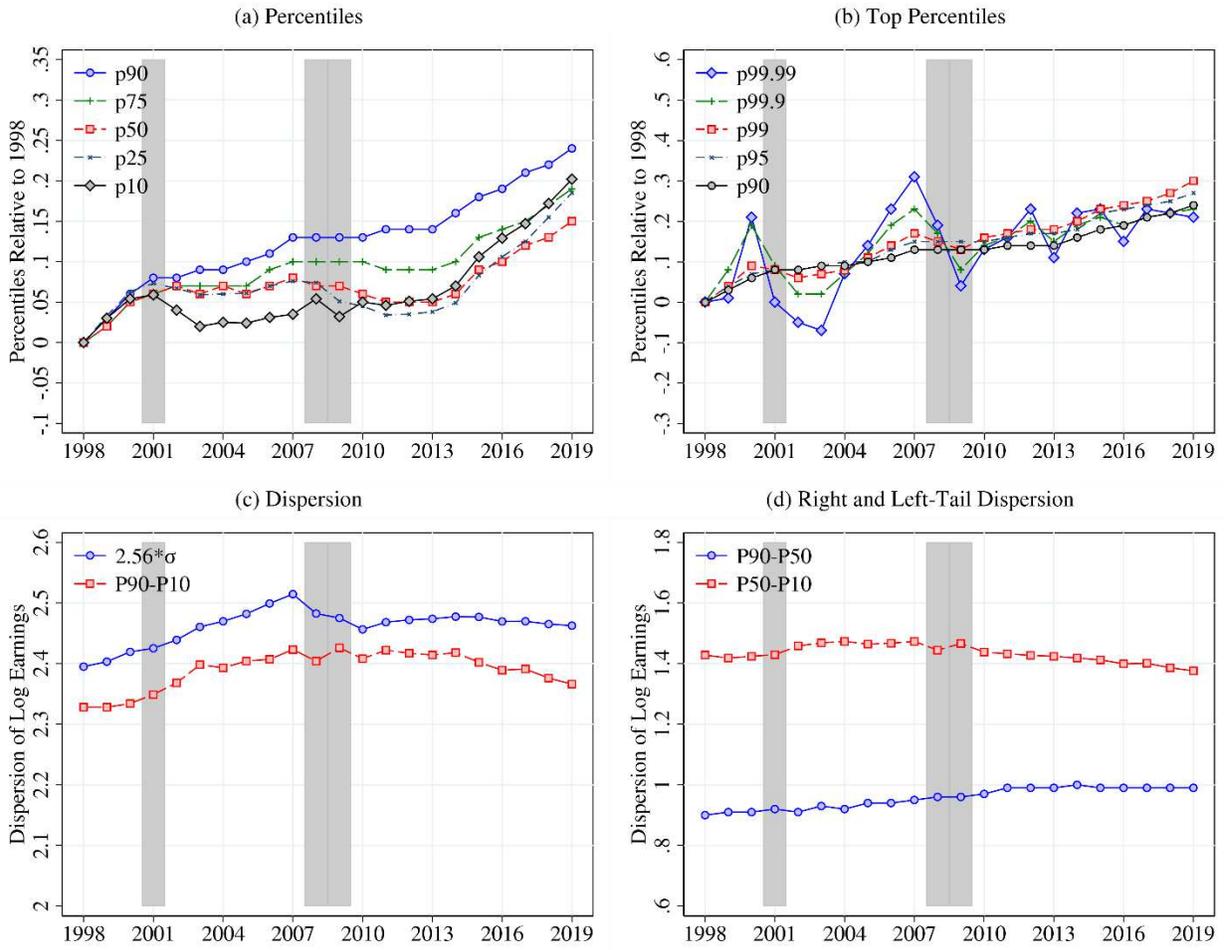

Notes: LEHD CS sample. Shaded areas are recessions. The analysis variable is the log of $y_{it}$. $y_{it}$ must be greater than 260*federal hourly minimum wage. The percentiles relative to 1998 are calculated as $PX_t - PX_{1998}$, where $X$ is the percentile. $2.56 * \sigma$ corresponds to $P90 - P10$ for the normal distribution.

## Figure A6: Distribtion of Residual Log Earnings (Age, All Workers)

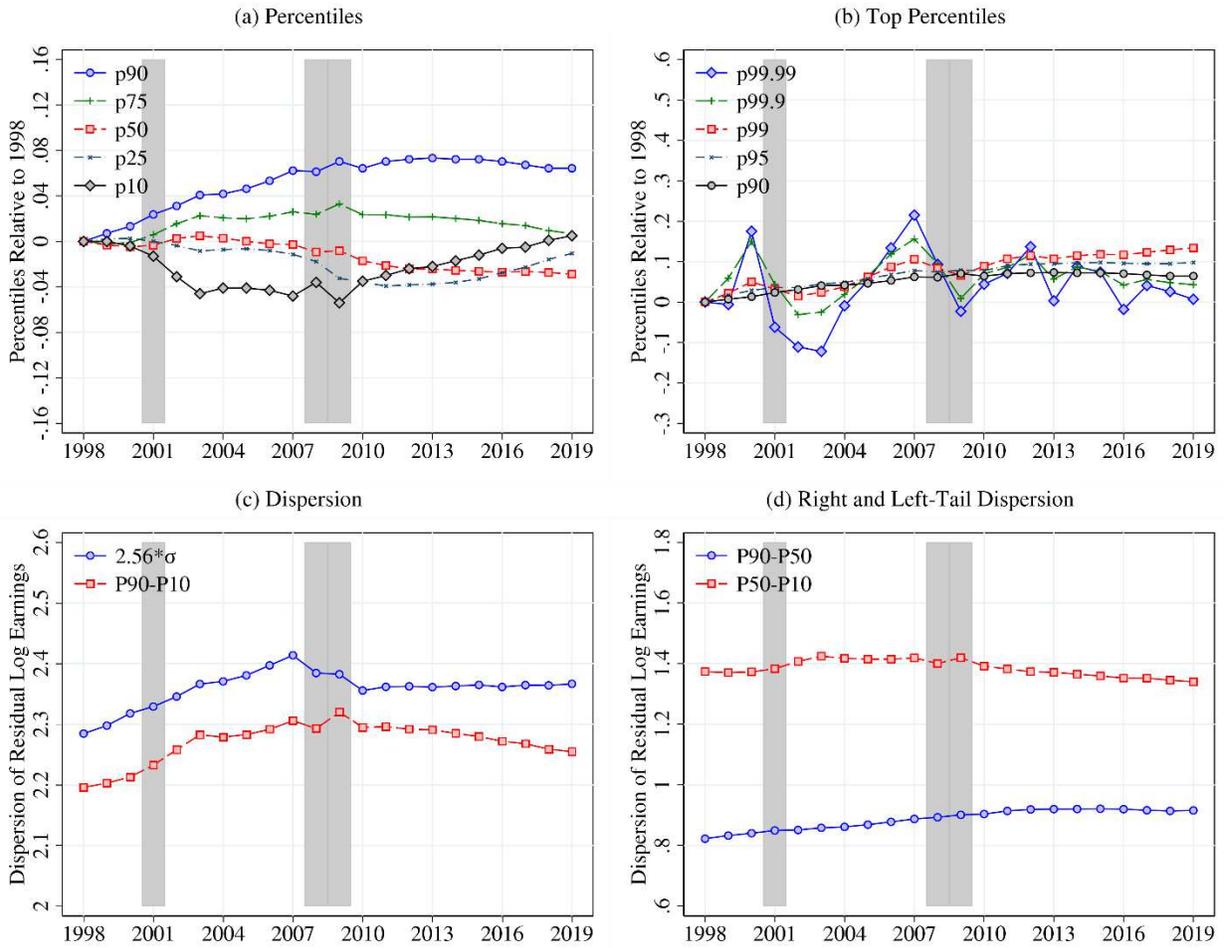

Notes: LEHD CS sample. Shaded areas are recessions. The analysis variable is the residual log earnings $\varepsilon_{it}$. $\varepsilon_{it}$ is the residual from a regression of log $y_{it}$ on a set of age indicator variables by sex and year. $y_{it}$ must be greater than 260*federal hourly minimum wage. The percentiles relative to 1998 are calculated as $PX_t - PX_{1998}$, where $X$ is the percentile. $2.56 * \sigma$ corresponds to $P90 - P10$ for the normal distribution.

Figure A7: Distribtion of Residual Log Earnings (Age and Education, All Workers)

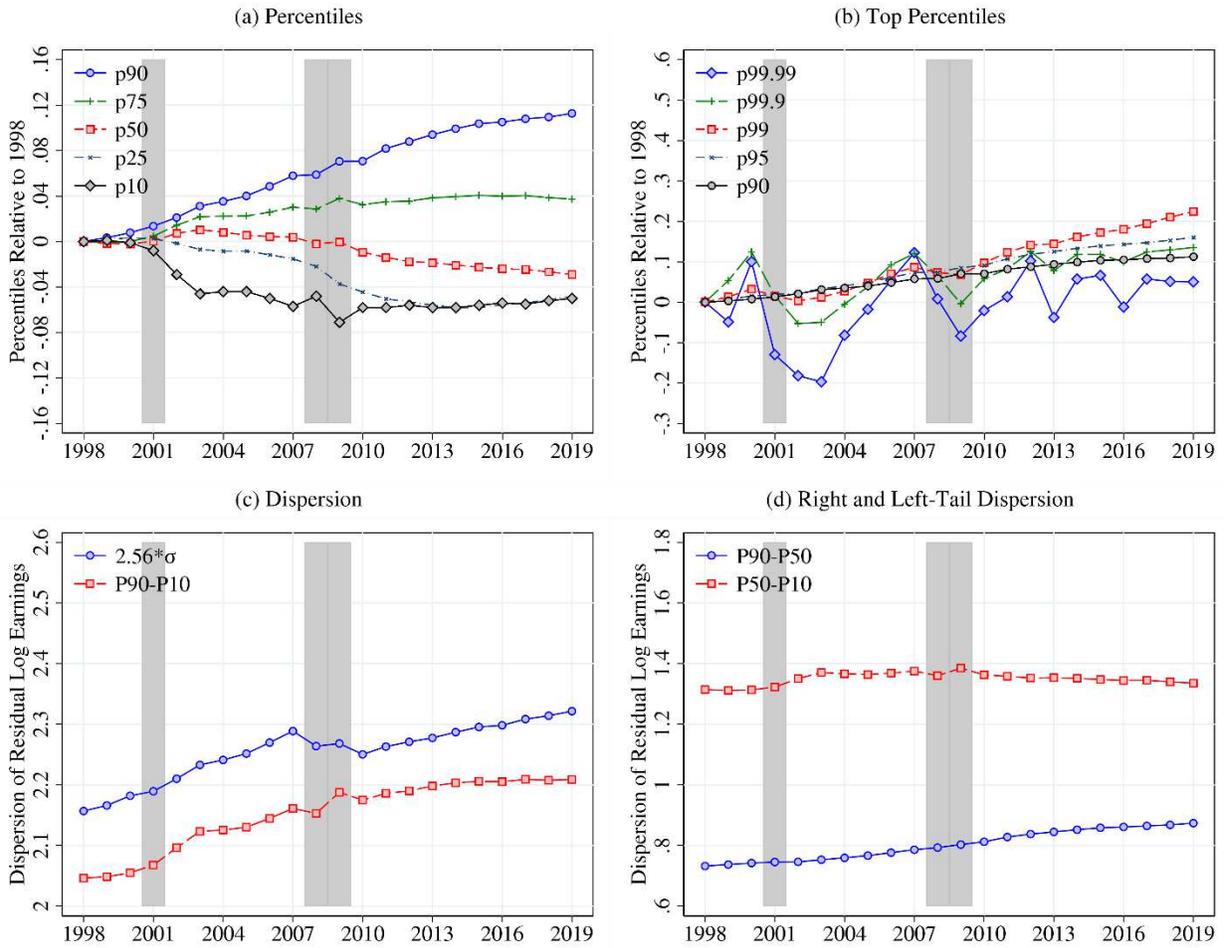

Notes: LEHD CS sample. Shaded areas are recessions. The analysis variable is the residual log earnings $\delta_{it}$. $\delta_{it}$ is the residual from a regression of log $y_{it}$ on a set of age and education indicator variables by sex and year. $y_{it}$ must be greater than 260*federal hourly minimum wage. The percentiles relative to 1998 are calculated as $PX_t - PX_{1998}$, where $X$ is the percentile. $2.56 * \sigma$ corresponds to $P90 - P10$ for the normal distribution.

Figure A8: Top Annual Earnings Inequality: Pareto Tail Log-Log Plot (Top 1%)

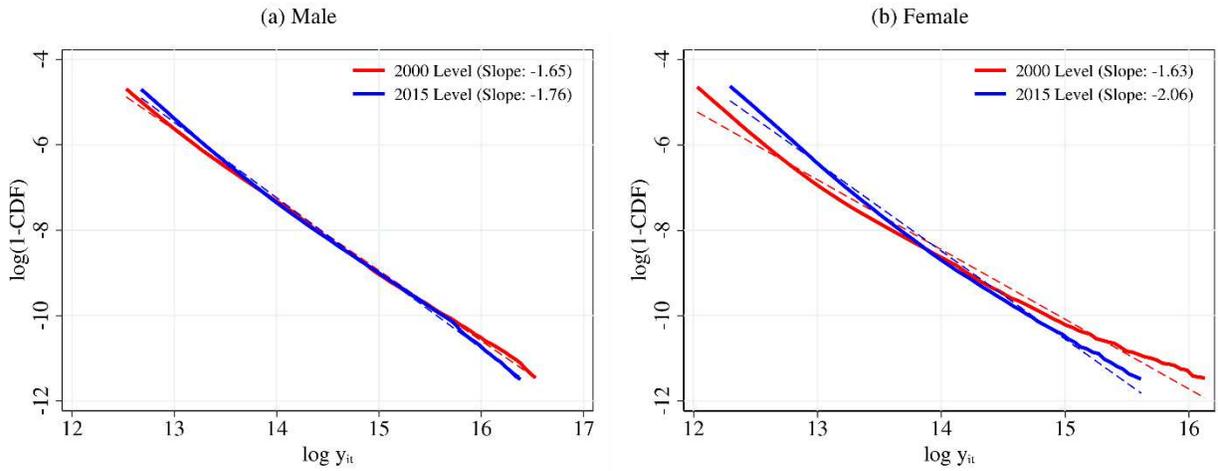

Notes: LEHD CS Sample. The analysis variable is $y_{it}$. $y_{it}$ must be greater than 260*federal hourly minimum wage. The top 0.001% of annual earnings in each year are excluded.

Figure A9: Top Annual Earnings Inequality: Pareto Tail Log-Log Plot (Top 5%)

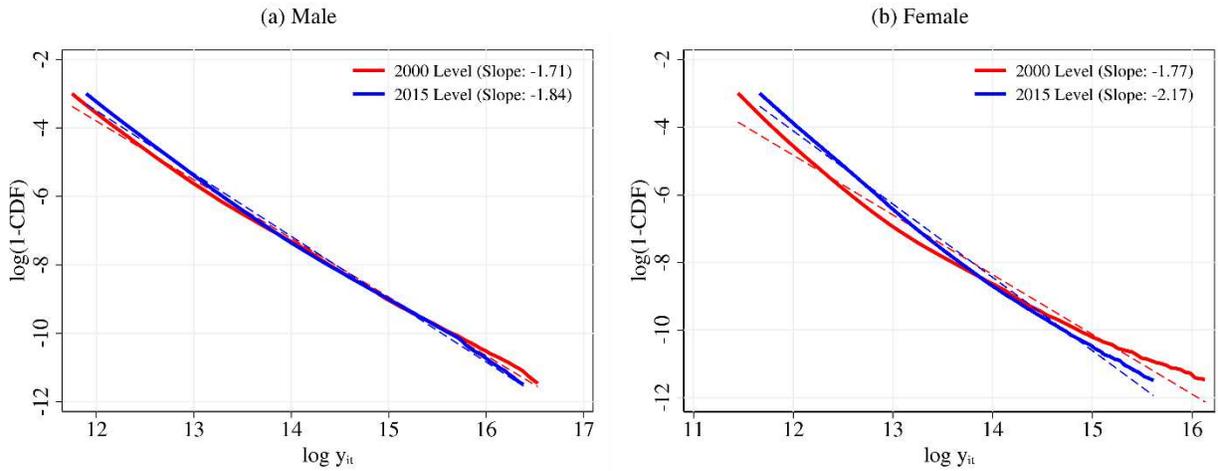

Notes: LEHD CS Sample. The analysis variable is $y_{it}$. $y_{it}$ must be greater than 260*federal hourly minimum wage. The top 0.001% of annual earnings in each year are excluded.

## Figure A10: Earnings Shares

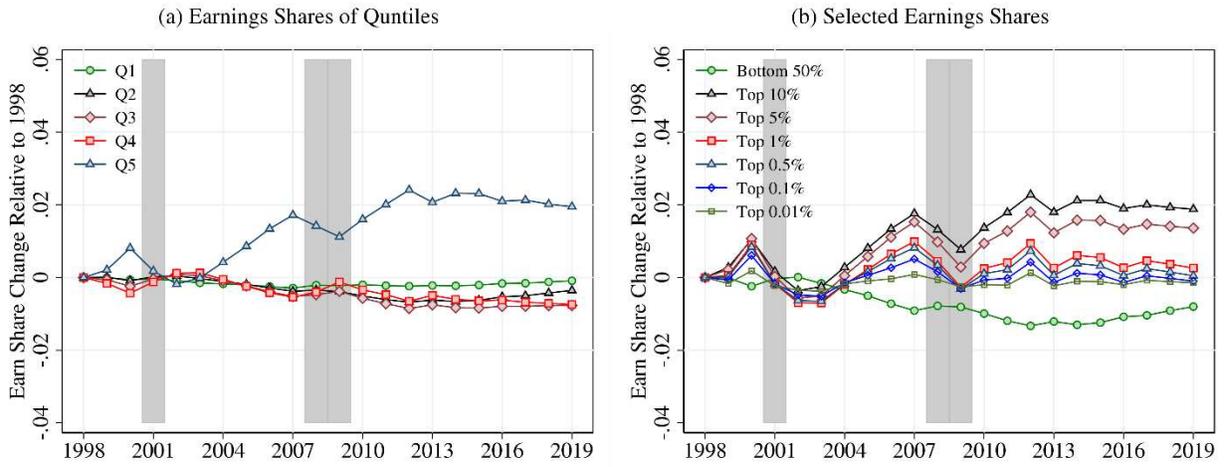

Notes: LEHD CS Sample. Shaded areas are recessions The analysis variable is $y_{it}$. $y_{it}$ must be greater than 260*federal hourly minimum wage. The earnings share change is calculated as $\Delta share = (shareYYYY - share1998)/100$.

Figure A11: Gini Coefficient

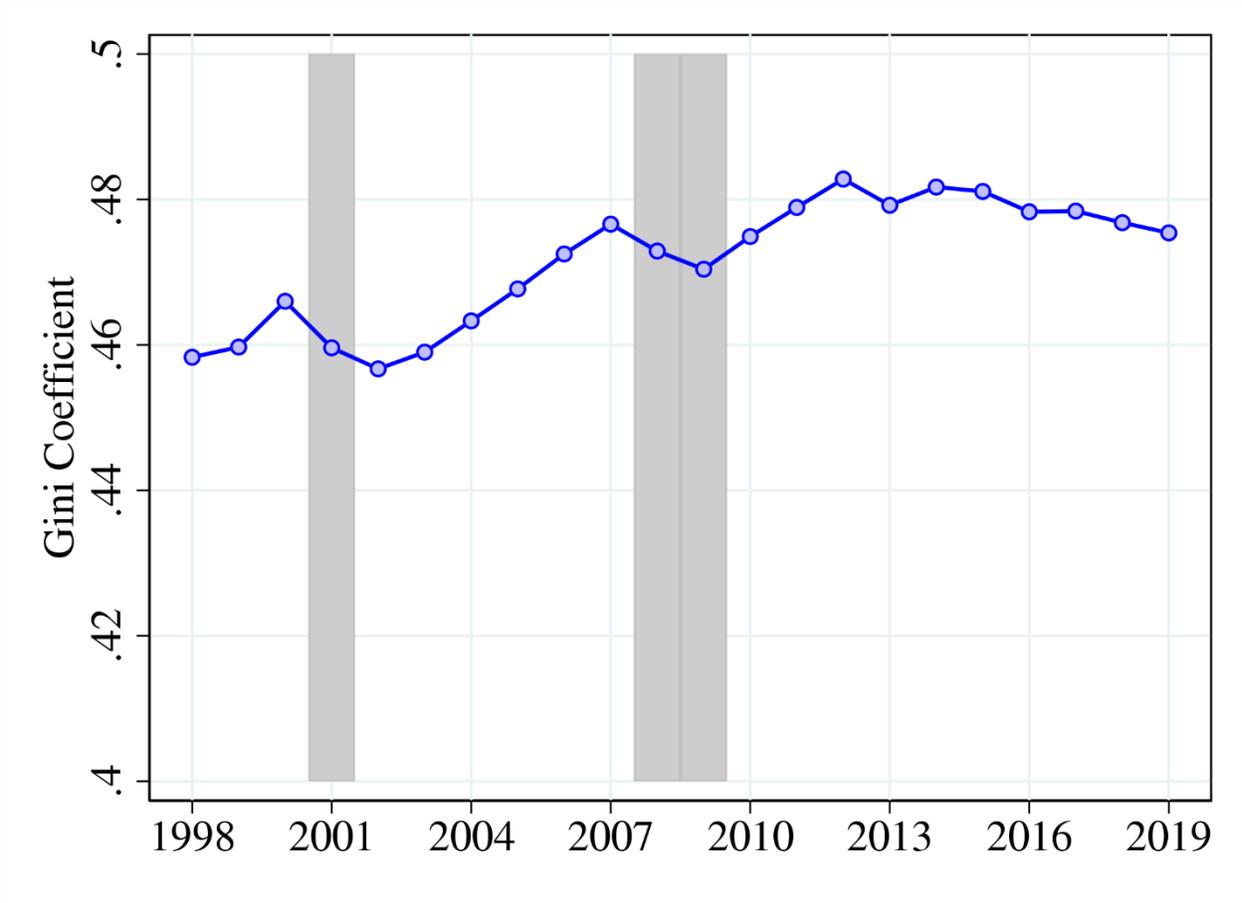

Notes: LEHD CS Sample. Shaded areas are recessions The analysis variable is $y_{it}$. $y_{it}$ must be greater than 260*federal hourly minimum wage.

Figure A12: Dispersion of Five-Year Residual Log Earnings Changes

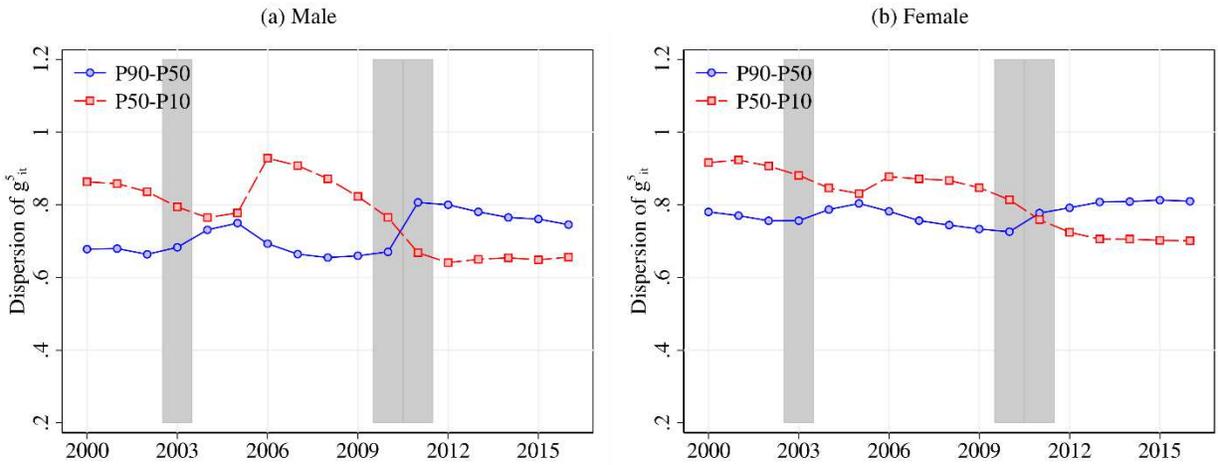

(a) Male          (b) Female

Notes: LEHD LX_5 sample. Shaded areas are recessions. Residual log earnings $g_{it}^5 = \varepsilon_{it+5} - \varepsilon_{it}$. $\varepsilon_{it}$ is the residual from a regression of log $y_{it}$ on a set of age indicator variables by sex and year. $y_{it}$ must be greater than 260*federal hourly minimum wage in $t$ and greater than 1/3*260*federal hourly minimum wage in $t+5$.

Figure A13: Skewness and Kurtosis of Five-Year Residual Log Earnings Changes

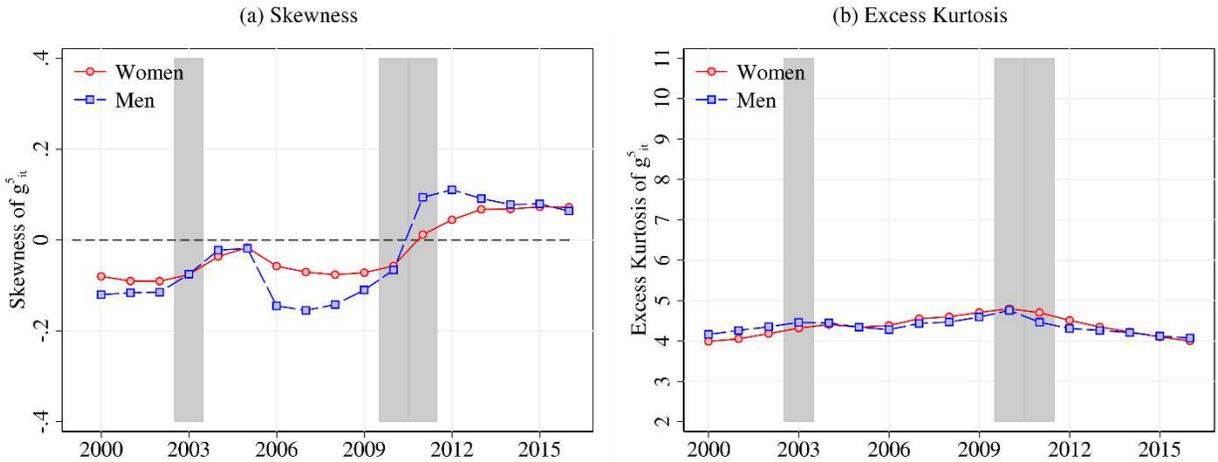

(a) Skewness          (b) Excess Kurtosis

Notes: LEHD LX_5 sample. Shaded areas are recessions. Residual Log Earnings $g_{it}^5 = \varepsilon_{it+5} - \varepsilon_{it}$. $\varepsilon_{it}$ is the residual from a regression of log $y_{it}$ on a set of age indicator variables by sex and year. $y_{it}$ must be greater than 260*federal hourly minimum wage in $t$ and greater than 1/3*260*federal hourly minimum wage in $t+5$. Kelley skewness is $\frac{(P90-P50)-(P50-P10)}{(P90-P10)}$. Excess Crow-Siddiqui kurtosis is $\frac{(P97.5-P2.5)}{(P75-P25)} - 2.91$. 2.91 is the kurtosis value for the normal distribution.

Figure A14: Dispersion, Skewness, and Kurtosis of Five-Year Residual Log Earnings Changes

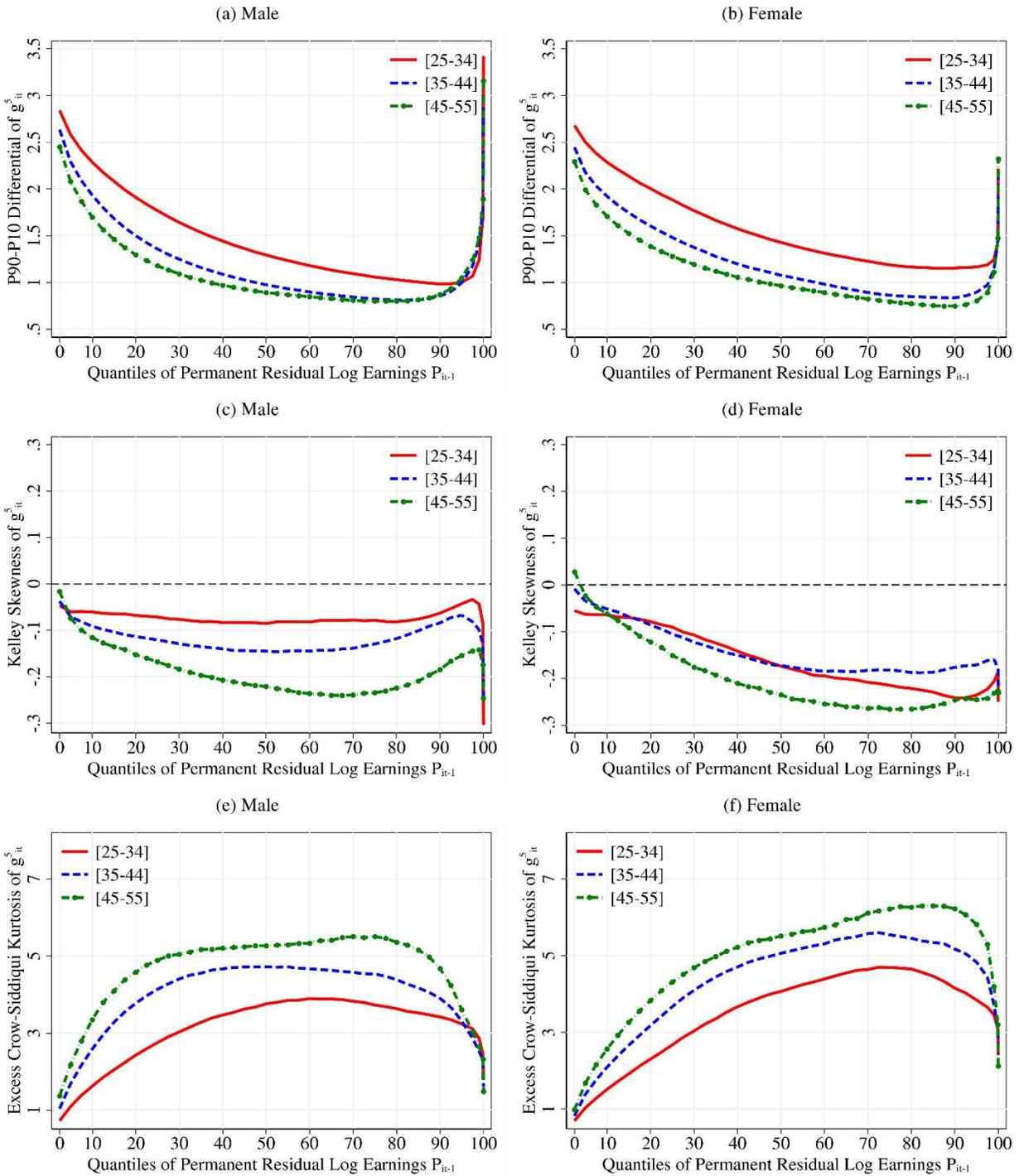

Notes: LEHD H_5 sample. Residual Log Earnings $g_{it}^5 = \varepsilon_{it+5} - \varepsilon_{it}$. $\varepsilon_{it}$ is the residual from a regression of log $y_{it}$ on a set of age indicator variables by sex and year. $y_{it}$ must be greater than 260*federal hourly minimum wage in $t$ and greater than 1/3*260*federal hourly minimum wage in $t + 5$. Kelley skewness is $\frac{(P90-P50)-(P50-P10)}{(P90-P10)}$. Excess Crow-Siddiqui kurtosis is $\frac{(P97.5-P2.5)}{(P75-P25)} - 2.91$. 2.91 is the kurtosis value for the normal distribution. $P_{it-1}$ is a three-year measure of permanent residual log earnings (see the text for more details).

Figure A15: Moments of the One-Year Residual Log Earnings Changes

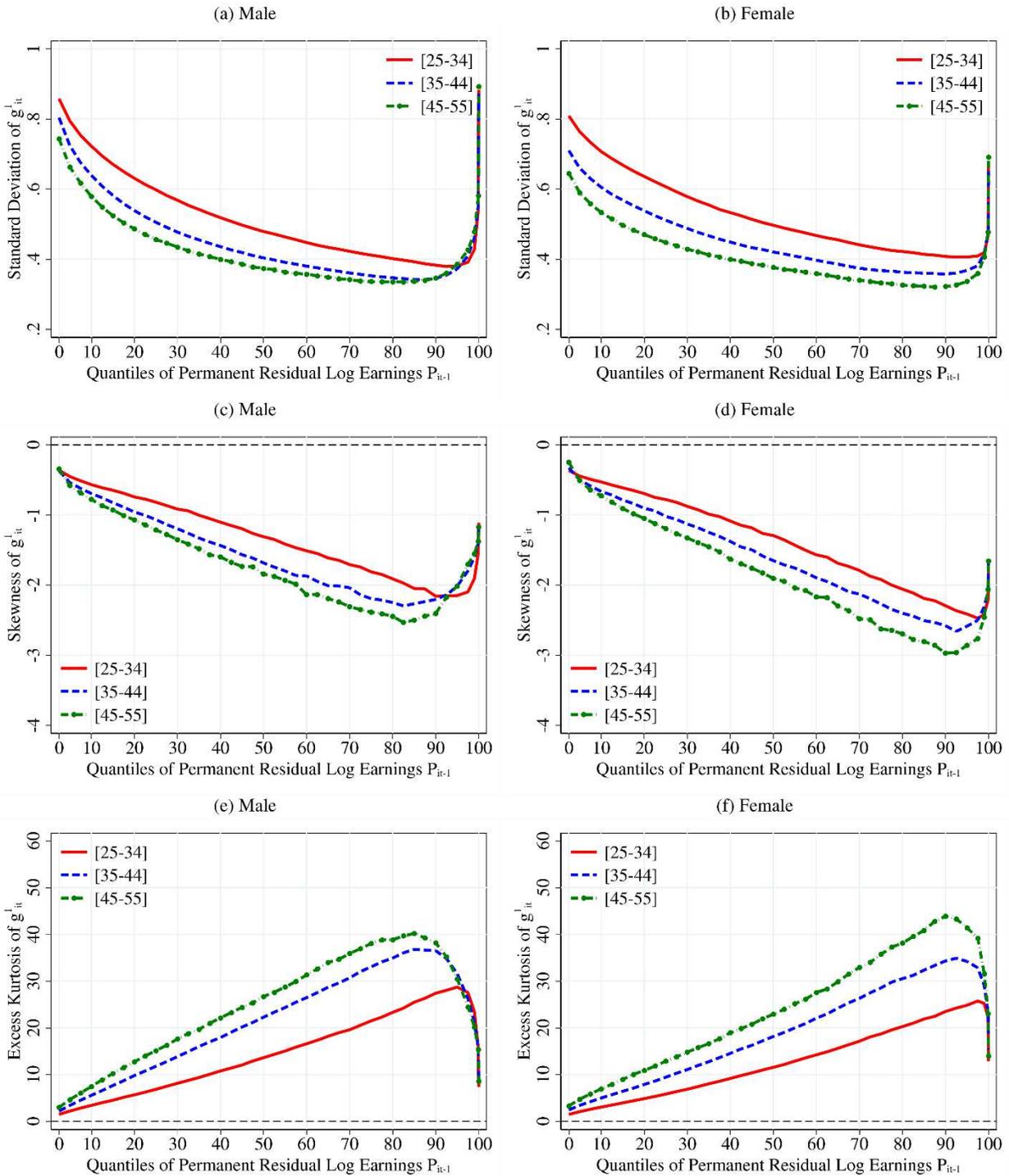

Notes: LEHD H_1 sample (2001-2014). Residual Log Earnings $g_{it}^1 = \varepsilon_{it+1} - \varepsilon_{it}$. $\varepsilon_{it}$ is the residual from a regression of log $y_{it}$ on a set of age indicator variables by sex and year. $y_{it}$ must be greater than 260*federal hourly minimum wage in $t$ and greater than 1/3*260*federal hourly minimum wage in $t + 1$.

Figure A16: Moments of the Five-Year Residual Log Earnings Changes

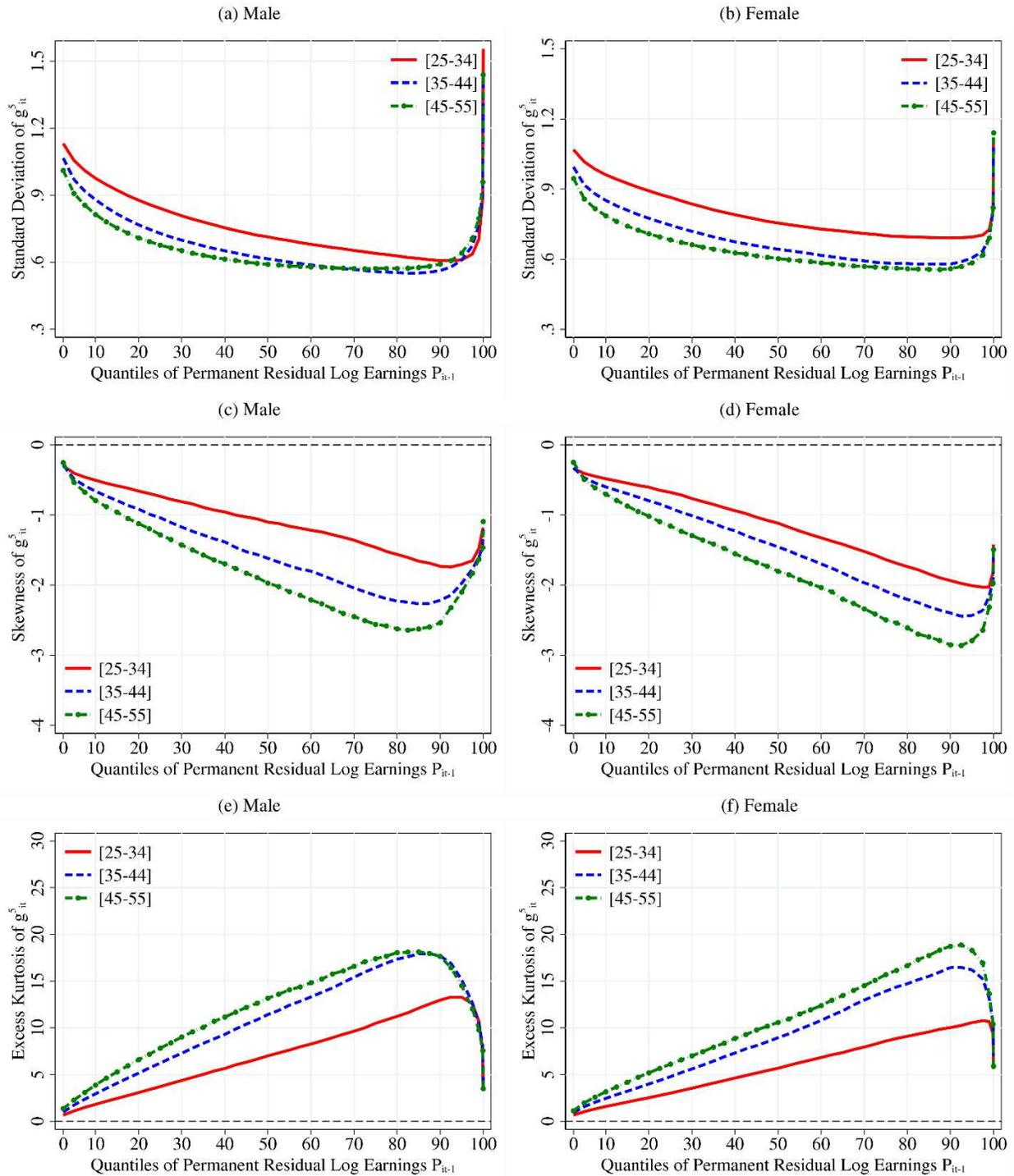

Notes: LEHD H_5 sample. Residual Log Earnings $g_{it}^5 = \varepsilon_{it+5} - \varepsilon_{it}$. $\varepsilon_{it}$ is the residual from a regression of log $y_{it}$ on a set of age indicator variables by sex and year. $y_{it}$ must be greater than 260*federal hourly minimum wage in $t$ and greater than 1/3*260*federal hourly minimum wage in $t + 5$.

## Figure A17: Five-Year Earnings Mobility by Age

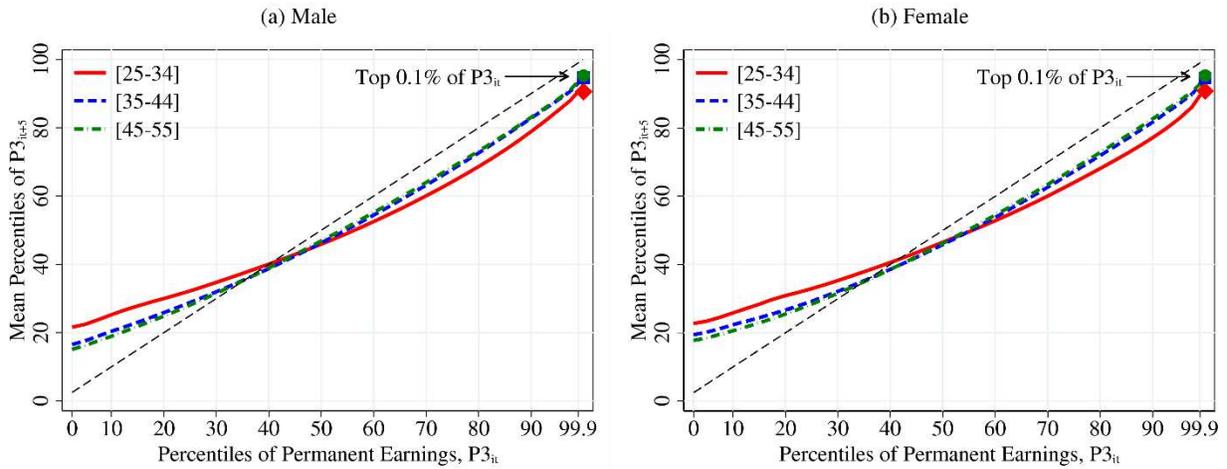

Notes: LEHD PA_5 sample. Permanent earnings $P3_{it} = \frac{(\sum_j e_{ijt-2} + \sum_j e_{ijt-1} + \sum_j e_{ijt})}{3}$. $P3_{it}$ is missing unless $\sum_j e_{ijt} > 260*$federal hourly minimum wage in at least one of the three years. In both $t$ and $t+5$ permanent earnings are ranked (0,100] separately by sex and age. The vertical axis shows the average rank in $t+5$ for workers of a given rank (percentile) in $t$.

## Figure A18: Five-Year Earnings Mobility by Selected Years

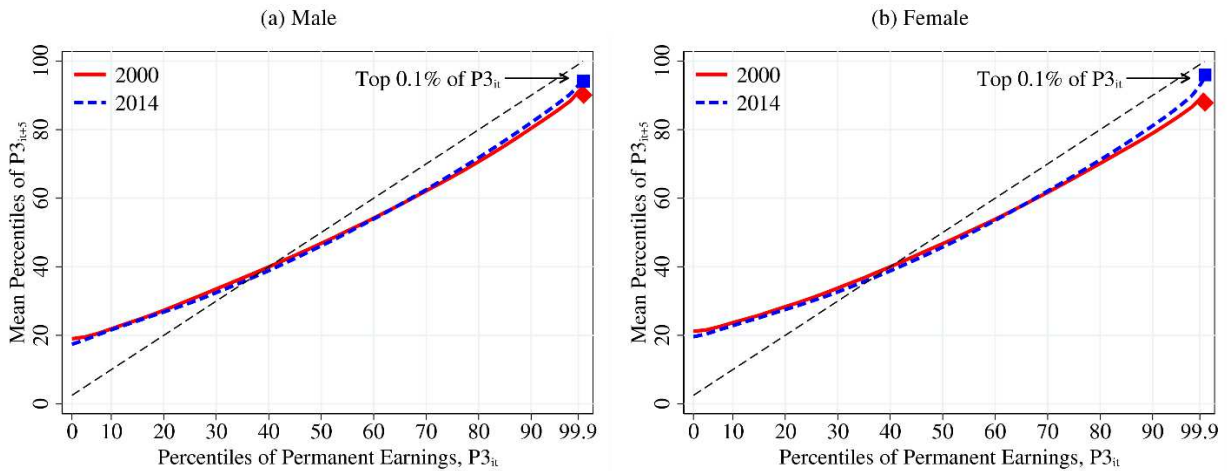

Notes: LEHD PA_5 sample. Permanent earnings $P3_{it} = \frac{(\sum_j e_{ijt-2} + \sum_j e_{ijt-1} + \sum_j e_{ijt})}{3}$. $P3_{it}$ is missing unless $\sum_j e_{ijt} > 260*$federal hourly minimum wage in at least one of the three years. In both $t$ and $t+5$ permanent earnings are ranked (0,100] separately by sex and age. The vertical axis shows the average rank in $t+5$ for workers of a given rank (percentile) in $t$.

## Figure A19: Density of the One-Year Residual Log Earnings Growth

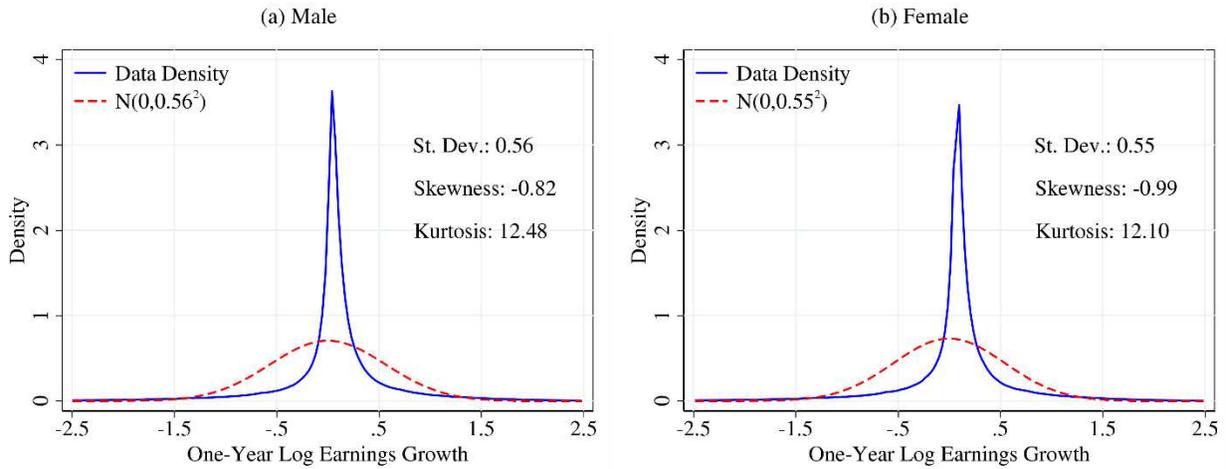

(a) Male                                (b) Female

Notes: LEHD LX_1 sample. Year 2010. Residual log earnings $g_{it}^1 = \varepsilon_{it+1} - \varepsilon_{it}$. $\varepsilon_{it}$ is the residual from a regression of $\log y_{it}$ on a set of age indicator variables by sex and year. $y_{it}$ must be greater than 260*federal hourly minimum wage in $t$ and greater than 1/3*260*federal hourly minimum wage in $t+1$.

## Figure A20: Density of the Five-Year Residual Log Earnings Growth

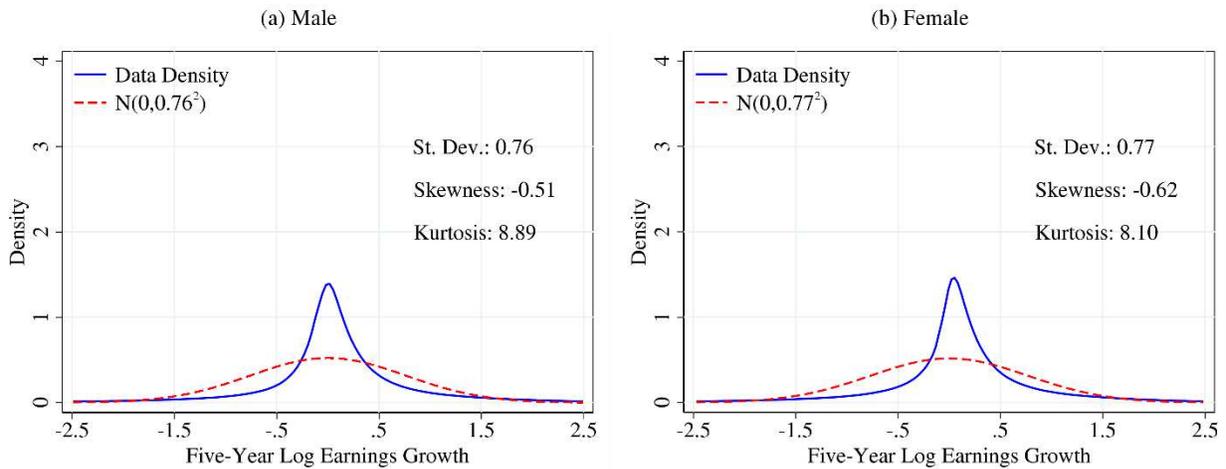

(a) Male                                (b) Female

Notes: LEHD LX_5 sample. Year 2010. Residual log earnings $g_{it}^5 = \varepsilon_{it+5} - \varepsilon_{it}$. $\varepsilon_{it}$ is the residual from a regression of $\log y_{it}$ on a set of age indicator variables by sex and year. $y_{it}$ must be greater than 260*federal hourly minimum wage in $t$ and greater than 1/3*260*federal hourly minimum wage in $t+5$.

Figure A21: Log Density of the One-Year Residual Log Earnings Growth

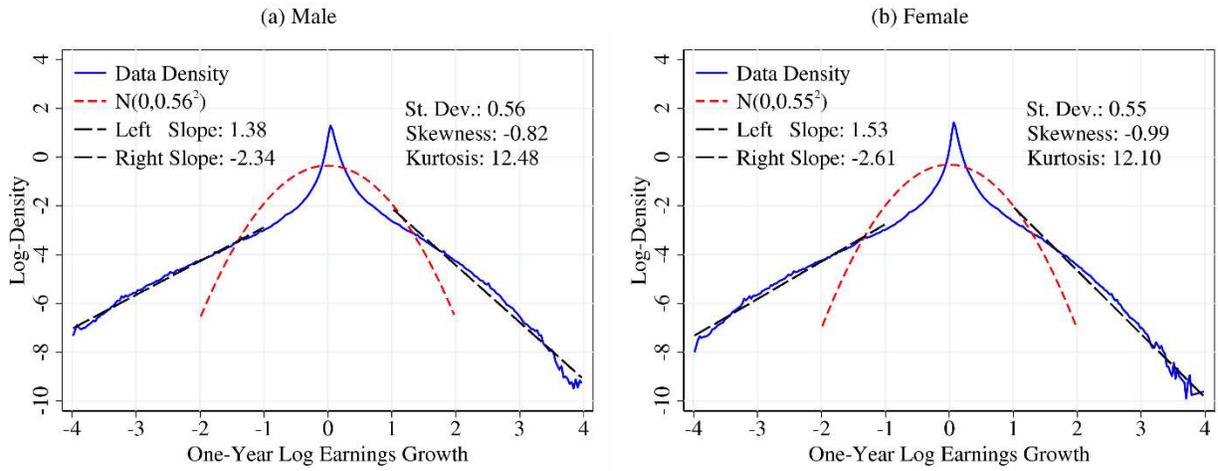

Notes: LEHD LX_1 sample. Year 2010. Residual log earnings $g_{it}^1 = \varepsilon_{it+1} - \varepsilon_{it}$. $\varepsilon_{it}$ is the residual from a regression of log $y_{it}$ on a set of age indicator variables by sex and year. $y_{it}$ must be greater than 260*federal hourly minimum wage in $t$ and greater than 1/3*260*federal hourly minimum wage in $t + 1$.

Figure A22: Log Density of the Five-Year Residual Log Earnings Growth

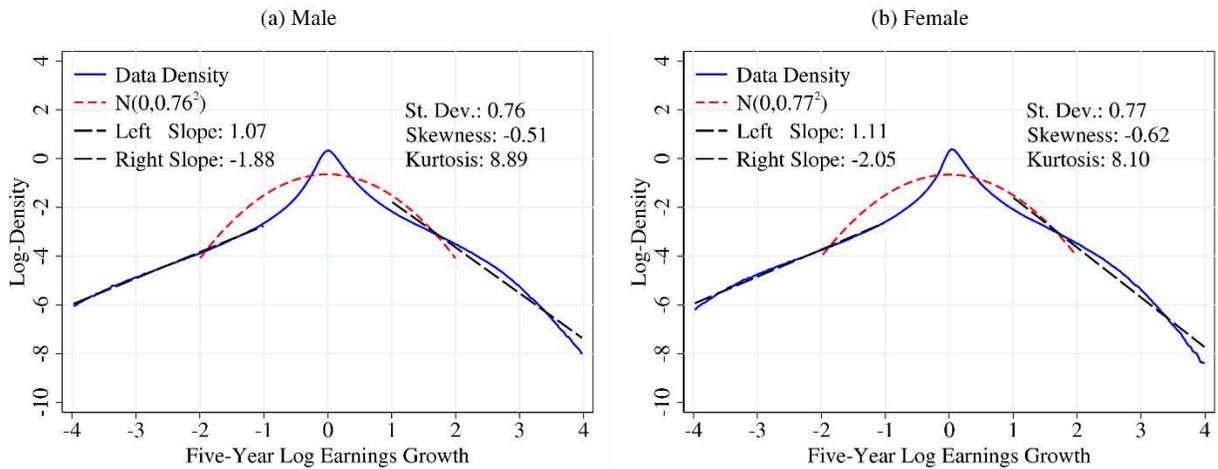

Notes: LEHD LX_5 sample. Year 2010. Residual log earnings $g_{it}^5 = \varepsilon_{it+5} - \varepsilon_{it}$. $\varepsilon_{it}$ is the residual from a regression of log $y_{it}$ on a set of age indicator variables by sex and year. $y_{it}$ must be greater than 260*federal hourly minimum wage in $t$ and greater than 1/3*260*federal hourly minimum wage in $t + 5$.

Appendix B: Supplemental Results for Long-Term Average Earnings

Table B1 - Age by Years in Sample 2

| **Years in Sample 2 / Calendar Year** | | | | | | | | | | | |
|---|---|---|---|---|---|---|---|---|---|---|---|
| 1 | 2 | 3 | 4 | 5 | 6 | 7 | 8 | 9 | 10 | 11 | 12 |
| 2004 | 2005 | 2006 | 2007 | 2008 | 2009 | 2010 | 2011 | 2012 | 2013 | 2014 | 2015 |
| 25 | 26 | 27 | 28 | 29 | 30 | 31 | 32 | 33 | 34 | 35 | 36 |
| 26 | 27 | 28 | 29 | 30 | 31 | 32 | 33 | 34 | 35 | 36 | 37 |
| 27 | 28 | 29 | 30 | 31 | 32 | 33 | 34 | 35 | 36 | 37 | 38 |
| 28 | 29 | 30 | 31 | 32 | 33 | 34 | 35 | 36 | 37 | 38 | 39 |
| 29 | 30 | 31 | 32 | 33 | 34 | 35 | 36 | 37 | 38 | 39 | 40 |
| 30 | 31 | 32 | 33 | 34 | 35 | 36 | 37 | 38 | 39 | 40 | 41 |
| 31 | 32 | 33 | 34 | 35 | 36 | 37 | 38 | 39 | 40 | 41 | 42 |
| 32 | 33 | 34 | 35 | 36 | 37 | 38 | 39 | 40 | 41 | 42 | 43 |
| 33 | 34 | 35 | 36 | 37 | 38 | 39 | 40 | 41 | 42 | 43 | 44 |
| 34 | 35 | 36 | 37 | 38 | 39 | 40 | 41 | 42 | 43 | 44 | 45 |
| 35 | 36 | 37 | 38 | 39 | 40 | 41 | 42 | 43 | 44 | 45 | 46 |
| 36 | 37 | 38 | 39 | 40 | 41 | 42 | 43 | 44 | 45 | 46 | 47 |
| 37 | 38 | 39 | 40 | 41 | 42 | 43 | 44 | 45 | 46 | 47 | 48 |
| 38 | 39 | 40 | 41 | 42 | 43 | 44 | 45 | 46 | 47 | 48 | 49 |
| 39 | 40 | 41 | 42 | 43 | 44 | 45 | 46 | 47 | 48 | 49 | 50 |
| 40 | 41 | 42 | 43 | 44 | 45 | 46 | 47 | 48 | 49 | 50 | 51 |
| 41 | 42 | 43 | 44 | 45 | 46 | 47 | 48 | 49 | 50 | 51 | 52 |
| 42 | 43 | 44 | 45 | 46 | 47 | 48 | 49 | 50 | 51 | 52 | 53 |
| 43 | 44 | 45 | 46 | 47 | 48 | 49 | 50 | 51 | 52 | 53 | 54 |
| 44 | 45 | 46 | 47 | 48 | 49 | 50 | 51 | 52 | 53 | 54 | 55 |
| 45 | 46 | 47 | 48 | 49 | 50 | 51 | 52 | 53 | 54 | 55 | 56 |
| 46 | 47 | 48 | 49 | 50 | 51 | 52 | 53 | 54 | 55 | 56 | 57 |
| 47 | 48 | 49 | 50 | 51 | 52 | 53 | 54 | 55 | 56 | 57 | 58 |
| 48 | 49 | 50 | 51 | 52 | 53 | 54 | 55 | 56 | 57 | 58 | 59 |
| 49 | 50 | 51 | 52 | 53 | 54 | 55 | 56 | 57 | 58 | 59 | 60 |
| 50 | 51 | 52 | 53 | 54 | 55 | 56 | 57 | 58 | 59 | 60 | 61 |
| 51 | 52 | 53 | 54 | 55 | 56 | 57 | 58 | 59 | 60 | 61 | 62 |
| 52 | 53 | 54 | 55 | 56 | 57 | 58 | 59 | 60 | 61 | 62 | 63 |
| 53 | 54 | 55 | 56 | 57 | 58 | 59 | 60 | 61 | 62 | 63 | 64 |
| 54 | 55 | 56 | 57 | 58 | 59 | 60 | 61 | 62 | 63 | 64 | 65 |

## Table B2: Geography Division and Industry Definitions

| Census Geography Divisions | | |
|---|---|---|
| Number | Name | States |
| 1 | New England | CT,ME,MA,NH,RI,VT |
| 2 | Middle Atlantic | NJ,NY,PA |
| 3 | East North Central | IN,IL,MI,OH,WI |
| 4 | West North Central | IA,KS,MN,MO,NE,ND,SD |
| 5 | South Atlantic | DL,DC,FL,GA,MD,NC,SC,VA,WV |
| 6 | East South Central | AL,KY,MS,TN |
| 7 | West South Central | AR,LA,OK,TX |
| 8 | Mountain | AZ,CO,ID,NM,MT,UT,NV,WY |
| 9 | Pacific | AK,CA,HI,OR,WA |

| Industry Sectors | | |
|---|---|---|
| Abbreviation | NAICS 2017 Code | Name |
| A | 11 | Agriculture, Forestry, Fishing and Hunting |
| B | 21 | Mining, Quarrying, and Oil and Gas Extraction |
| C | 22 | Utilities |
| D | 23 | Construction |
| E | 31-33 | Manufacturing |
| F | 42 | Wholesale Trade |
| G | 44-45 | Retail Trade |
| H | 48-49 | Transportation and Warehousing |
| I | 51 | Information |
| J | 52 | Finance and Insurance |
| K | 53 | Real Estate and Rental and Leasing |
| L | 54 | Professional, Scientific, and Technical Services |
| M | 55 | Management of Companies and Enterprises |
| N | 56 | Administrative and Support and Waste Management and Remediation Services |
| O | 61 | Educational Services |
| P | 62 | Health Care and Social Assistance |
| Q | 71 | Arts, Entertainment, and Recreation |
| R | 72 | Accomotation and Food Services |
| S | 81 | Other Services (exc. Public Administration) |
| T | 92 | Public Administration |

Figure B1: Census Geography Regions and Divisions

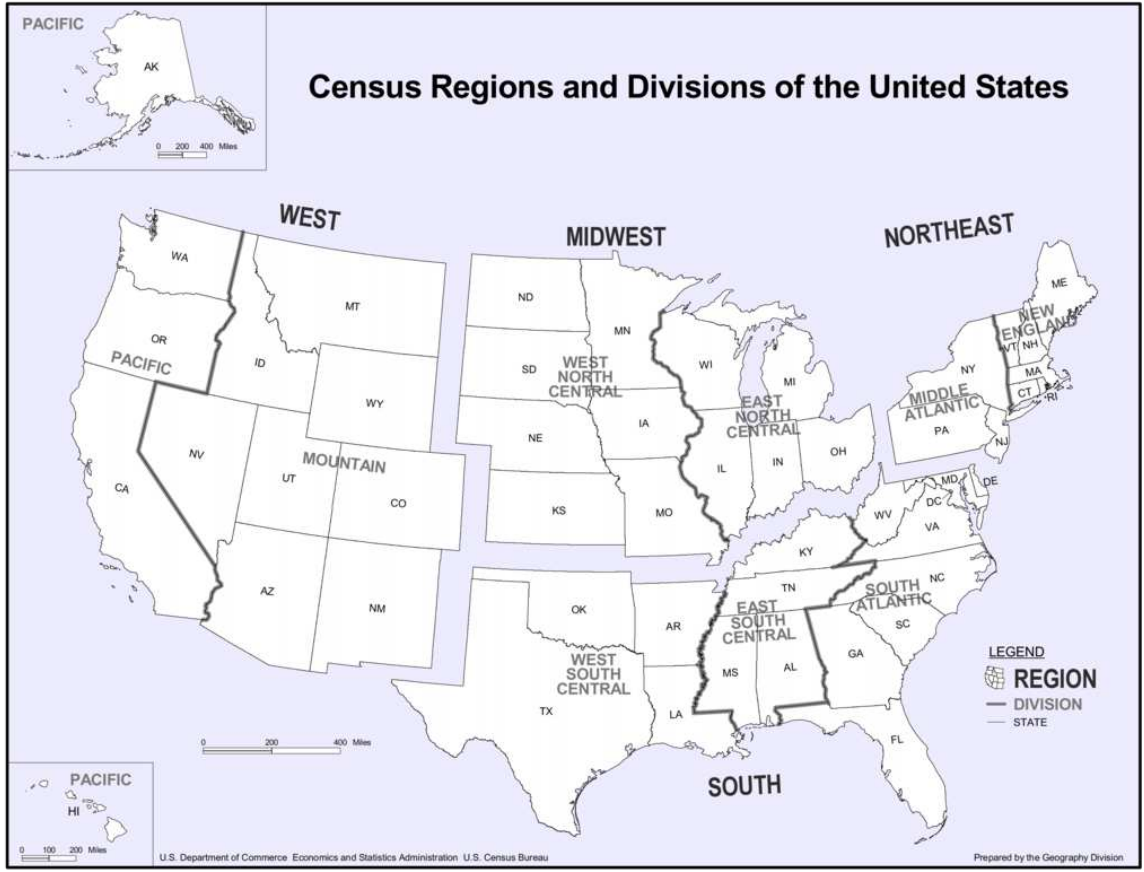